\documentclass[fleqn,10pt]{wlscirep}

\usepackage[utf8]{inputenc}
\usepackage{graphicx}
\usepackage{times}
\usepackage{soul}
\usepackage{url}
\usepackage{caption}
\usepackage{amsmath}
\usepackage{amsthm}
\usepackage{booktabs}
\usepackage{algorithm}
\usepackage{algorithmic}
\usepackage[switch]{lineno}
\usepackage{verbatim,wrapfig}
\usepackage{multirow}

\usepackage[T1]{fontenc}
\usepackage{microtype}
\usepackage{cite}

\usepackage{placeins}

\usepackage{float}
\title{A vision foundation model for single-cell biology via spatial gene cartography}

\author[1]{Ridvan Yesiloglu}
\author[2]{Sakib Mostafa}
\author[1,3,4]{James Zou}
\author[5]{Ash Alizadeh}
\author[4]{Jiajun Wu}
\author[1,2,6]{Lei Xing}
\author[3,4,7,*]{Ehsan Adeli}
\author[2,*]{Md Tauhidul Islam}
\affil[1]{Department of Electrical Engineering, Stanford University, Stanford, California, USA}
\affil[2]{Department of Radiation Oncology, Stanford University, Stanford, California, USA}
\affil[3]{Department of Biomedical Data Science, Stanford University School of Medicine, Stanford, California, USA}
\affil[4]{Department of Computer Science, Stanford University, Stanford, California, USA}
\affil[5]{Department of Medicine, Division of Oncology, Stanford University, Stanford, California, USA}
\affil[6]{Institute of Computational and Mathematical Engineering, Stanford University, Stanford, California, USA}
\affil[7]{Department of Psychiatry and Behavioral Sciences, Stanford University, Stanford, California, USA}
\affil[*]{correspondence: tauhid@stanford.edu and eadeli@stanford.edu}

\keywords{single-cell transcriptomics, foundation model, vision transformer, optimal transport, gene co-expression}

\begin{abstract}
\noindent
Single-cell transcriptomics has made it possible to measure gene expression in tens of millions of individual cells, revealing cellular diversity that bulk profiling cannot resolve. Foundation models aim to learn general representations from these large datasets that can be reused across many biological tasks.  However, most current single-cell foundation models are adapted from language models and represent each cell as a set or sequence of gene tokens. This design has two limitations. It treats genes as largely unordered inputs, even though genes act together in coordinated programs, and it often requires expression values to be discretized or ranked, losing quantitative information about expression magnitude.  Here we present scVision, a vision foundation model for single-cell biology. Instead of converting genes into tokens, scVision represents each cell as a continuous gene-expression image. It assigns genes to fixed spatial positions using optimal transport, so that genes with related expression patterns are placed near one another and coordinated gene programs form local image regions. The resulting image preserves both the quantitative expression level of each gene and the biological relationships among genes. We pretrain a vision transformer with masked image modelling on 72 million human cells, creating one of the largest pretrained models for single-cell analysis. In zero-shot evaluations across six independent, held-out studies, frozen scVision representations outperform existing foundation models and classical baselines in cell-type annotation and gene-program discovery, without task-specific retraining. On multi-study integration, scVision matches the strongest token-based foundation model on the combined benchmark score and conserves more biological structure than any method tested. The spatial organization of scVision also improves interpretability: image regions correspond to groups of co-expressed genes, and attention maps can be read as gene-program activity. This structure also enables spatial masking experiments, in which a neighborhood of related genes is perturbed as a single unit, an operation with no direct counterpart in token-based foundation models. By preserving continuous gene-expression values and giving genes biologically meaningful positions, scVision reframes single-cell representation learning as a vision problem. This approach retains more of the original transcriptomic signal while opening a direct path for applying modern computer vision methods to single-cell biology.
\end{abstract}

\begin{document}

\flushbottom
\maketitle

\thispagestyle{empty}

\begin{figure*}[t]
\centering
\includegraphics[width=\linewidth]{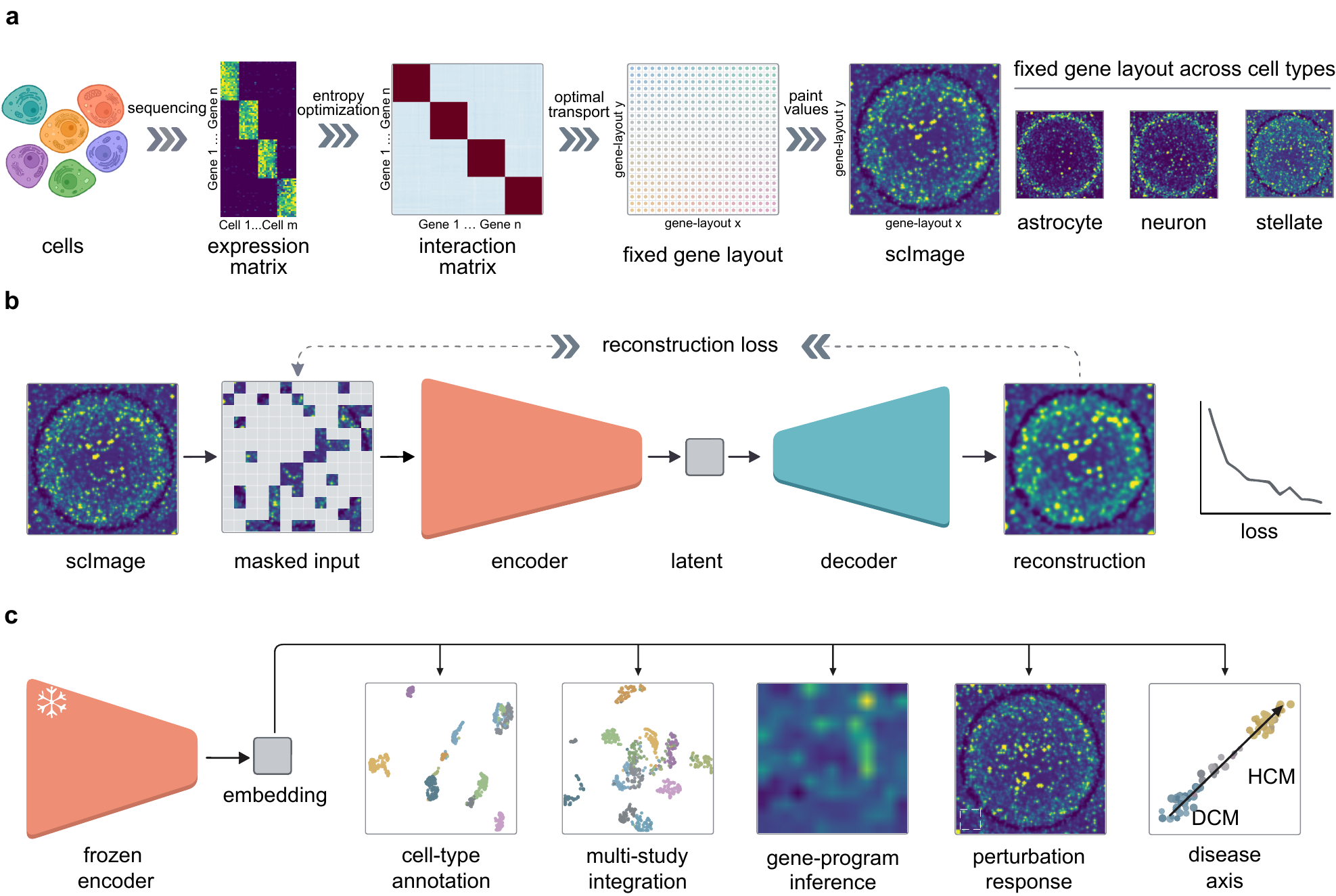}
\caption{\textbf{scVision renders each cell as an image and learns one frozen representation by masked image modelling.}
\textbf{a}, Construction of the scImage. The 10,816 most informative genes define a gene-gene co-expression matrix, and Gromov--Wasserstein optimal transport places the genes on a fixed $104\times104$ lattice so that co-expressed genes become spatial neighbours. The layout is computed once, pan-tissue, and shared across all cells; painting a single cell's expression onto it renders the transcriptome as one image, in which co-regulation appears as local texture and cellular identity as global pattern.
\textbf{b}, Pretraining. A ViT-base encoder (approximately 86 million parameters) and a lightweight decoder are trained by masked image modelling, reconstructing the 75\% of scImage patches hidden at the input, on 72 million of a roughly 94 million human-cell dataset.
\textbf{c}, From the single frozen encoder, a cell embedding supports, with no fine-tuning, cell-type annotation, multi-study integration, gene-program inference read from attention, perturbation response probed by masking a gene neighbourhood (64 co-regulated genes at once, one $8\times8$ patch), and a disease axis that transfers across assay and subtype, from dilated to hypertrophic cardiomyopathy.}
\label{fig:overview}
\end{figure*}

\section*{Introduction}

Single-cell RNA sequencing (scRNA-seq) has changed the scale at which biology can be studied. Instead of measuring average gene expression across mixed tissues, it now allows individual cells to be profiled across organs, developmental stages, disease states, treatments and patient populations\cite{tanay2017scaling,macosko2015dropseq,klein2015indrops,cao2019organogenesis,tabulamuris2018,tabulasapiens2022}. This shift makes it possible to examine biology at the level where important events first arise. Rare immune populations expand within a tissue. Progenitor cells move through intermediate states toward differentiation. Drug-resistant malignant states emerge before they are detectable clinically\cite{tirosh2016melanoma}. Large cellular atlases built using scRNA-seq serve as reference maps of human biology\cite{sikkema2023hlca,eraslan2022crosstissue}. They can define the molecular identity of cell types across tissues, reveal rare or previously unrecognized cell populations, reconstruct developmental trajectories, and map how cells change during ageing, infection, inflammation, cancer and treatment response\cite{litvinukova2020heart,elmentaite2021gut,tadross2025hypomap}. In disease, such atlases can uncover pathogenic immune and stromal states, identify cell-type-specific gene programs, distinguish primary disease mechanisms from secondary tissue responses, and show how abnormal cells interact with their microenvironment\cite{stephenson2021covid,combat2022covid,chaffin2022cardiomyopathy,reichart2022cardiomyopathies}. In translational settings, they provide a framework for comparing patient samples with healthy references, discovering biomarkers, prioritizing therapeutic targets, and understanding why only some cells or patients respond to a given intervention\cite{yazar2022onek1k}. As these atlases grow, they offer the possibility of building computational models that learn general principles of cellular organization from millions of cells and then apply that knowledge to new biological problems. Such foundation models could help annotate newly generated datasets, harmonize studies collected across different laboratories and technologies, detect rare or transitional cell states, identify disease-associated programs in small patient cohorts, and predict how cells respond to genetic perturbations, drugs or environmental signals. 

 Most current single-cell foundation models borrow their design from natural-language processing. In these models, a cell is treated like a sentence and genes are treated like words. Models such as scBERT, Geneformer, scGPT, scFoundation and Universal Cell Embeddings are trained by masking or predicting gene tokens across large collections of cells\cite{yang2022scbert,theodoris2023geneformer,cui2024scgpt,hao2024scfoundation,rosen2023uce}. These approaches have shown that self-supervised learning can be scaled to single-cell data and have established foundation modelling as an important direction for the field. However, independent evaluations have also raised important questions. In common benchmarks such as cell-type classification, several token-based models do not consistently outperform simple regularized linear models; in zero-shot settings, their embeddings can underperform standard dimensionality-reduction methods for clustering or integration; and across multi-task evaluations, they do not always exceed task-specific tools\cite{boiarsky2023,kedzierska2023,liu2023sceval}. A key limitation of these token based models lies in how the transcriptome is represented. These models make two simplifying assumptions. First, they usually do not give genes a biologically meaningful arrangement. A transcriptome is represented as a set or sequence of gene tokens, but genes that participate in the same pathway, regulon or co-expression program are not necessarily placed near one another. Second, many token-based approaches require expression values to be ranked, discretized or otherwise converted into tokens, which can reduce the quantitative information contained in continuous gene-expression measurements. Yet cellular identity is shaped by both of these features: genes act together in coordinated programs, and the strength of those programs depends on expression magnitude. A representation that removes gene-gene structure and compresses continuous expression values may therefore fail to learn the underlying biological relationships and lead to suboptimal results.

Here we introduce scVision, a vision foundation model for single-cell biology that preserves both gene relationships and expression magnitude. Instead of representing a cell as a sequence of tokens, scVision represents each cell as a continuous-valued image. We first assign informative genes to fixed positions on a two-dimensional lattice using optimal transport\cite{schiebinger2019waddingtonot}, so that genes with similar expression patterns across cells are placed near one another. Each cell’s expression profile is then painted onto this shared layout, producing an scImage in which each pixel corresponds to one gene and pixel intensity corresponds to its expression level. In this representation, co-regulated genes form local image regions, gene-program activity appears as spatial texture, and global cellular identity appears as an image-level pattern. This formulation turns single-cell representation learning into a computer-vision problem\cite{lecun2015deeplearning}. It enables the use of local feature extraction, spatial attention and masked image modelling while retaining the continuous nature of gene expression. We pretrain scVision, a vision transformer with approximately 86 million parameters, as a masked autoencoder on 72 million human cells from a dataset of approximately 94 million cells assembled from public atlases\cite{tabulasapiens2022,eraslan2022crosstissue,sikkema2023hlca}. During pretraining, the model learns to reconstruct masked regions of the scImage, encouraging it to capture both local gene-program structure and global cellular state.

We evaluate scVision in a zero-shot setting, where cells from held-out studies are encoded using the frozen pretrained backbone, without fine-tuning, and downstream tasks are performed using the learned representation. Under this cross-study evaluation, scVision provides the strongest cell-type annotation of the methods tested and identifies gene-expression programs without pathway supervision. It also matches the strongest existing methods on multi-study integration and enables spatial masking of gene neighborhoods, an operation with no direct counterpart in token-based models\cite{korsunsky2019harmony,lopez2018scvi,luecken2022scib,replogle2022perturbseq,norman2019perturb,adamson2016perturb}. Because the same spatial gene layout is used across all cells, the representation also supports direct biological interpretation. Contiguous image regions correspond to groups of co-expressed genes, attention maps highlight active gene programs, and spatial masking allows coordinated gene neighborhoods to be perturbed together in silico.

Representation may be a central bottleneck in single-cell foundation modelling. By giving genes biologically meaningful positions while preserving continuous expression values, scVision offers an alternative to gene tokenization and connects single-cell biology to modern computer vision\cite{lecun2015deeplearning}. This spatial view of the transcriptome enables both transferable representations and interpretable analyses of gene programs, opening a path toward vision foundation models for cellular biology\cite{theodoris2023geneformer,cui2024scgpt,hao2024scfoundation,lopez2018scvi}.

\section*{Results}

\begin{figure*}[p]
  \centering
  \captionsetup{font=footnotesize}
  \includegraphics[width=\textwidth,height=0.96\textheight,keepaspectratio]{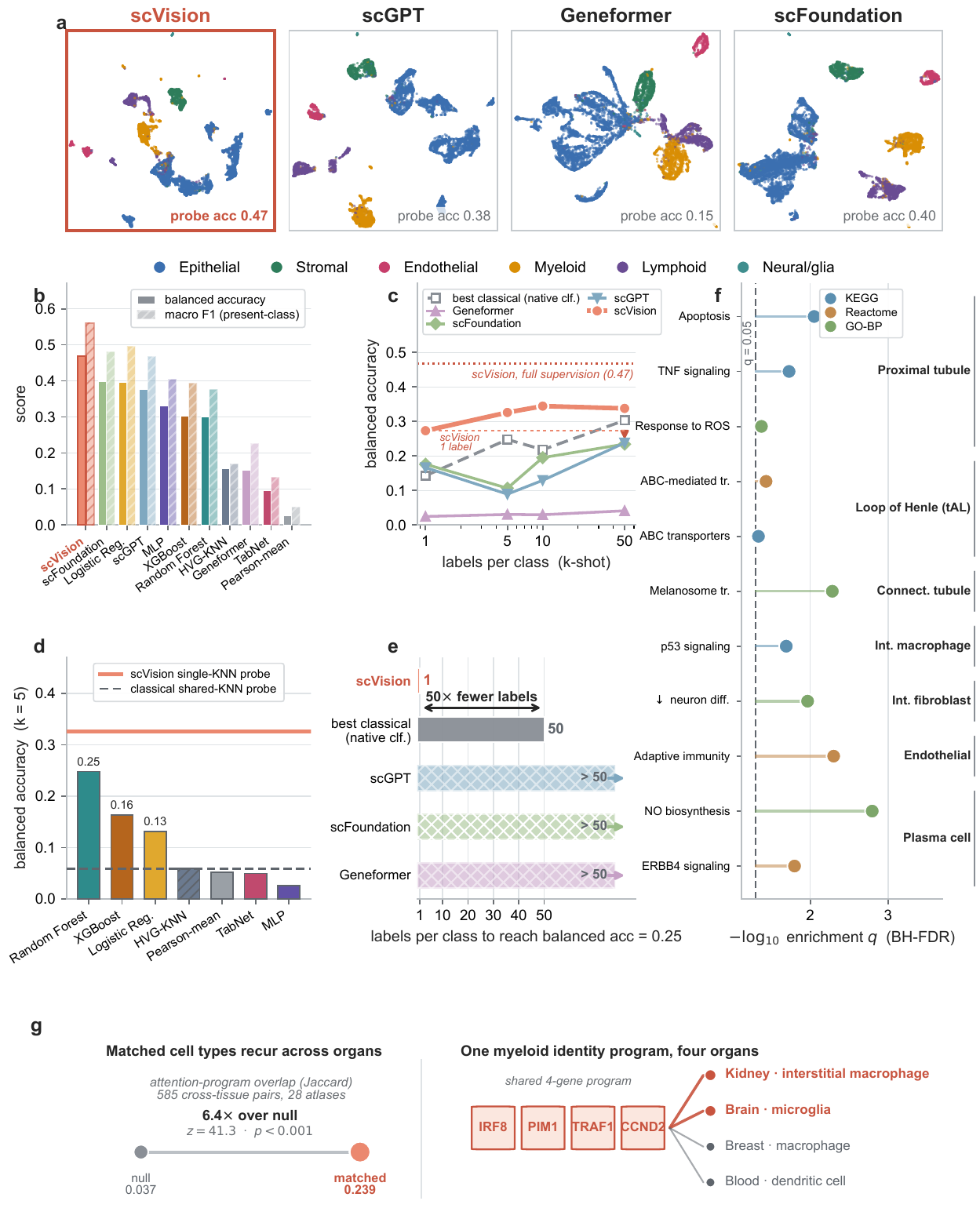}
  \caption{\textbf{A spatial representation annotates the held-out kidney cortex
  most accurately, learns to do so from a single labelled cell, and recovers
  nephron biology without supervision.}}
  \label{fig:kidney}
\end{figure*}

\begin{figure*}[tp]
  \ContinuedFloat
  \captionsetup{font=footnotesize}
  \caption{Held-out kidney cortex under the
  cross-study holdout (paradigm~A); every method reads a frozen representation and
  is scored on the full external test pool against an 87-type reference bank.
  \textbf{a}, Per-method embedding atlas: UMAP of the frozen representation for
  scVision (coral border) and three single-cell foundation models, computed with
  one matched routine and coloured by lineage (twenty-six native cortical cell
  types grouped into six lineages). Each panel is annotated with the embedding's
  matched 20-nearest-neighbour probe balanced accuracy, the common cross-atlas
  identifiability metric (Fig.~\ref{fig:xatlas_purity}): scVision $0.47$,
  scFoundation $0.40$, scGPT $0.38$, Geneformer $0.15$. scVision yields the most
  label-transferable cortical geometry, consistent with the matched probe
  (\textbf{b}).
  \textbf{b}, Annotation accuracy under a matched full-bank 20-nearest-neighbour
  probe: balanced accuracy (solid) and present-class macro-F1 (hatched) for
  scVision, three single-cell foundation models and seven classical and tabular
  baselines. scVision (coral) leads both metrics; we report these present-class
  metrics, which reward correct calls on the minority cortical types, rather than
  threshold-free areas under the curve, which are dominated by the
  abundant classes.
  \textbf{c}, Label efficiency: balanced accuracy versus labelled cells per type
  ($k$-shot, log axis). The four foundation models share the same
  20-nearest-neighbour probe, isolating representation quality; scVision leads at
  every budget and, with one labelled cell per type, already exceeds every other
  foundation model given fifty (horizontal guide). Dotted line, scVision's fully
  supervised balanced accuracy; grey dashed curve, the best of seven classical
  baselines each fitting its own native classifier.
  \textbf{d}, Representation versus classifier at $k=5$: each classical baseline
  fits its own best classifier on the identical support (bars); the dashed line is
  the shared nearest-neighbour probe (identical across classical methods, which
  share the raw highly-variable-gene space), the coral line scVision's single
  probe. The hatched bar ($k$-nearest neighbours) coincides with the shared probe
  by construction.
  \textbf{e}, The same efficiency read as an annotation budget: labelled cells per
  type needed to reach a balanced accuracy of $0.25$. scVision needs one, the best
  steelmanned classical baseline needs fifty ($50\times$ fewer), and the other
  foundation models do not reach the target within fifty (hatched, off-axis).
  \textbf{f}, scVision's attention recovers nephron-segment biology: per-cell-type
  top-attention gene programs, mapped to genes through the scImage, tested for
  pathway enrichment (hypergeometric, Benjamini--Hochberg FDR) against KEGG,
  Reactome and GO biological process, with no pathway supervision. Points mark
  significant cell-type$\rightarrow$pathway hits by $-\log_{10}q$; dashed line,
  $q=0.05$; cell types grouped at right.
  \textbf{g}, These attention programs are not kidney-private. Across $585$
  cross-tissue cell-type pairs spanning twenty-eight atlases, attention-program
  overlap (top-attended-gene Jaccard) for matched cell types exceeds a
  label-permuted null $6.4$-fold (mean $0.239$ versus $0.037$; $z=41.3$,
  $p<0.001$; left). At right, a single four-gene myeloid program (IRF8, PIM1,
  TRAF1, CCND2) supplies the top-attended patch for the kidney interstitial
  macrophage, brain microglia, breast macrophage and autoimmune-blood dendritic
  cell: one model, one program, four organs.}
\end{figure*}

\subsection*{Label-efficient zero-shot annotation of held-out kidney cortex}

We first tested scVision on an external kidney cortex dataset, where the model never sees kidney cells during training. The task is to assign cortical cell types using a frozen embedding and a reference bank containing 87 cell types from other studies. In two-dimensional projections, all four foundation models separated the kidney cortex into coherent lineage structure, but the scVision embedding was the most identifiable: annotating each panel with its matched 20-nearest-neighbour probe balanced accuracy, the common geometry metric we report for every held-out atlas (Fig.~\ref{fig:xatlas_purity}), ranked scVision first, ahead of scFoundation, scGPT and Geneformer (Fig.~\ref{fig:kidney}a). Using a nearest-neighbour probe on the frozen scVision embedding, scVision achieved the highest performance on the external cortex, with 0.47 balanced accuracy and 0.56 present-class macro-F1 (Fig.~\ref{fig:kidney}b). These metrics emphasize performance across all cell types, including rare cortical populations. scVision outperformed the next-best model in balanced accuracy, scFoundation\cite{hao2024scfoundation} (0.40), and the next-best method in macro-F1, logistic regression (0.50). scGPT\cite{cui2024scgpt} reached 0.38 balanced accuracy, whereas Geneformer\cite{theodoris2023geneformer} reached 0.15. We report balanced accuracy and present-class macro-F1 because these metrics better reflect annotation performance on rare but biologically important cell types than threshold-free area-under-the-curve scores, which can be dominated by abundant classes.

We then inquired how many labelled examples are needed to annotate a new cell type. For each method, we sampled $k\in\{1,5,10,50\}$ labelled cells per type from the reference bank and evaluated performance on the full external cortex, using the same support cells for all models (Fig.~\ref{fig:kidney}c). With only one labelled cell per type, scVision reached 0.27 balanced accuracy. With ten labelled cells per type, it reached 0.34, which is 72\% of its fully supervised performance of 0.47. Performance then changed little with additional labels. Under the same nearest-neighbour readout, scVision performed best at every label budget. Remarkably, one labelled cell per type with scVision outperformed fifty labelled cells per type with scGPT, scFoundation and Geneformer (0.27 versus 0.24, 0.23 and 0.04, respectively). This shows that scVision reduces the required labelling effort by more than fifty-fold in this setting.

To test whether this advantage came only from the nearest-neighbour classifier, we also gave each classical baseline its own preferred classifier. These included logistic regression, a multilayer perceptron, gradient-boosted trees, a random forest, $k$-nearest neighbours, a correlation classifier and TabNet, all trained on the same support cells (Fig.~\ref{fig:kidney}d). The $k$-nearest-neighbour baseline reproduced the shared-probe result, confirming that the comparison isolated the effect of the classifier. Allowing each method to use its own classifier improved the strongest classical baselines, but none matched scVision. At five labelled cells per type, the best classical method reached 0.25 balanced accuracy, compared with 0.33 for scVision. scVision remained best at all label budgets, with the gap narrowing only at fifty labelled cells per type (0.34 versus 0.30). In practical terms, scVision needed only one labelled cell per type to reach 0.25 balanced accuracy, whereas the best classical method needed fifty; the other foundation models did not reach this level within fifty labels (Fig.~\ref{fig:kidney}e). The label efficiency therefore reflects the learned spatial representation rather than a particular classifier.

Finally, we examined whether the same frozen representation also captured interpretable kidney biology. We read scVision’s last-block attention as gene-expression programs by mapping the most-attended image patches back to genes through the scImage. This analysis used no pathway supervision. The resulting programs matched known nephron biology (Fig.~\ref{fig:kidney}f). Proximal-tubule epithelial cells, which are highly susceptible to injury, were enriched for Apoptosis and TNF signalling (KEGG $q=0.009$ and $0.019$) and for the reactive-oxygen-species response (GO biological process $q=0.042$), consistent with canonical tubular-injury programs. The thin ascending limb of the loop of Henle, which is involved in solute handling, was enriched for ABC transporters (Reactome $q=0.037$). Interstitial macrophages were enriched for p53 signalling ($q=0.020$).

These programs were not specific to kidney tissue alone. The same four-gene myeloid program, IRF8, PIM1, TRAF1 and CCND2, formed the top-attended patch in both kidney interstitial macrophages and brain microglia. The model recovered a shared myeloid program across two organs without pathway labels. Across 585 matched cell-type pairs from 28 atlases, attention-derived programs from the same cell type overlapped 6.4-fold more than expected under a label-permuted null model (mean Jaccard 0.239 versus 0.037; $z=41.3$; Fig.~\ref{fig:kidney}g). These results suggest that scVision learns reusable cell-type identity programs that can be recognized across tissues.

\FloatBarrier
\begin{figure*}[p]
  \centering
  \captionsetup{font=footnotesize}
  \includegraphics[width=\textwidth,height=0.90\textheight,keepaspectratio]{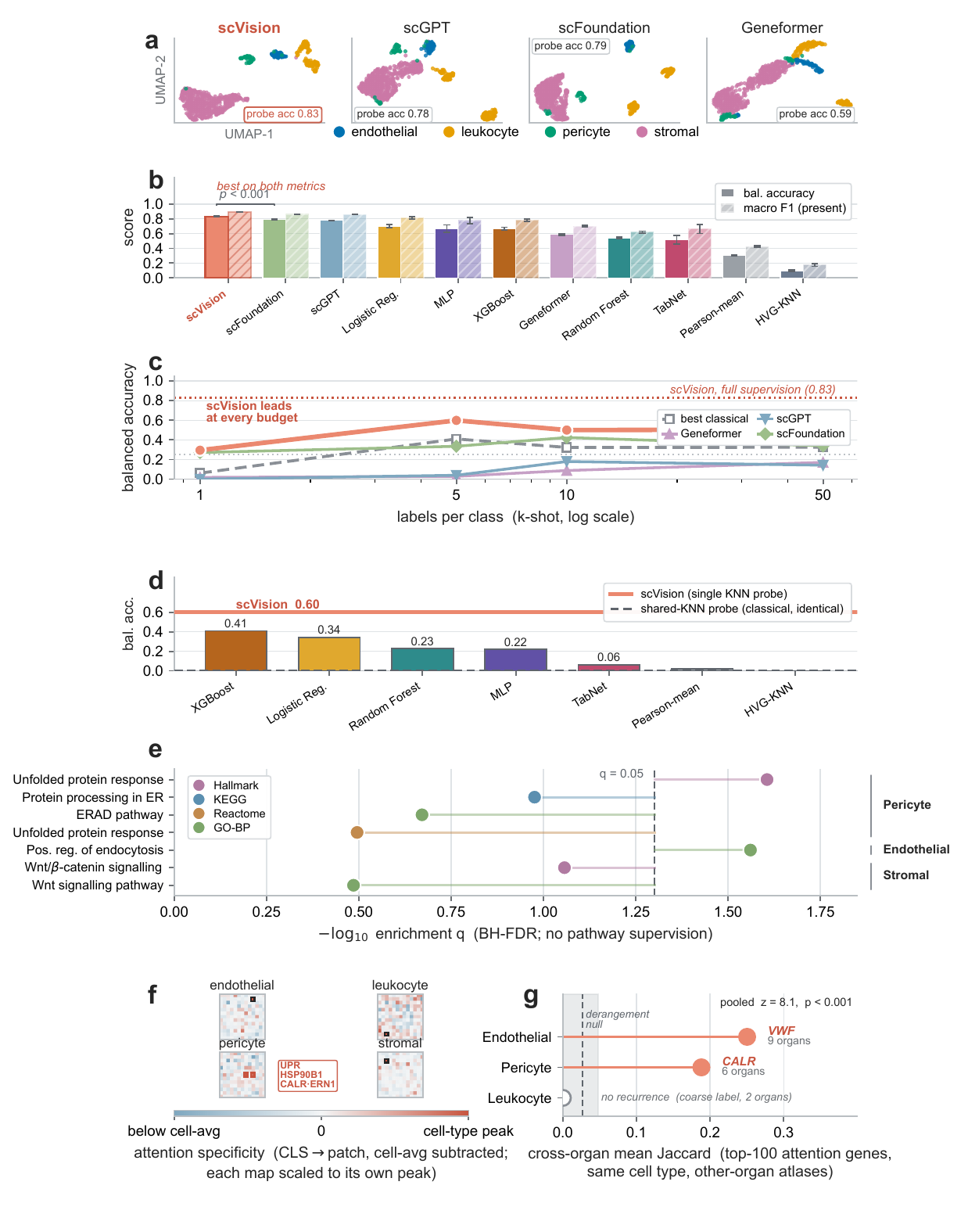}
  \caption{
  \textbf{A spatial representation annotates the held-out ovary most accurately, leads at
  every labelling budget, and recovers a coherent pericyte stress program without
  supervision.} Held-out ovary, cross-study holdout (paradigm~A); methods read a frozen
  representation scored against the 87-type bank (present pool: endothelial, leukocyte,
  pericyte, stromal; four-class chance $0.25$). Panels \textbf{a}--\textbf{g} are described
  in the continued caption overleaf.}
  \label{fig:ovary}
\end{figure*}

\begin{figure*}[tp]
  \ContinuedFloat
  \captionsetup{font=footnotesize}
  \caption{
  \textbf{a}, Embedding geometry: UMAP of each frozen representation coloured by reference
  type, annotated with the matched 20-nearest-neighbour probe balanced accuracy, the
  common cross-atlas identifiability metric (Fig.~\ref{fig:xatlas_purity}). scVision is the
  most identifiable embedding ($0.83$), ahead of scFoundation ($0.79$), scGPT ($0.78$) and
  Geneformer ($0.59$); this ordering reproduces the accuracy ordering in (b). \textbf{b}, Matched full-bank 20-NN probe: balanced accuracy (solid) and present-class
  macro-F1 (hatched) for scVision, three foundation models and seven classical baselines
  (bars, five-fold mean; whiskers $\pm1$\,s.e.m.). scVision (coral) leads both; its
  $+3.6$-point balanced-accuracy margin over scFoundation is significant (paired $t$,
  $\mathrm{df}=4$, $p<0.001$; versus scGPT $p<0.01$). ROC-AUC exceeds $0.98$ throughout, so
  present-class metrics are reported. \textbf{c}, Label efficiency ($k$-shot, log axis),
  the four foundation models sharing the 20-NN probe; scVision leads at every budget and
  saturates by five shots. Dotted, scVision fully supervised ($0.83$); grey dashed, best
  classical; faint guide, chance. \textbf{d}, Representation versus classifier at $k=5$:
  each classical baseline fits its own best classifier (bars); dashed, the shared NN probe;
  coral, scVision's. Native classifiers lift the strongest baselines (gradient-boosted trees
  $0.41$), yet none reaches scVision ($0.60$). \textbf{e}, Attention recovers ovarian
  biology: per-cell-type top-attention programs tested for pathway enrichment
  (hypergeometric, Benjamini--Hochberg FDR). A pericyte unfolded-protein-response program is
  significant (Hallmark $q=0.025$), an endothelial endocytosis program likewise (GO
  biological process $q=0.027$). Points, $-\log_{10}q$; dashed line $q=0.05$. \textbf{f},
  The programs are spatially localised: per-cell-type attention maps (the $13\times13$ grid
  minus the cell-type average) give each population a distinct fingerprint, the pericyte
  signature concentrating on one region carrying HSP90B1, CALR and ERN1 from (e).
  \textbf{g}, The programs recur across organs: mean Jaccard overlap of each ovarian type's
  100 top-attention genes with the same type across twenty-eight atlases. Endothelial and
  pericyte programs recur far above a thousand-permutation null (grey band; pooled $z=8.1$,
  $p<0.001$): VWF across kidney, brain, blood and fallopian-tube endothelium, CALR across
  prostate, breast and intestinal pericytes; the coarse leukocyte label does not.}
\end{figure*}

\subsection*{A pericyte stress program in held-out ovary}

We next tested scVision on an external ovary dataset. This was a smaller transfer test than the kidney cortex analysis. After restricting the dataset to cell types present in the 87-type reference bank, four ovarian populations remained: endothelial cells, leukocytes, pericytes and ovarian stromal cells. The model had not seen ovarian cells during training, so it had to assign these cell types using only the frozen embedding and the reference bank. Chance performance for this four-class task is 0.25.

Using a nearest-neighbour probe on the frozen scVision embedding, scVision achieved the best performance, with 0.83 balanced accuracy and 0.89 present-class macro-F1 (Fig.~\ref{fig:ovary}b). scFoundation\cite{hao2024scfoundation} and scGPT\cite{cui2024scgpt} were the closest foundation-model baselines, with balanced accuracies of 0.79 and 0.78, respectively. Across the five matched folds, scVision’s improvement over scFoundation was significant (paired $t$-test, $p<0.001$), as was its improvement over scGPT ($p<0.01$). Logistic regression was the strongest classical baseline, reaching 0.70 balanced accuracy, while Geneformer\cite{theodoris2023geneformer} reached 0.59. We focus on balanced accuracy and present-class macro-F1 because they measure performance across all four cell types. In contrast, ROC AUC was above 0.98 for every method and therefore did not distinguish the models in this coarse four-class setting.

We then tested how well each representation performed with very few labelled examples. For each label budget, $k\in\{1,5,10,50\}$ labelled cells per type were sampled from the reference bank and shared across methods (Fig.~\ref{fig:ovary}c). With only one labelled cell per type, scVision reached 0.30 balanced accuracy. With five labelled cells per type, it reached 0.60, which is 72\% of its fully supervised value of 0.83. Performance then changed little with more labels. Under the same nearest-neighbour readout, scVision led at every label budget. The gap was smaller than in the kidney cortex task, but scVision still performed best. One labelled cell per type with scVision outperformed fifty labelled cells per type with scGPT and Geneformer (0.30 versus 0.14 and 0.17). We do not make a fifty-fold label-efficiency claim here, because scFoundation with fifty labels reached 0.34, slightly above scVision with one label.

To test whether this advantage was caused by the nearest-neighbour classifier, we also allowed each classical baseline to use its own classifier. These included logistic regression, a multilayer perceptron, gradient-boosted trees, a random forest, $k$-nearest neighbours, a correlation classifier and TabNet, all trained on the same support cells (Fig.~\ref{fig:ovary}d). The $k$-nearest-neighbour baseline reproduced the shared-probe result, confirming that the comparison was controlled. Allowing each method to use its own classifier improved some baselines. For example, gradient-boosted trees reached 0.41 balanced accuracy at five labelled cells per type. However, this remained below scVision’s 0.60. scVision retained the lead at every label budget. At fifty labelled cells per type, scVision reached 0.51, compared with 0.32 for the best classical method. Again, the label efficiency reflects the learned spatial representation rather than a particular classifier.

We also asked whether the same frozen representation captured interpretable ovarian biology. We read scVision’s last-block attention as gene-expression programs by mapping the most-attended image patches back to genes through the scImage. This analysis used no pathway supervision. The recovered programs were cell-type specific and biologically coherent (Fig.~\ref{fig:ovary}e). In pericytes, the strongest program was enriched for the unfolded-protein response (MSigDB Hallmark $q=0.025$), an endoplasmic-reticulum stress pathway marked by CALR, ERN1 and HSP90B1. Related endoplasmic-reticulum stress annotations appeared below significance in KEGG, GO and Reactome, suggesting a consistent signal across databases. The endothelial program was enriched for positive regulation of endocytosis (GO biological process $q=0.027$), consistent with endothelial vesicular transport. The ovarian stromal program showed enrichment for Wnt/$\beta$-catenin signalling (Hallmark $q=0.088$), a pathway linked to ovarian stromal and follicular biology. The number of recovered programs was smaller than in the kidney analysis, as expected for a four-cell-type dataset, but the programs were tissue appropriate and were obtained without pathway labels.

Two additional analyses showed that these programs were not artefacts of the readout and were not specific only to ovary. First, when attention was displayed as cell-type-specific maps over the image grid, each ovarian population showed a distinct spatial pattern (Fig.~\ref{fig:ovary}f). The pericyte signal was concentrated in a contiguous image region containing the unfolded-protein-response genes HSP90B1, CALR and ERN1, showing that the program was spatially localized rather than diffuse. Second, we compared each ovarian attention program with programs from the same cell type across a 28-atlas attention bank, using the overlap of the 100 most-attended genes. The endothelial and pericyte programs recurred far above a permutation null (pooled $z=8.1$, $p<0.001$; Fig.~\ref{fig:ovary}g). The shared programs were biologically meaningful: VWF anchored the endothelial program across kidney, brain, blood and fallopian-tube endothelium, while CALR recurred in pericytes from prostate, breast and intestine. These results suggest that scVision’s unsupervised attention captures conserved cell-type programs rather than tissue-specific noise.

Finally, we examined whether the annotation differences were already visible in the geometry of the frozen embeddings. When the four ovarian populations were projected to two dimensions, scVision separated endothelial cells, leukocytes, pericytes and stromal cells into compact and distinct clusters (Fig.~\ref{fig:ovary}a). Annotating each embedding with its matched 20-nearest-neighbour probe balanced accuracy, the common identifiability metric we report across all held-out atlases (Fig.~\ref{fig:xatlas_purity}), placed scVision first (0.83), followed by scFoundation (0.79), scGPT (0.78) and Geneformer (0.59), reproducing the annotation ranking in (b). Label-free clustering scores agreed: among the foundation embeddings, scVision also attained the highest 15-nearest-neighbour purity (0.97) and Moran’s $I$ (0.93). Even before annotation, the frozen scVision embedding organizes held-out ovarian cells into the most readable cell-type geometry.

\FloatBarrier
\begin{figure*}[p]
  \centering
  \includegraphics[width=\textwidth,height=0.88\textheight,keepaspectratio]{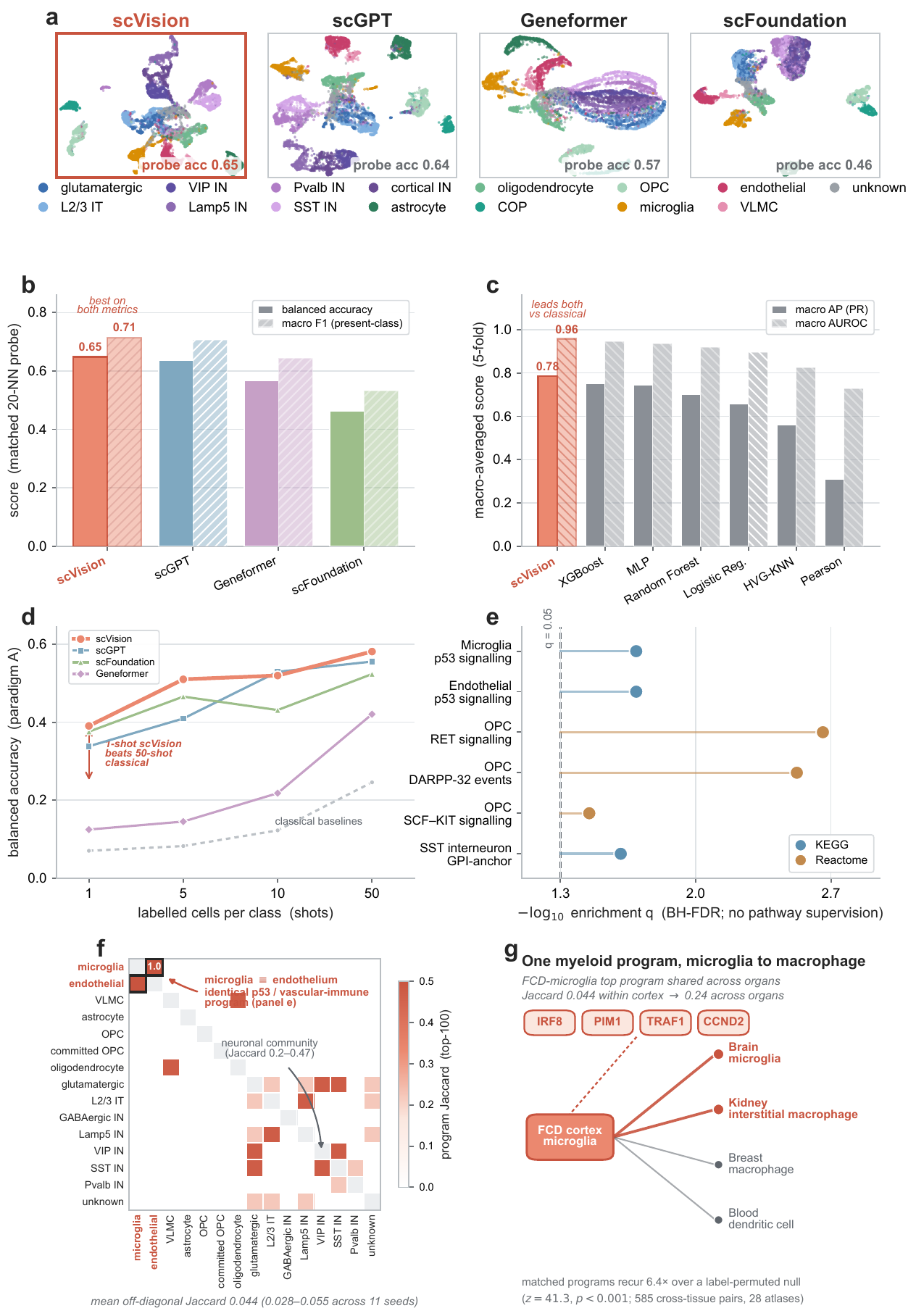}
  \caption{
  \textbf{scVision annotates the held-out focal cortical dysplasia cortex most accurately
  and reads a microglial identity program it shares with macrophages across organs.}
  Held-out focal cortical dysplasia cortex, cross-study holdout (paradigm~A); every method
  reads a frozen representation scored on the full external test pool.
  \emph{(Panel-by-panel description \textbf{a}--\textbf{g} continued on the next page.)}}
  \label{fig:fcd}
\end{figure*}

\begin{figure*}[t]
  \caption*{\textbf{Figure~\ref{fig:fcd} (continued).}
  \textbf{a}, Per-method
  frozen-embedding UMAP of the same held-out external test pool, coloured by the fifteen native
  cortical cell types (shared legend) and annotated with each model's matched
  20-nearest-neighbour probe balanced accuracy, the common cross-atlas identifiability
  metric (Fig.~\ref{fig:xatlas_purity}); higher means a more readily annotated embedding. All
  four panels use identical UMAP settings ($30$ neighbours, minimum distance $0.3$, cosine
  metric, and a PCA-$50$ pre-step where the embedding exceeds $100$ dimensions), so only the
  representation being visualised differs; scVision is the most identifiable embedding ($0.65$,
  against $0.64$/$0.57$/$0.46$ for scGPT/Geneformer/scFoundation). \textbf{b}, The matched
  full-bank 20-nearest-neighbour probe reading the same four frozen representations: balanced
  accuracy (solid) and present-class macro-F1 (hatched), all sharing the identical probe so only
  the embedding differs. scVision (coral) leads both, and the panel-a badge is exactly this
  matched-probe balanced accuracy. \textbf{c}, Calibrated detection against six
  classical and tabular baselines: macro-averaged average precision (solid) and AUROC
  (hatched), five-fold. \textbf{d}, Label
  efficiency ($k=1,5,10,50$), each method read through the same 20-nearest-neighbour probe.
  scVision (coral) leads the foundation models at one, five and fifty shots ($0.39$, $0.51$,
  $0.58$) and is edged only by scGPT at ten ($0.53$ versus $0.52$). Every foundation model
  sits far above the seven classical probes (grey), whose fifty-shot accuracy ($0.25$)
  scVision exceeds with one labelled cell per class. \textbf{e}, Attention recovers
  cortical biology: per-cell-type top-attention programs tested for pathway enrichment
  (hypergeometric, Benjamini--Hochberg FDR against KEGG and Reactome). Points, significant
  cell-type$\rightarrow$pathway hits by $-\log_{10}q$; dashed line, $q=0.05$. \textbf{f},
  Programs are cell-type-specific: pairwise Jaccard overlap of the top-$100$ attention
  programs across the fifteen cortical cell types (mean off-diagonal $0.044$; $0.028$--$0.055$
  across eleven cell-resampling seeds). The dominant cross-lineage overlap is the
  microglia--endothelium pair (Jaccard $1.0$; boxed), the same p53-linked vascular--immune
  module scoring in (e) and the most reproducible off-diagonal feature, an identical
  top-$100$ program in nine of eleven seeds, sharing at least $64$ of $100$ genes in all
  eleven (mean Jaccard $0.90$). Excitatory and inhibitory neurons form a looser community
  (Jaccard $0.2$--$0.47$). \textbf{g}, The dysplastic-cortex microglia's
  top four-gene program (IRF8, PIM1, TRAF1, CCND2) is also top-attended in healthy-brain
  microglia, kidney interstitial macrophage, breast macrophage and autoimmune-blood
  dendritic cell, one myeloid identity program across five organs (matched-type Jaccard
  $0.24$ versus $0.044$ within-cortex). Across $585$ cross-tissue cell-type pairs spanning
  twenty-eight atlases, matched programs recur $6.4$-fold over a label-permuted null
  ($z=41.3$, $p<0.001$).}
\end{figure*}

\subsection*{A microglial program recurs across organs in held-out dysplastic cortex}

We next tested scVision on an external focal cortical dysplasia cortex dataset. Focal cortical dysplasia is a developmental abnormality of the cerebral cortex and a common cause of drug-resistant focal epilepsy. This dataset provides a stringent transfer test because the model had not seen dysplastic-cortex cells during training. The task was to assign cortical cell types using a frozen embedding and a reference bank built from other studies.

Using the same nearest-neighbour probe for all foundation models, scVision achieved the best performance, with 0.65 balanced accuracy and 0.71 present-class macro-F1 (Fig.~\ref{fig:fcd}b). It narrowly outperformed scGPT\cite{cui2024scgpt} on both metrics: scGPT reached 0.64 balanced accuracy and 0.706 macro-F1, against 0.65 and 0.714 for scVision. scVision more clearly outperformed Geneformer\cite{theodoris2023geneformer} (0.57 and 0.64) and scFoundation\cite{hao2024scfoundation} (0.46 and 0.53). Because all models used the same readout, these differences reflect the quality of the learned representations.

This performance was also visible in the embedding geometry (Fig.~\ref{fig:fcd}a). Under identical UMAP settings, scVision separated the 15 cortical cell types into the clearest clusters, and each panel is annotated with its matched 20-nearest-neighbour probe balanced accuracy, the common identifiability metric we report across held-out atlases (Fig.~\ref{fig:xatlas_purity}), on which scVision ranks first. Label-free clustering agreed: scVision also formed the tightest neighbourhoods, with a 15-nearest-neighbour purity of 0.87 against 0.83 for scGPT, 0.77 for Geneformer and 0.73 for scFoundation. This ranking matched the annotation performance, showing that scVision produced the most readable cell-type structure.

We also compared scVision with classical and tabular baselines using threshold-free metrics. In this focal cortical dysplasia cortex, scVision achieved the highest macro-averaged average precision (0.78) and AUROC (0.96), ahead of gradient-boosted trees (0.75 and 0.94), a multilayer perceptron (0.74 and 0.94), and other classical baselines (Fig.~\ref{fig:fcd}c). Its balanced accuracy of 0.65 was also higher than the best classical method, which reached 0.51. Across all measured metrics, scVision was the most accurate annotator.

We then tested label efficiency. Each embedding was kept frozen, and each method was evaluated using $k=1,5,10,50$ labelled cells per class. scVision was the most label-efficient foundation model at one, five and fifty labels per class, reaching 0.39, 0.51 and 0.58 balanced accuracy, respectively (Fig.~\ref{fig:fcd}d). Geneformer and scFoundation performed lower across the label sweep. The difference from classical baselines was larger: the seven tabular probes reached only 0.25 balanced accuracy even with fifty labelled cells per class, whereas scVision exceeded this level with a single labelled cell per class. The same representation that produced the best full-reference annotation also required the fewest labels.

We next examined whether scVision’s attention captured interpretable disease biology. We mapped the most-attended image patches from the last transformer block back to genes through the scImage. This produced cell-type-specific gene programs without using pathway labels (Fig.~\ref{fig:fcd}e). Microglia and endothelial cells were enriched for p53 signalling (KEGG $q=0.020$), consistent with immune and vascular stress in the epileptogenic lesion. Oligodendrocyte precursor cells were enriched for growth-factor and differentiation pathways, including RET signalling (Reactome $q=0.002$), DARPP-32 events ($q=0.003$) and SCF--KIT signalling ($q=0.036$), consistent with oligodendroglial abnormalities in dysplastic cortex. Somatostatin interneurons were enriched for glycosylphosphatidylinositol-anchor biosynthesis (KEGG $q=0.024$), relevant to neuronal cell-surface biology. Re-evaluated across the same eleven cell-resampling seeds used in (f), this p53 vascular--immune enrichment was the most reproducible of these programs, significant (BH-FDR $q<0.05$) in all eleven endothelial and nine of eleven microglial seeds, with oligodendrocyte-precursor RET signalling holding in nine of eleven, whereas the remaining precursor and interneuron annotations were more seed-dependent.

These attention programs were mostly specific to individual cell types. Across the 15 cortical cell types, the mean overlap between top-attention programs was low, with a mean Jaccard index of 0.044 (Fig.~\ref{fig:fcd}f). This value was stable across eleven cell-resampling seeds (range 0.028--0.055), indicating that the model did not simply learn one broad lesion-wide stress signal. The main exception was the microglia--endothelium pair, whose top programs were nearly identical (Jaccard 1.0 in nine of eleven seeds; at least 64 of 100 genes shared in all eleven, mean Jaccard 0.90), reflecting a shared p53-linked vascular--immune module. This was the only such overlap between distinct lineages; neuronal subtypes showed moderate overlap, as expected for related excitatory and inhibitory cell populations, whereas most other cell-type pairs remained distinct.

Finally, we asked whether the microglial program was specific to focal cortical dysplasia or reflected a broader myeloid identity. The top-attended program in dysplastic-cortex microglia contained four genes: IRF8, PIM1, TRAF1 and CCND2. The same four-gene program was also the top-attended program in healthy-brain microglia, kidney interstitial macrophages, breast macrophages and autoimmune-blood dendritic cells (Fig.~\ref{fig:fcd}g). scVision recovered the same myeloid program across five organs without pathway supervision.

This cross-tissue signal was stronger than the overlap between different cell types within the same lesion. Different cortical cell types shared only 0.044 of their attention programs on average, whereas the same cell type across organs shared 0.24, about fivefold higher. Across 585 matched cell-type pairs from 28 atlases, attention programs from the same cell type overlapped 6.4-fold more than expected under a label-permuted null model ($z=41.3$, $p<0.001$). These results suggest that scVision learns reusable cell-type identity programs that can be recognized across tissues, including activated microglia in diseased human cortex.

\FloatBarrier
\begin{figure*}[p]
  \centering
  \includegraphics[width=\textwidth]{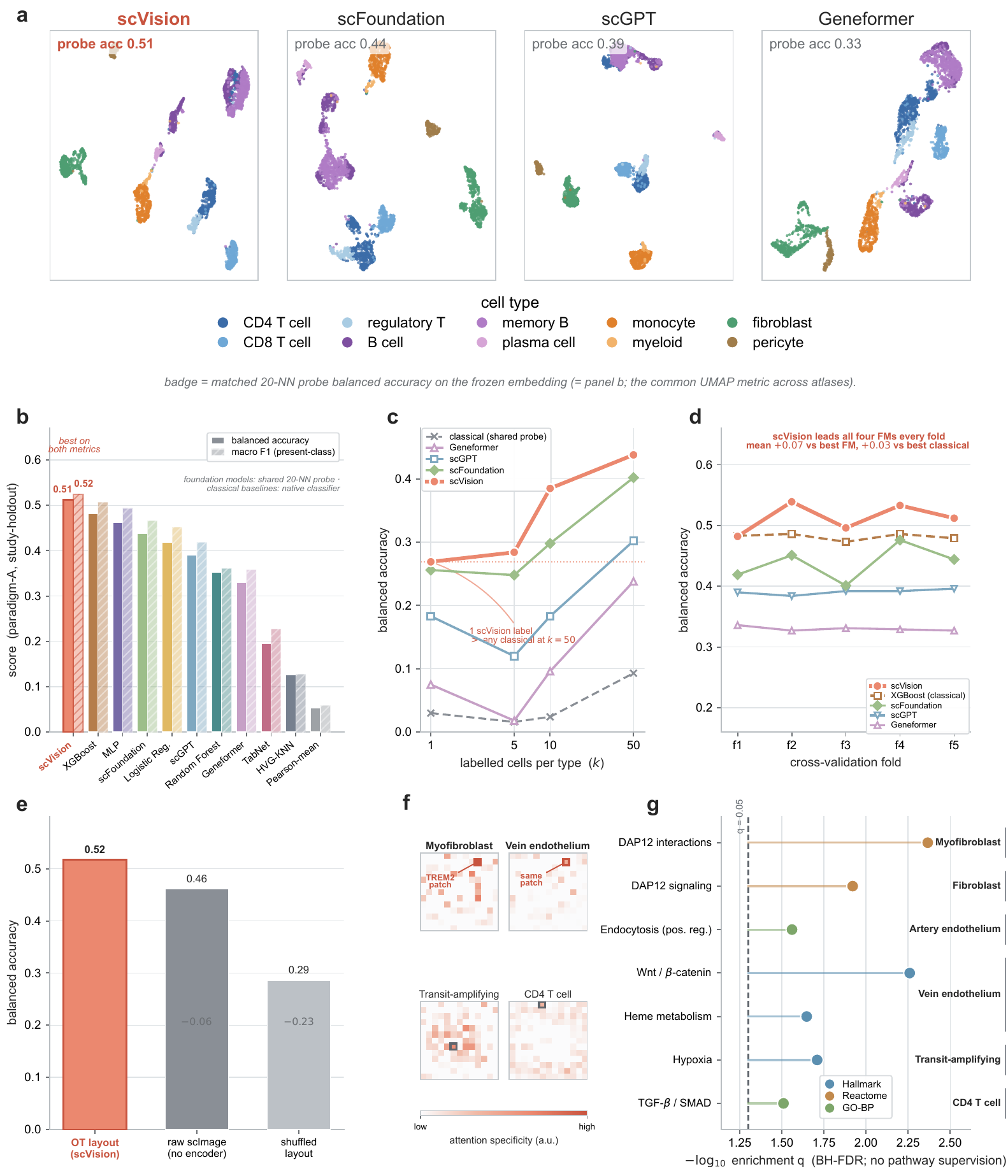}
  \caption{
  \textbf{scVision is the most accurate and most label-efficient annotator of the held-out
  Crohn's-disease ileum, and its frozen attention reads the inflamed-stroma biology.}
  Held-out Crohn's-disease ileum, cross-study holdout (paradigm~A); every method reads a
  frozen representation scored on the full external test pool.
  \emph{(Panel-by-panel description \textbf{a}--\textbf{g} continued on the next page.)}}
  \label{fig:crohn}
\end{figure*}

\begin{figure*}[t]
  \caption*{\textbf{Figure~\ref{fig:crohn} (continued).}
  \textbf{a}, Embedding
  geometry: UMAP of each model's frozen embedding of the external test cells, coloured by
  expert type; each badge is the matched 20-nearest-neighbour probe balanced accuracy, the
  common cross-atlas identifiability metric (Fig.~\ref{fig:xatlas_purity}): scVision $0.51$,
  scFoundation $0.44$, scGPT $0.39$, Geneformer $0.33$. scVision is the most identifiable
  embedding; within-test cluster tightness (kNN purity) separates the foundation
  embeddings far less than cross-study transfer does, so the decisive margin is cross-study
  transfer (b), not within-test geometry.
  \textbf{b}, Matched
  read-out: foundation models share the same $20$-NN probe, classical and tabular baselines
  use their own native classifier. Balanced accuracy (solid) and present-class macro-F1
  (hatched) for scVision, three foundation models and seven classical baselines; scVision
  (coral) leads both and alone among single-cell embeddings reaches the strong
  non-foundation baselines. \textbf{c}, Label efficiency at $k\in\{1,5,10,50\}$, foundation
  models read through the same $20$-NN probe, classical through the shared raw-gene probe.
  One scVision label exceeds every classical baseline given fifty. \textbf{d}, Per-fold
  robustness across five method-matched folds. scVision leads every foundation model on all
  five and leads or ties the best classical (gradient-boosted trees) on every fold (fold~1 a
  $0.001$ dead heat); mean $+0.07$ over the best foundation model (scFoundation, paired
  $p{=}5\times10^{-4}$), $+0.03$ over the best classical. \textbf{e}, The advantage is in
  the spatial representation: balanced accuracy for the optimal-transport scImage layout
  (native head $0.52$; $20$-NN probe $0.51$), the same encoder on a permuted layout
  (``shuffled''), and a perceptron on the flattened raw scImage (``no encoder''). Destroying
  the layout costs $0.23$, removing the encoder $0.06$. \textbf{f}, Frozen attention
  concentrates on compartment-specific patches: specificity-corrected last-block attention
  for four cell types on the $13\times13$ grid; myofibroblast and vein endothelium peak on
  the same TREM2/DAP12 inflamed-stroma patch (outlined). \textbf{g}, Attention recovers
  inflamed-gut biology: per-cell-type top-attention programs tested for pathway enrichment
  (hypergeometric, Benjamini--Hochberg FDR; Hallmark, Reactome, GO biological process).
  Points, significant hits by $-\log_{10}q$; dashed line, $q=0.05$.}
\end{figure*}

\subsection*{Inflamed-stroma programs in held-out Crohn's-disease ileum}

We next tested scVision on an external Crohn's-disease ileum dataset. This was a demanding transfer test because the model had not seen ileal cells during training. The task was to assign each held-out cell to a cell type using a frozen embedding and a fixed reference bank of 87 cell types from other studies. The original ileum dataset contains about 30 expert-annotated populations, including epithelial, stromal, endothelial and immune cells. Of these, 10 cell types were also present in the reference bank and could therefore be evaluated in the matched annotation task (Fig.~\ref{fig:crohn}a).

We compared all foundation models using the same 20-nearest-neighbour probe on their frozen embeddings. This controlled comparison isolates the quality of the representation rather than the classifier used for readout. Under this matched readout, scVision achieved the best performance, with 0.51 balanced accuracy and 0.52 present-class macro-F1 (Fig.~\ref{fig:crohn}b). It also matched or exceeded the strongest non-foundation baselines, including gradient-boosted trees (0.48 balanced accuracy and 0.51 macro-F1) and a multilayer perceptron (0.46 and 0.49). Other foundation models performed lower under the same readout: scFoundation\cite{hao2024scfoundation} reached 0.44 balanced accuracy and 0.47 macro-F1, scGPT\cite{cui2024scgpt} reached 0.39 and 0.42, and Geneformer\cite{theodoris2023geneformer} reached 0.33 and 0.36. We focus on balanced accuracy and present-class macro-F1 because these metrics evaluate performance across all scored cell types, including rarer mucosal populations that are important for Crohn's disease biology.

We then tested how many labelled examples were needed for annotation. For each method, we sampled $k\in\{1,5,10,50\}$ labelled cells per type and evaluated performance on the full external dataset (Fig.~\ref{fig:crohn}c). scVision was the most label-efficient method. With only one labelled cell per type, it reached 0.27 balanced accuracy, which was higher than any classical baseline reached even with 50 labelled cells per type (0.09). scVision also led all other foundation models at every label budget. Because the same readout was used across embeddings, this improvement reflects the learned representation rather than a more powerful classifier.

The five cross-validation folds were fixed and shared across methods, allowing a fold-by-fold comparison (Fig.~\ref{fig:crohn}d). At the matched probe, scVision outperformed the best competing foundation model, scFoundation, on all five folds. The mean improvement was 0.07 balanced accuracy (paired $p=5\times10^{-4}$; 95\% confidence interval $[+0.05,+0.10]$). Compared with the strongest non-foundation baseline, gradient-boosted trees, the margin was smaller but still positive, with a mean improvement of 0.03 balanced accuracy across four of five folds. On this difficult atlas, scVision clearly outperformed the other foundation models and performed at least as well as the strongest classical baselines.

We next tested whether the spatial scImage layout was important. When the gene-to-pixel assignment was randomly permuted, balanced accuracy dropped from 0.52 to 0.29 (Fig.~\ref{fig:crohn}e). This shows that the learned spatial arrangement of genes is essential. Removing the vision encoder and classifying the flattened raw scImage with a multilayer perceptron caused a smaller drop, from 0.52 to 0.46. Both the vision backbone and the spatial gene layout together contribute to performance, but the gene arrangement carries the larger share of the signal.

The embedding geometry supported these results (Fig.~\ref{fig:crohn}a). In two-dimensional projections, scVision organized the external mucosa into compact and well-separated cell-type clusters, and its matched 20-nearest-neighbour probe balanced accuracy on the frozen embedding was the highest of the foundation models (0.51; the common cross-atlas identifiability metric, Fig.~\ref{fig:xatlas_purity}). A complementary label-free view was more nuanced: within-test $k$-nearest-neighbour purity was 0.92 for scVision, compared with about 0.70 for the raw-gene space used by classical learners, but scFoundation (0.91), Geneformer (0.88) and scGPT (0.87) nearly matched it. These local purity scores did not fully predict annotation accuracy. For example, scFoundation nearly matched scVision in within-dataset purity but performed worse when cells had to be matched to the external reference bank. This suggests that scVision’s main advantage is not only better clustering within the held-out tissue, but stronger transfer across studies.

Finally, we examined whether scVision’s attention captured disease biology. We mapped last-block attention back onto the scImage and found that attention concentrated on a small number of cell-type-specific patches rather than spreading diffusely (Fig.~\ref{fig:crohn}f). Most cell types had distinct peak patches, but myofibroblasts and venous endothelial cells shared the same peak patch. Mapping these patches back to genes, without pathway supervision, recovered programs consistent with Crohn's disease pathophysiology (Fig.~\ref{fig:crohn}g).

The shared myofibroblast--endothelium patch was anchored by TREM2. Both myofibroblasts and fibroblasts were enriched for TREM2--DAP12 signalling (Reactome $q=0.004$ and $0.012$), suggesting a shared inflamed-stroma program in the gut wall. Other cell types showed distinct programs. Venous endothelial cells were enriched for Wnt/$\beta$-catenin signalling and heme metabolism (Hallmark $q=0.006$ and $0.023$), consistent with vascular remodelling. Arterial endothelial cells were enriched for positive regulation of endocytosis (GO biological process $q=0.028$), consistent with activated vessel-wall transport. Transit-amplifying crypt epithelial cells were enriched for hypoxia (Hallmark $q=0.020$), consistent with the metabolic stress of inflamed and regenerating gut tissue. CD4 T cells were enriched for TGF-$\beta$/SMAD regulation (GO biological process $q=0.031$), a pathway linked to the Treg--Th17 immune axis in Crohn's disease. These programs were recovered without pathway labels, showing that the frozen scVision representation captures both transferable cell identity and disease-relevant biology.

\FloatBarrier  
\begin{figure*}[p]
  \centering
  \includegraphics[width=\textwidth]{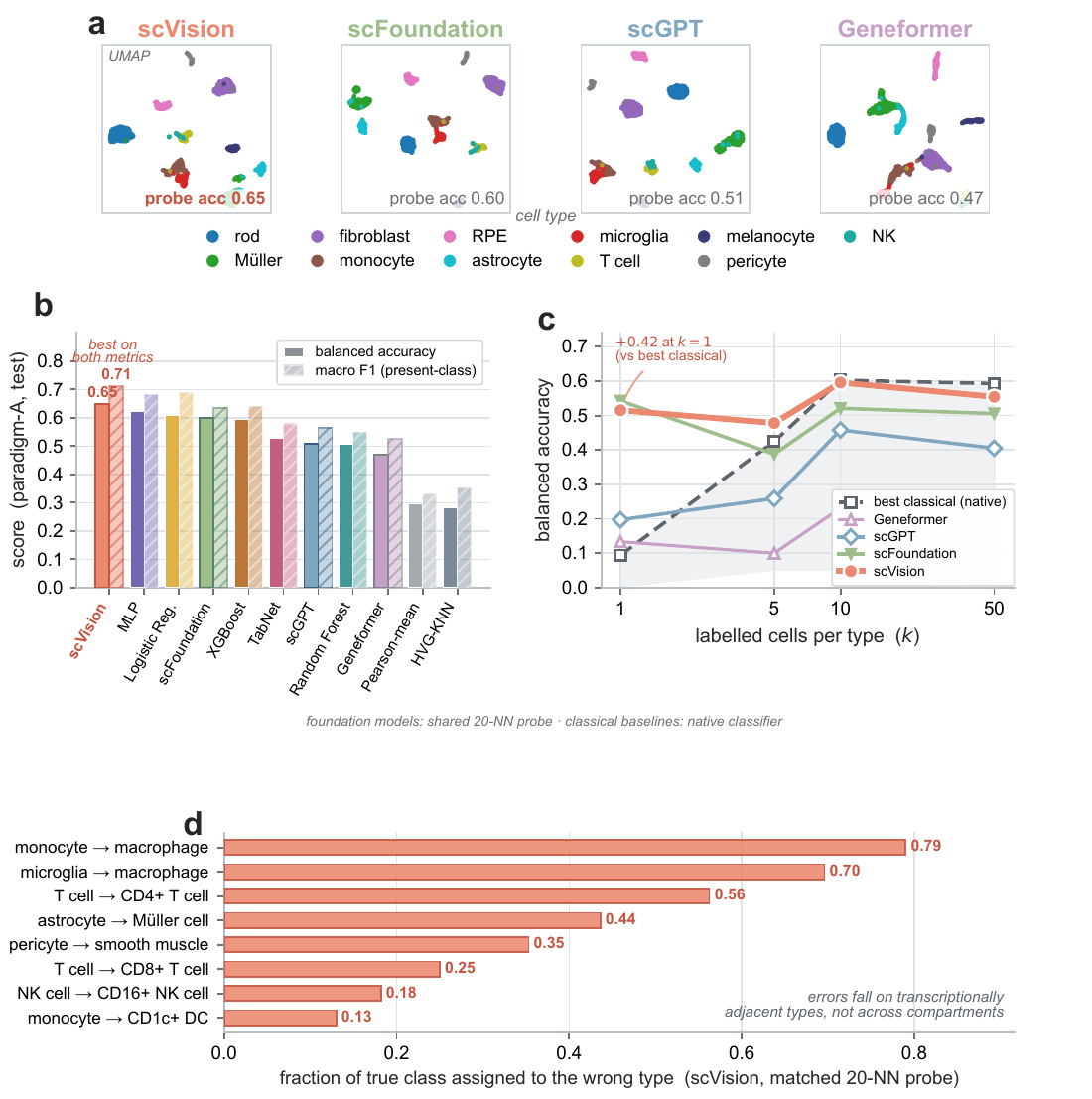}
  \caption{
  \textbf{scVision annotates the held-out human retina most accurately and is the strongest
  foundation-model embedding from a handful of labels per type, while classical learners
  close the gap as labels accumulate.} Held-out human retina, cross-atlas transfer
  (paradigm~A); every method reads a frozen representation scored on the full external test
  pool. Panels \textbf{a}--\textbf{d} are described in the continued caption overleaf.}
  \label{fig:retina}
\end{figure*}

\begin{figure*}[tp]
  \ContinuedFloat
  \caption{\textbf{a}, Frozen-embedding structure: UMAP (balanced per-class subsample)
  coloured by the eleven coarse retinal cell types, annotated with each embedding's matched
  20-nearest-neighbour probe balanced accuracy, the common cross-atlas identifiability
  metric (Fig.~\ref{fig:xatlas_purity}). scVision, scFoundation and scGPT separate the
  neuronal, glial, immune and vascular compartments into distinct islands, and scVision is the
  most identifiable embedding ($0.65$, ahead of scFoundation $0.60$, scGPT $0.51$ and
  Geneformer $0.47$); Geneformer recovers the same structure more diffusely. None saw a
  retinal cell in training. \textbf{b}, Matched read-out: foundation models share the same $20$-NN probe,
  classical and tabular baselines native. Balanced accuracy (solid) and present-class
  macro-F1 (hatched); scVision (coral) leads both, by a slim margin over the strongest
  tabular learners. \textbf{c}, Label efficiency ($k\in\{1,5,10,50\}$), foundation models
  through the same $20$-NN probe, classical at their native classifier (best-of-seven
  envelope; band, full spread). At one label per type the foundation embeddings beat the
  best classical by $0.43$, the regime where the frozen embeddings help most. \textbf{d}, Where the annotator's residual errors fall: the
  eight most frequent confusions of the matched-read-out annotator. Errors fall on
  transcriptionally adjacent types (monocytes and microglia as macrophages, T-cell
  subsets split among themselves, astrocytes as M\"uller glia, pericytes as smooth
  muscle), not on cells from a different compartment.}
\end{figure*}

\subsection*{Label-efficient annotation of held-out human retina}

We next tested scVision on an external human retina dataset. This is a challenging transfer setting because the retina contains many closely related neuronal and support cell types, including rods, cones, bipolar cells, amacrine cells, horizontal cells, ganglion cells, M\"uller glia, astrocytes, microglia and vascular cells. Many retinal subclasses differ by only a small number of transcripts. The model had not seen retinal cells during training, so the task was to assign more than 20 retinal cell types using a frozen embedding and a reference bank of 87 cell types from other atlases.

We compared all foundation models using the same 20-nearest-neighbour probe on their frozen embeddings. This allows a controlled comparison of the representations rather than the classifiers used on top of them. Under this matched readout, scVision achieved the best performance, with 0.65 balanced accuracy and 0.71 present-class macro-F1 (Fig.~\ref{fig:retina}b). The margin was modest because several methods performed well on this atlas. A multilayer perceptron reached 0.62 balanced accuracy and 0.68 macro-F1, logistic regression reached 0.61 and 0.69, scFoundation\cite{hao2024scfoundation} reached 0.60 and 0.64, and gradient-boosted trees reached 0.59 and 0.64. scGPT\cite{cui2024scgpt} and Geneformer\cite{theodoris2023geneformer} performed lower under the same matched readout, reaching 0.51 and 0.47 balanced accuracy, respectively. We focus on balanced accuracy and present-class macro-F1 because these metrics evaluate performance across all retinal cell types, including rare populations, rather than being dominated by abundant rods and cones.

This ranking was also visible in the embedding geometry. When each frozen representation was projected to two dimensions, scVision, scFoundation and scGPT separated the main retinal compartments into clear groups, including rod-dominated neurons, M\"uller glia, immune cells and vascular support cells (Fig.~\ref{fig:retina}a). Geneformer recovered the same broad structure but with more diffuse separation. The UMAP layouts themselves are qualitative and not aligned across methods, but each panel is annotated with the embedding's matched 20-nearest-neighbour probe balanced accuracy, the common cross-atlas identifiability metric (Fig.~\ref{fig:xatlas_purity}), on which scVision ranked first (0.65), ahead of scFoundation (0.60), scGPT (0.51) and Geneformer (0.47). Cell types that form clearer groups in the embedding are more easily annotated by the 20-nearest-neighbour probe.

We then tested performance when labels were scarce. For each method, we sampled $k\in\{1,5,10,50\}$ labelled cells per type and evaluated annotation on the full external dataset (Fig.~\ref{fig:retina}c). Foundation-model embeddings performed especially well in the one-shot setting: with only one labelled cell per type, scVision reached 0.52 balanced accuracy, far above the best classical learner (0.09), making it the strongest representation in the low-label regime. On the near-saturated, threshold-free ROC-AUC, scVision reached 0.998.

Finally, we examined the errors made by scVision (Fig.~\ref{fig:retina}d). The most frequent mistakes occurred between biologically related cell types. For example, monocytes and microglia were sometimes called macrophages, CD4 and CD8 T-cell subsets were confused with each other, astrocytes were sometimes called M\"uller glia, and pericytes were sometimes called smooth muscle cells. These errors occurred within related lineages, such as myeloid cells, T cells, glial cells and perivascular support cells, rather than across unrelated compartments. When scVision made mistakes, they were usually between transcriptionally similar neighbouring cell states.

\FloatBarrier  
\begin{figure*}[p]
  \centering
  \includegraphics[width=\textwidth]{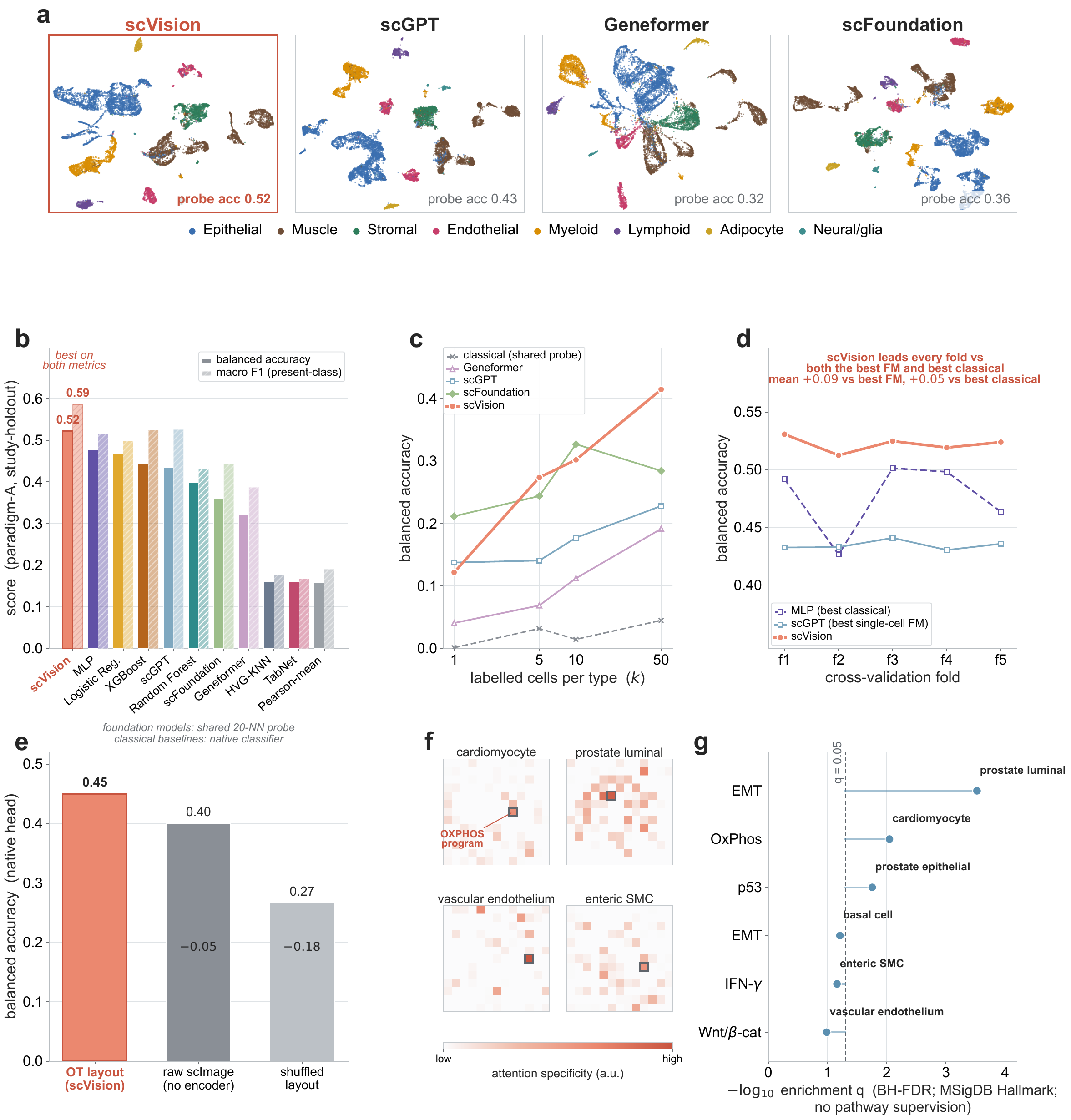}
  \caption{\textbf{scVision most accurately annotates a held-out, multi-organ
  single-nucleus reference, is the only single-cell embedding to clear the classical
  baselines, and its frozen attention assigns each fine-grained cell type its own
  program.}
  Held-out multi-organ single-nucleus reference under the cross-study holdout
  (paradigm~A); every method reads a frozen representation and is scored on the full
  external test pool against the reference bank.
  \emph{(Panel-by-panel description \textbf{a}--\textbf{g} continued on the next page.)}}
  \label{fig:crosstissue}
\end{figure*}

\begin{figure*}[t]
  \caption*{\textbf{Figure~\ref{fig:crosstissue} (continued).}
  \textbf{a}, Per-method frozen-embedding UMAP of the same held-out external test pool,
  coloured by the eight lineages the fifty-three native cell types collapse into (shared
  legend) and annotated with each model's matched 20-nearest-neighbour probe balanced
  accuracy, the common cross-atlas identifiability metric (Fig.~\ref{fig:xatlas_purity}),
  computed at the full native resolution; higher means a more readily annotated embedding.
  All four panels use one matched UMAP routine ($30$ neighbours, minimum distance $0.3$,
  cosine metric, and a PCA-$50$ pre-step where the embedding exceeds $100$ dimensions), so
  only the representation differs. scVision is the most identifiable embedding on this
  fifty-three-type reference ($0.52$), ahead of scGPT ($0.43$), scFoundation ($0.36$) and
  Geneformer ($0.32$). Raw cluster tightness tells a different story: by label-free
  15-nearest-neighbour purity scFoundation packs marginally the tightest islands ($0.80$),
  ahead of scVision ($0.78$), yet scFoundation annotates the same reference far worse
  than scVision (panel~b), so raw embedding geometry is not a proxy for cross-study transfer
  accuracy.
  \textbf{b}, Annotation accuracy at a matched read-out: every foundation model is read
  with the same $20$-nearest-neighbour probe on its frozen embedding (so the bars
  isolate representation quality rather than the choice of classifier), while the
  classical and tabular baselines use their own native classifier (their best case).
  Balanced accuracy (solid) and present-class macro-F1 (hatched) for scVision, three
  single-cell foundation models and seven classical and tabular baselines; scVision
  (coral) leads both metrics and is the only single-cell embedding to clear the strong
  non-foundation baselines, which here outrank all three single-cell foundation models.
  \textbf{c}, Label efficiency: balanced accuracy at $k\in\{1,5,10,50\}$ labelled cells
  per type for the four single-cell foundation models, each read through the matched probe,
  against the classical baselines (which share the raw-gene probe and collapse to a single
  floor band that never clears $0.05$). All four foundation embeddings sit above that floor
  from a single label, and scVision scales best across the budget range, reaching the
  highest balanced accuracy of the foundation models by $k{=}50$ ($0.41$).
  \textbf{d}, Per-fold robustness across the five deterministic, method-matched folds:
  scVision and the best foundation model (scGPT) at the matched probe, the best classical
  learner (a multilayer perceptron) native. scVision leads both on all five folds, a
  mean $+0.09$ over the best foundation model (paired $p{=}9\times10^{-6}$; $95\%$
  confidence interval $[{+}0.08,{+}0.10]$) and $+0.05$ over the best classical (paired
  $p{=}0.02$; $[{+}0.01,{+}0.08]$).
  \textbf{e}, The advantage is in the spatial representation. Balanced accuracy for the
  optimal-transport scImage layout (scVision's native head, $0.45$; the matched $20$-NN
  probe of panel~b reads it higher, at $0.52$), a multilayer-perceptron probe on the
  flattened raw scImage (``no encoder'', $0.40$), and the same encoder reading a randomly
  permuted layout (``shuffled'', $0.27$). Destroying the layout costs $0.18$ balanced
  accuracy; removing the encoder costs $0.05$; the learned gene layout carries most
  of the signal.
  \textbf{f}, scVision's frozen attention concentrates on compartment-specific scImage
  patches: specificity-corrected last-block attention for four cell types drawn from
  different compartments of the reference (a cardiomyocyte, a prostate luminal cell,
  a vascular endothelial cell and an enteric smooth-muscle cell) on the
  $13\times13$ patch grid. The cardiomyocyte peaks on its oxidative-phosphorylation patch
  (outlined); each type peaks on a patch of its own.
  \textbf{g}, That attention recovers cell-type-appropriate biology: per-cell-type
  top-attention gene programs, mapped to genes through the scImage, tested for pathway
  enrichment (hypergeometric, Benjamini--Hochberg FDR) against the Hallmark collection
  with no pathway supervision. Points mark cell-type$\rightarrow$pathway hits by
  $-\log_{10}q$, each labelled with its cell type; dashed line, $q=0.05$. Three programs
  clear significance (prostate luminal cells with epithelial--mesenchymal
  transition, cardiomyocytes with oxidative phosphorylation, prostate epithelium with the
  p53 pathway), and three more corroborate just below threshold.}
\end{figure*}

\subsection*{Zero-shot annotation across a held-out 53-type multi-organ reference}

We next tested scVision on a held-out multi-organ single-nucleus reference. This was the most demanding transfer test because the dataset contains many fine-grained cell types from several organs, including cardiomyocytes and vascular endothelial cells from heart, luminal, basal and epithelial cells from prostate, and smooth muscle cells from gut. The model had not seen this test study during training. The task was to assign each cell type using a frozen embedding and a reference bank built from other atlases.

We compared all foundation models using the same 20-nearest-neighbour probe on their frozen embeddings. This controlled comparison isolates the quality of the representation rather than the classifier placed on top of it. Under this matched readout, scVision achieved the best performance, with 0.52 balanced accuracy and 0.59 present-class macro-F1 (Fig.~\ref{fig:crosstissue}b). Other foundation models performed lower under the same readout: scGPT\cite{cui2024scgpt} reached 0.43 balanced accuracy and 0.53 macro-F1, scFoundation\cite{hao2024scfoundation} reached 0.36 and 0.44, and Geneformer\cite{theodoris2023geneformer} reached 0.32 and 0.39. The strongest non-foundation baselines were a multilayer perceptron, with 0.48 balanced accuracy and 0.51 macro-F1, and gradient-boosted trees, with 0.44 and 0.52. On this multi-organ reference, scVision was the only single-cell foundation-model embedding that exceeded the strongest classical baselines. We focus on balanced accuracy and present-class macro-F1 because these metrics evaluate performance across rare and abundant cell types more evenly.

We then tested label efficiency by varying the number of labelled examples per type. For each method, we sampled $k\in\{1,5,10,50\}$ support cells per type and evaluated performance on the full external dataset, while keeping the readout fixed (Fig.~\ref{fig:crosstissue}c). The foundation-model embeddings clearly separated from the classical baselines. The classical methods, which read the same raw-gene representation, remained near a shared floor and did not exceed 0.05 balanced accuracy even with 50 labelled cells per type. In contrast, a single labelled cell per type already lifted the foundation-model embeddings above that floor, and scVision scaled best across the budget range.

Among the foundation models, scVision's lead widened as labels accrued: by 50 labels per type it was clearly strongest, reaching 0.41 balanced accuracy, compared with 0.28 for scFoundation, 0.23 for scGPT and 0.19 for Geneformer. These results show that label efficiency comes from the pretrained representation, and that scVision benefits most as additional labels become available.

Because the five cross-validation folds were fixed and shared across methods, we also compared performance fold by fold (Fig.~\ref{fig:crosstissue}d). At the matched probe, scVision outperformed the best competing foundation model, scGPT, on all five folds. The mean improvement was 0.09 balanced accuracy (paired $p=9\times10^{-6}$; 95\% confidence interval $[+0.08,+0.10]$). scVision also outperformed the strongest non-foundation baseline, the multilayer perceptron, on all five folds, with a smaller mean improvement of 0.05 balanced accuracy (paired $p=0.02$; 95\% confidence interval $[+0.01,+0.08]$). The advantage over classical learning was modest but consistent across splits.

We next tested whether the spatial scImage layout was important (Fig.~\ref{fig:crosstissue}e). With its native head, the optimal-transport scImage reached 0.45 balanced accuracy, while the matched 20-nearest-neighbour readout in panel b reached 0.52. Replacing the encoder with a multilayer perceptron on the flattened scImage reduced performance by 0.05. Randomly permuting the gene-to-pixel layout before encoding reduced performance by 0.18. The learned spatial arrangement of genes therefore carries much of the signal, more than the encoder alone.

Finally, we examined scVision’s attention programs across organs. We mapped last-block attention back to the $13\times13$ scImage grid for cell types from different compartments, including cardiomyocytes, prostate luminal cells, vascular endothelial cells and enteric smooth muscle cells (Fig.~\ref{fig:crosstissue}f). Each cell type showed a distinct peak attention patch. The genes in these patches matched relevant cell-type biology. Without using pathway labels, pathway enrichment against the Hallmark collection identified significant programs for prostate luminal cells, enriched for epithelial--mesenchymal transition ($q=3\times10^{-4}$), cardiomyocytes, enriched for oxidative phosphorylation ($q=0.009$), and prostate epithelial cells, enriched for the p53 pathway ($q=0.02$) (Fig.~\ref{fig:crosstissue}g). Additional programs, including basal-cell epithelial--mesenchymal transition, enteric smooth-muscle interferon-$\gamma$ response and vascular endothelial Wnt signalling, were just below the significance threshold.

Together, these results show that scVision transfers across organs, improves with limited labels, and assigns distinct attention programs to fine-grained cell types. The pathway signal is modest and unsupervised, but it shows that the model often attends to genes with biology appropriate for the cell type being annotated.

\FloatBarrier

\subsection*{Embedding geometry predicts annotation accuracy across atlases}

\begin{figure*}[tp]
  \centering
  \includegraphics[width=\textwidth]{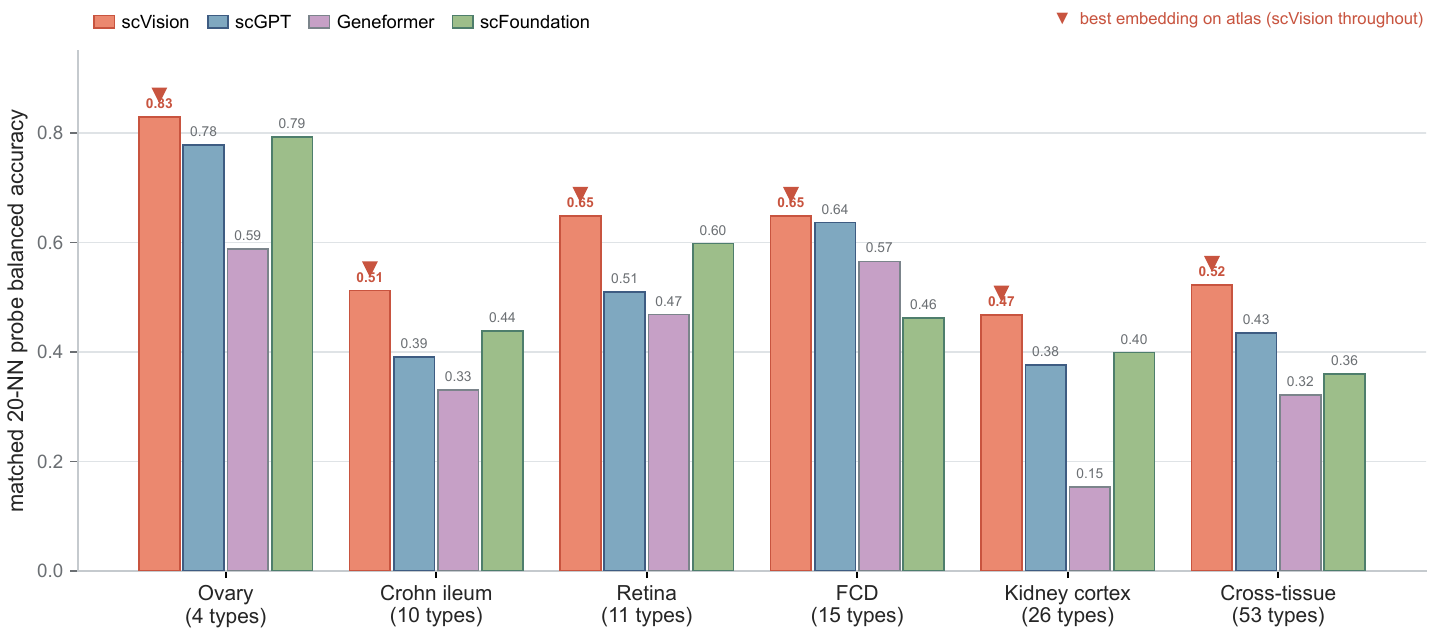}
  \caption{\textbf{The annotation ordering is written into the embedding geometry across
  every atlas: scVision yields the most identifiable frozen embedding on all six held-out
  references.} For each held-out atlas (native cell-type count in parentheses, spanning 4
  to 53 types), every embedding is scored with one common metric, the balanced accuracy
  of a matched $20$-nearest-neighbour probe read directly from the frozen embedding at native
  cell-type resolution, computed identically for scVision and the three single-cell
  foundation models. A triangle marks the best embedding on each atlas. scVision is the most
  identifiable embedding on every atlas (ovary $0.83$, Crohn's ileum $0.51$, retina $0.65$,
  focal cortical dysplasia cortex $0.65$, kidney cortex $0.47$, multi-organ reference
  $0.52$); scFoundation is second on four atlases but falls to third or fourth on the other
  two, scGPT is second on two and third on the rest, and Geneformer is third or fourth
  throughout.}
  \label{fig:xatlas_purity}
\end{figure*}

The link between annotation accuracy and embedding structure held across every held-out atlas. For each frozen embedding we measured a single common identifiability score: the balanced accuracy of a matched 20-nearest-neighbour probe read directly from the embedding at native cell-type resolution, the same metric annotated on every atlas's UMAP panels. Across all six held-out references, spanning 4 to 53 native cell types, scVision produced the most identifiable embedding on every atlas (Fig.~\ref{fig:xatlas_purity}).

scVision was the most identifiable embedding on all six atlases: the four-type ovary (0.83), the ten-type Crohn's ileum (0.51), the eleven-type retina (0.65), the fifteen-type focal cortical dysplasia cortex (0.65), the twenty-six-type kidney cortex (0.47) and the fifty-three-type multi-organ reference (0.52). No other embedding led any atlas. scFoundation was the most frequent runner-up, ranking second on the ovary (0.79), kidney cortex (0.40), retina (0.60) and Crohn's ileum (0.44), but it fell to third on the multi-organ reference and fourth on the dysplastic cortex. scGPT was second on the dysplastic cortex (0.64) and the multi-organ reference (0.43) but third on the other four atlases. Geneformer ranked third or fourth throughout. scVision was therefore the only frozen single-cell embedding that was most identifiable on every held-out atlas, from a simple four-type ovary to a complex fifty-three-type multi-organ reference.

These embeddings can be inspected directly: we projected the two largest references, the kidney cortex with 26 native cell types and the multi-organ atlas with 53 native cell types, into two dimensions using the same UMAP settings as for the dysplastic cortex (Fig.~\ref{fig:kidney}a and Fig.~\ref{fig:crosstissue}a). Each panel carries its matched-probe badge, and scVision is the most identifiable embedding in both, separating the major epithelial, immune, stromal and endothelial compartments into coherent groups while the other embeddings are progressively more diffuse.

Because this identifiability score reads each embedding the same way the annotation probe does, it tracks annotation accuracy directly: scVision is both the most identifiable embedding and the most accurate annotator on every atlas. Label-free clustering scores tell a subtler story: by raw 15-nearest-neighbour purity, scGPT forms marginally the tightest neighbourhoods on the kidney cortex and scFoundation on the multi-organ reference, confirming that placing identical labels closest together is not sufficient for accurate cross-study annotation. scVision instead organizes cell states so that a simple matched probe reads them most reliably across all six held-out atlases.

\begin{figure*}[p]
  \centering
  \includegraphics[width=\textwidth,height=0.80\textheight,keepaspectratio]{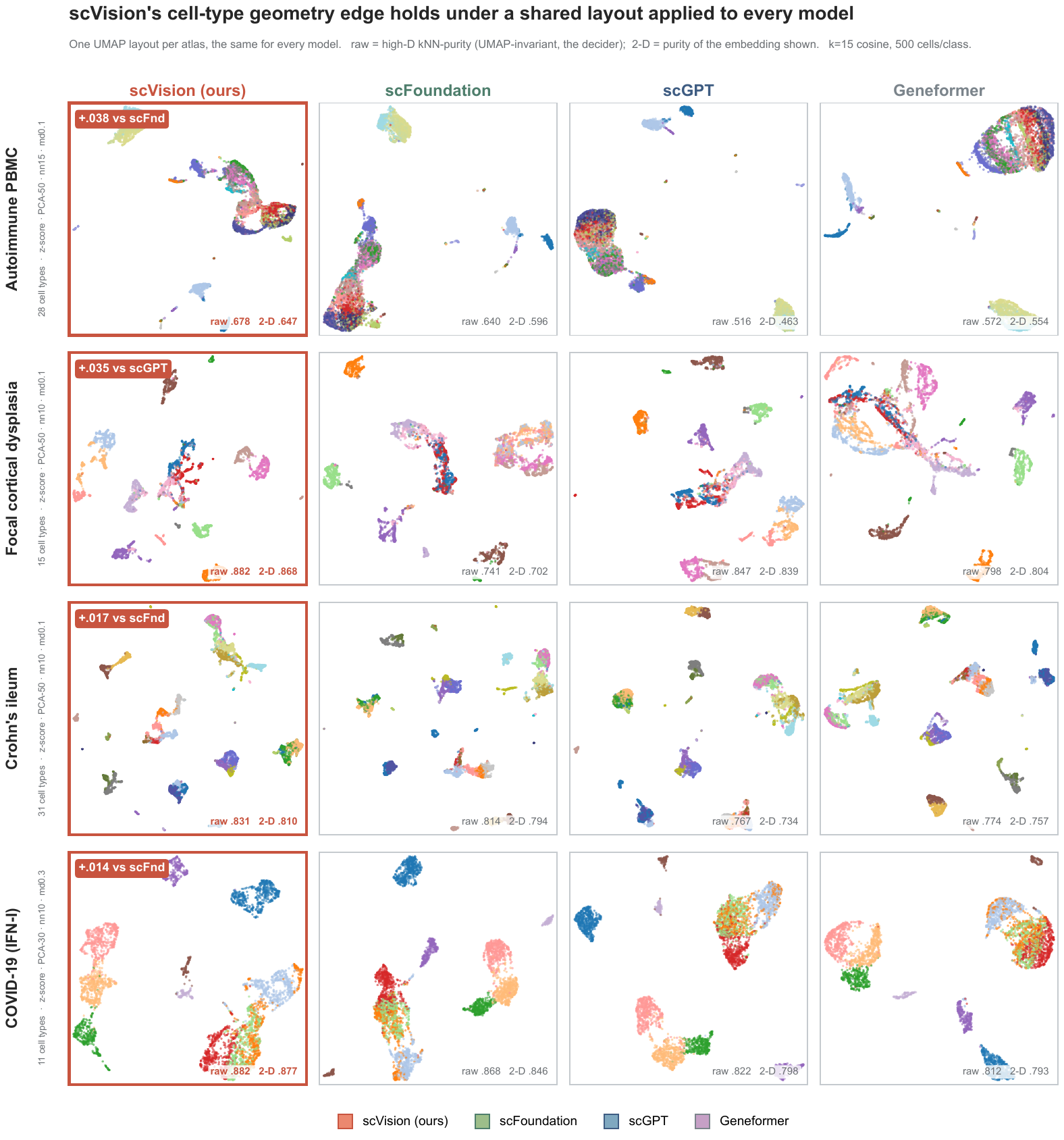}
  \caption{\textbf{On four held-out atlases, a single UMAP layout shared by every model shows
  scVision's native cell-type geometry edge is not a projection artifact.} Rows, four held-out atlases, the
  focal cortical dysplasia cortex and Crohn's ileum (also Figs.~\ref{fig:fcd}a
  and~\ref{fig:crohn}a) and two further held-out atlases, autoimmune blood (PBMC) and
  type-I-interferon COVID-19; columns, the four single-cell foundation models. For each atlas
  one UMAP layout (normalisation, PCA dimension, neighbours and minimum distance; annotated at
  left) was chosen and applied identically to all four models, so only the representation
  differs across a row; rows are ordered by scVision's high-dimensional margin. Points are
  individual cells coloured by native cell type ($11$--$31$ types per atlas). Each panel
  carries two $k$NN label-purity values ($k=15$ cosine, balanced $500$ cells per class, seed
  $0$): \emph{raw}, on the frozen high-dimensional features (UMAP-invariant, the genuine
  label-geometry decider), and \emph{2-D}, on the embedding shown. scVision (coral frame)
  gives the tightest native neighbourhoods on every atlas in both values; the coral flag is
  its raw-space margin over the closest competitor (scFoundation on three atlases, scGPT on
  the dysplastic cortex), from $+0.014$ to $+0.038$. scGPT and Geneformer are looser on every
  atlas. These modest margins match the main-text finding that within-test purity separates
  the foundation embeddings less sharply than the matched cross-study probe
  (Fig.~\ref{fig:xatlas_purity}); the point here is that the ordering purity does show is
  stable to the layout, not a product of per-model tuning.}
  \label{fig:leads_umap}
\end{figure*}

This layout-free purity is also the comparison most easily dismissed as a projection artifact, so we tested its robustness directly. On four held-out atlases on which scVision does lead label-free purity\,---\,the focal cortical dysplasia cortex and Crohn's ileum, plus two further held-out atlases, autoimmune blood (PBMC) and type-I-interferon COVID-19\,---\,we chose one UMAP layout per atlas and applied it identically to all four foundation models, with no per-model tuning (Fig.~\ref{fig:leads_umap}). scVision retained the tightest native cell-type neighbourhoods on all four, both on the UMAP-invariant high-dimensional features and on the two-dimensional embedding shown. The high-dimensional margins over the closest competitor were modest, from $+0.014$ to $+0.038$ (scFoundation on three atlases, scGPT on the dysplastic cortex), consistent with purity separating the foundation embeddings less sharply than the matched probe does; but they were stable across the shared layout rather than the product of a favourable projection.

\FloatBarrier

\subsection*{The frozen embedding integrates cells across studies without batch labels}

We next asked whether the same frozen embedding could integrate cells across studies. Annotation tests whether a representation separates cell types. Integration tests a complementary property: whether cells from different studies can be placed into a shared space while preserving biological structure and reducing technical batch effects. This is a standard benchmark for single-cell foundation models\cite{luecken2022scib}. Because scVision is pretrained only by masked image modelling and is never trained with batch labels, this integration task tests whether its frozen embedding generalizes beyond its training objective.

\begin{figure*}[tp]
  \centering
  \includegraphics[width=\textwidth]{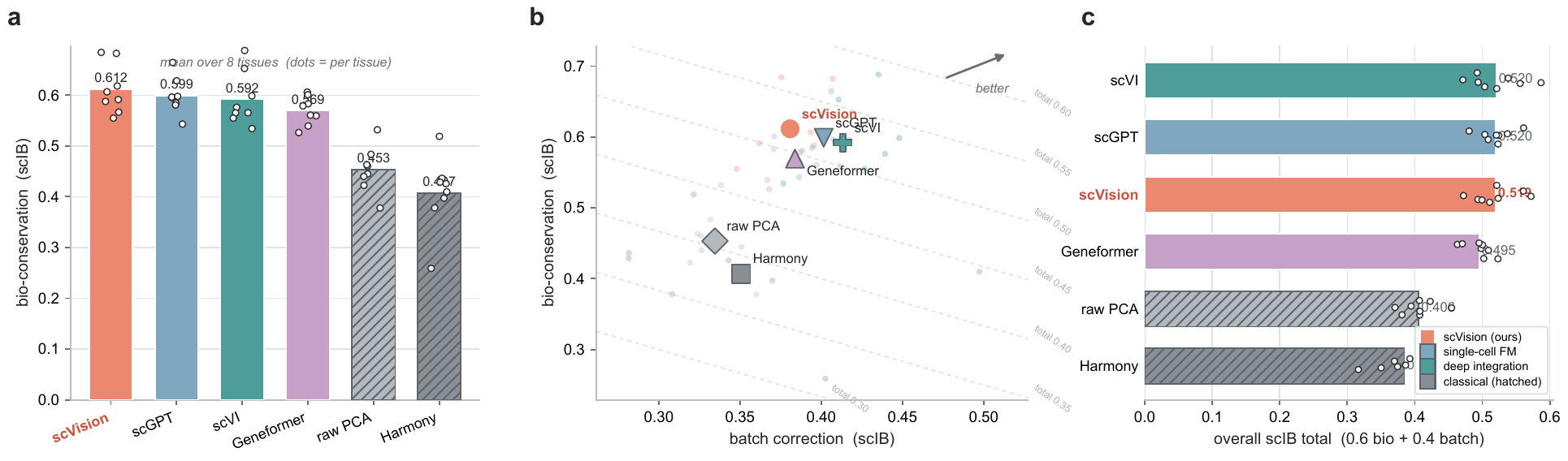}
  \caption{\textbf{A frozen scVision embedding integrates multi-study atlases: it best
  conserves biology and matches the strongest token model on the combined score,
  without being trained for the task.} Eight held-out atlases, each a balanced
  $\sim$20{,}000-cell pool spanning multiple studies (batches); six representations
  scored with the scIB v2 panel (bio-conservation $=$ mean of NMI, ARI,
  cell-type and isolated-label silhouette and cLISI; batch correction $=$ mean of batch
  silhouette, iLISI, graph connectivity and PCR; total $=0.6\,\mathrm{bio}+0.4\,\mathrm{batch}$).
  \textbf{a}, Bio-conservation per method (bar $=$ mean over tissues; white dots $=$ the
  eight individual atlases); scVision is highest and rank-1 in five of eight tissues.
  \textbf{b}, The bio--batch trade-off: batch correction ($x$) versus bio-conservation
  ($y$), faint dots per tissue and large markers per-method means; grey iso-total
  contours ($0.6\,\mathrm{bio}+0.4\,\mathrm{batch}=c$) show that scGPT's greater batch
  mixing offsets scVision's greater biological conservation, leaving their totals tied.
  \textbf{c}, Overall scIB total (horizontal bars, white dots per tissue): scVI $0.520$,
  scGPT $0.520$ and scVision $0.519$ are statistically tied (paired Wilcoxon $p\ge0.32$),
  all ahead of Geneformer and the classical baselines. Coral $=$ scVision; the token
  foundation models and scVI in solid colour; classical baselines (Harmony, PCA) hatched.
  Across-tissue confidence intervals are from $10{,}000$ bootstrap resamples.}
  \label{fig:integration}
\end{figure*}

For each of eight held-out atlases, including blood, hypothalamus, fovea, heart, kidney, lung, ovary and jejunum, we embedded a balanced pool of about 20,000 cells from multiple studies. We compared six methods: scVision, scGPT\cite{cui2024scgpt}, Geneformer\cite{theodoris2023geneformer}, scVI\cite{lopez2018scvi}, Harmony\cite{korsunsky2019harmony} and principal-component analysis. We evaluated each method using the scIB v2 benchmark, which combines five biological-conservation metrics and four batch-correction metrics into a weighted total score (Fig.~\ref{fig:integration}). This comparison is especially challenging because scVI is trained separately on each atlas with dataset identity provided as a batch covariate, whereas the foundation-model embeddings are used frozen and do not receive batch labels.

scVision preserved biological structure best overall (Fig.~\ref{fig:integration}a). It had the highest mean biological-conservation score, 0.61, with a 95\% bootstrap confidence interval of 0.58 to 0.64, and ranked first in five of eight atlases. scVision preserved biology significantly better than Geneformer (paired difference $+0.042$, $p=0.008$), scVI (paired difference $+0.020$, 95\% confidence interval $[+0.005,+0.036]$, $p=0.04$), raw PCA (difference $+0.16$, $p=0.004$) and Harmony (difference $+0.20$, $p=0.004$). scVision also slightly exceeded scGPT, although this difference was not statistically significant (paired difference $+0.013$, 95\% confidence interval $[-0.001,+0.028]$, $p=0.07$). The comparison with scVI is particularly important because scVI is designed and trained for integration, while scVision is used as a frozen embedding without batch labels.

On the combined scIB total score, the three learned embeddings were statistically tied. scVI scored 0.520, scGPT scored 0.520 and scVision scored 0.519, with all paired differences from scVision within $\pm0.001$ and not significant (paired Wilcoxon $p\ge0.32$). All three outperformed Geneformer, which scored 0.495 ($p=0.008$), and the classical baselines, raw PCA and Harmony, which scored 0.41 and 0.38, respectively ($p=0.004$).

The main trade-off was batch mixing. scVI, which is explicitly trained to remove batch effects, achieved the strongest batch-mixing score, 0.413. scVision had the lowest batch-mixing score among the foundation models (Fig.~\ref{fig:integration}b), and this was also supported by a post-hoc kBET acceptance test\cite{luecken2022scib} in which scVision scored 0.355, compared with 0.401 for scGPT and 0.417 for Geneformer. scVI and scGPT thus gained more batch mixing, whereas scVision preserved more biological structure. These effects nearly balanced in the total scIB score.

Together, these results show that scVision provides a frozen representation that preserves cell identity strongly across studies and reaches the same overall integration frontier as scGPT and scVI. Although it is not explicitly trained for batch correction and does not mix batches as strongly as scVI, it matches the best learned integration methods in total score while outperforming classical integration baselines.

\FloatBarrier

\subsection*{The frozen embedding is far cheaper to deploy than token foundation models, and its zero-shot readout is stable}

A representation meant to be used zero-shot is only useful if it can be computed
quickly and read out without per-dataset tuning. Because scVision encodes a cell
as a single fixed-size image rather than a variable-length sequence of gene
tokens, its end-to-end cost is low. Measured on one NVIDIA RTX 2080 Ti from raw
highly-variable-gene counts to the cell embedding (including each model's
required preprocessing), scVision rendered an scImage and encoded it with the
frozen vision transformer at $528$ cells per second, against $14.2$ for
Geneformer, $12.9$ for scFoundation and $1.8$ for scGPT, a $37$ to nearly
$300\times$ throughput advantage over the token models, whose per-cell gene
ranking and tokenization dominate their runtime. It
reached this throughput at a peak memory of $2.55$\,GB, below Geneformer
($8.6$\,GB) and scFoundation ($4.2$\,GB), with a parameter count ($85$\,M) in the
middle of the field. The spatial formulation therefore turns a frozen
general-purpose embedding into one that can be applied to atlas-scale data on a
single commodity GPU.

The zero-shot readout is likewise insensitive to its few inference-time choices.
Pooling the patch tokens by their mean yielded a more discriminative embedding
than the prepended class token across the held-out atlases (present-class
macro-F1 $0.498$ versus $0.471$, higher in seven of eight atlases), which is why
the mean-pooled embedding is used throughout. Annotation accuracy varied smoothly
with the nearest-neighbour probe size and was highest at small neighbourhoods,
decreasing monotonically as $k$ grew (for example on ovary from macro-F1 $0.74$
at $k{=}5$ to $0.40$ at $k{=}100$); rather than tune $k$ to maximize any method's
score we fixed a single conservative neighbourhood ($k{=}20$) a priori for every
method and atlas. Finally, re-rendering each scImage from raw counts reproduced
the stored embedding almost exactly (median cosine similarity $1.00$), so the
representation is deterministic and reproducible from the released layout and
weights.

\begin{table*}[t]
\centering
\caption{\textbf{Efficiency and scale.} Trainable parameters, embedding dimension, end-to-end encoding throughput and peak GPU memory, all measured on a single NVIDIA RTX 2080 Ti (11\,GB) for a fixed sample of 512 cells with the frozen encoder. Throughput is wall-clock from raw highly-variable-gene counts to the cell embedding and includes each model's required preprocessing (scImage rendering for scVision; gene tokenization for the token models). Parameter counts are read from each released checkpoint's state dict.}
\label{tab:efficiency}
\begin{tabular}{lrrrr}
\toprule
Model & Params (M) & Embed.\ dim & Throughput (cells/s) & Peak GPU mem (GB) \\
\midrule
\textbf{scVision} & 85.2 & 768 & 528 & 2.55 \\
scGPT & 51.3 & 512 & 1.8 & 1.36 \\
Geneformer & 104.4 & 768 & 14.2 & 8.60 \\
scFoundation & 119.4 & 3072 & 12.9 & 4.20 \\
\bottomrule
\end{tabular}
\end{table*}

\subsection*{The frozen embedding is robust to missing genes and degrades gracefully with sequencing depth}

\begin{figure*}[tp]
  \centering
  \includegraphics[width=\textwidth]{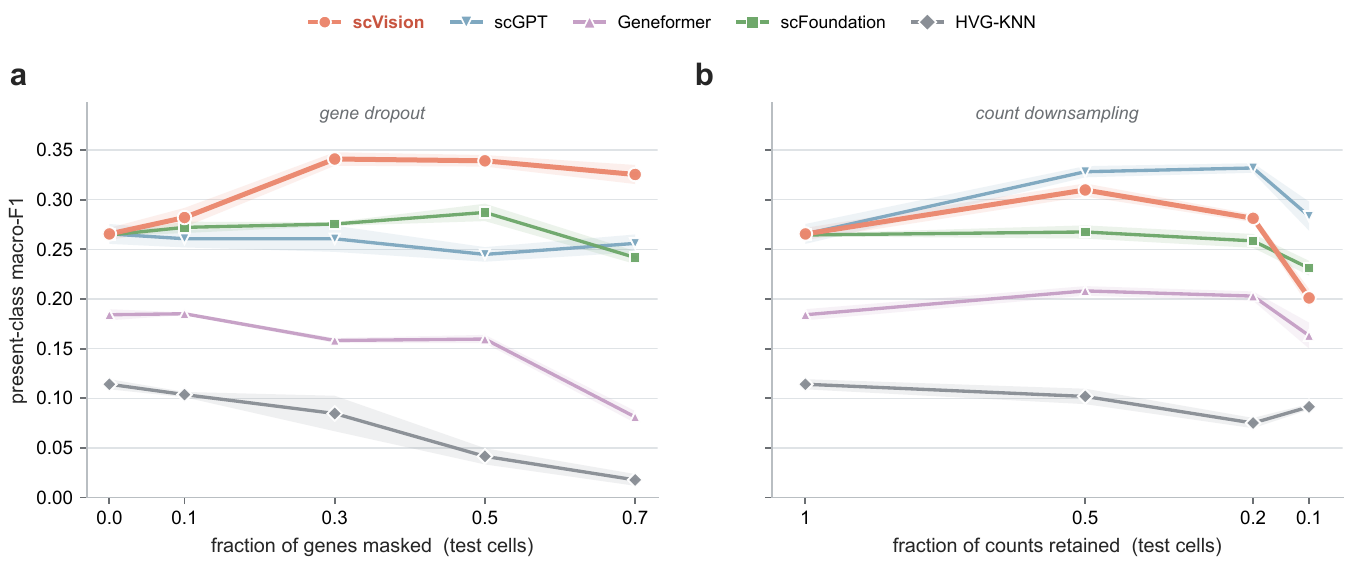}
  \caption{\textbf{The frozen scVision embedding is robust to missing genes and degrades
  gracefully with sequencing depth.} Held-out query cells of a cystic-fibrosis atlas are
  corrupted while the labelled reference bank stays clean; present-class macro-F1 of the
  shared $20$-nearest-neighbour probe is plotted for all five methods (mean over three
  folds; shaded band $=\pm1$ s.d.). \textbf{a}, Random gene dropout (a fraction of each
  query cell's genes is masked): scVision (coral), scGPT and scFoundation start tied
  ($\approx0.27$), but scVision separates as the clear leader as masking grows, rising to
  $0.34$ and holding $0.33$ at $70\%$ masked, the only method still above its clean-atlas score at this masking, while Geneformer drops to $44\%$ of its clean value and the raw
  highly-variable-gene baseline (HVG-KNN) falls to $16\%$. \textbf{b}, Count downsampling
  (counts thinned to the indicated fraction; axis reversed, depth decreasing rightward):
  the rank-based token models are the more robust, with scGPT leading at every reduced
  depth; scVision holds to half-depth but retains $76\%$ at one-tenth depth ($0.20$, below
  scGPT $0.28$ and scFoundation $0.23$), where the scImage's intensity-encoded image moves
  off the frozen encoder's input distribution.}
  \label{fig:robustness}
\end{figure*}

Real measurements are corrupted in ways a curated atlas is not: genes drop out and
libraries are sequenced shallowly. We stressed the zero-shot annotator accordingly,
corrupting only the held-out query cells of a cystic-fibrosis atlas (masking a
growing fraction of each cell's genes, or downsampling its counts), while the
labelled reference bank stayed clean, and re-ran the same $20$-nearest-neighbour probe
used throughout for every method. scVision re-rendered its scImage from the corrupted
counts; the token foundation models scGPT, scFoundation and Geneformer re-tokenized the
corrupted counts; and a raw highly-variable-gene baseline (HVG-KNN) read the corrupted
genes directly (Fig.~\ref{fig:robustness}).

Against random gene dropout the methods split sharply. The three modern foundation
models began statistically tied on the clean atlas (present-class macro-F1
$\approx0.27$), well above Geneformer ($0.18$) and the raw-gene baseline ($0.11$). As
genes were masked scVision separated as the clear leader: its score \emph{rose} to
$0.34$ by $30\%$ masking and held $0.33$ even with $70\%$ of genes removed, the only
method still above its clean score at this masking, whereas scGPT and scFoundation
stayed near their clean level ($0.24$--$0.26$), Geneformer fell to $44\%$ of its clean
value ($0.08$ from $0.18$) and the raw-gene baseline collapsed to $16\%$ ($0.02$ from
$0.11$, Fig.~\ref{fig:robustness}a). Fixing each gene to an scImage coordinate makes
masking punch holes in the image while leaving its global layout (and the encoder's
read of it) largely intact, a redundancy a flat expression vector lacks. Count
downsampling tells a different and more pointed story we report plainly
(Fig.~\ref{fig:robustness}b): here the rank-based token models were the more robust.
scGPT led at every reduced depth, reaching $0.33$ at one-fifth of counts and still $0.28$
at one-tenth, essentially unaffected by depth; scVision held through half-depth ($0.31$)
but degraded under heavier thinning, retaining $76\%$ of its clean score at one-tenth
depth ($0.20$), below scGPT ($0.28$) and scFoundation ($0.23$) though still above
Geneformer ($0.16$) and the raw-gene baseline. This is what the representations predict:
the scImage encodes expression as image intensity, so uniform downsampling dims the whole
image and pushes it off the frozen encoder's input distribution, whereas the token
models' rank-based tokenization is by construction near-invariant to total depth. The
frozen spatial embedding is thus markedly robust to \emph{which} genes are
observed (where it leads every baseline) and degrades gracefully with
\emph{how deeply} they are sequenced, which localizes its one sensitivity to sequencing
depth and points to count augmentation during rendering as a natural route to close the
remaining gap.

\FloatBarrier

\section*{Discussion}

scVision frames single-cell representation learning as a vision problem. Instead of treating a cell as an unordered set of gene tokens, scVision places each gene at a fixed position using optimal transport. Genes that are often co-expressed are placed close together, so each transcriptome can be represented as an image. A vision transformer is then pretrained by masked image modelling on 72 million human cells to produce a frozen cell representation.

In a strict study-level holdout setting, where entire studies are withheld from training, this frozen representation performed strongly across all six external atlases tested: kidney cortex, ovary, focal cortical dysplasia cortex, Crohn's-disease ileum, human retina and a multi-organ single-nucleus reference. scVision achieved the highest balanced accuracy and present-class macro-F1 on each atlas. It was also highly label-efficient. In many cases, one labelled cell per type with scVision performed better than other foundation models given many more labelled examples. When another foundation model was stronger at the smallest label budget, scVision often improved more as additional labels were added.

Several controls show that this advantage comes from the spatial gene layout. When the gene-to-pixel arrangement was randomly permuted before encoding, performance dropped substantially. This loss was larger than the drop caused by removing the vision encoder and applying a probe directly to the raw scImage. The learned spatial arrangement of genes is therefore not a cosmetic transformation; it carries much of the useful signal. The model benefits from the fact that genes with related expression patterns are placed near one another.

The results also show that scVision’s advantage is mainly in cross-study transfer. Within a single held-out tissue, some token-based foundation models can organize cells into locally clean neighbourhoods. However, those neighbourhoods do not always transfer well when cells must be assigned to an external reference bank built from other studies. In other words, an embedding can look well organized within one dataset but still fail across the study gap. scVision appears to preserve cell-type structure more reliably across studies that the model never saw during training.

Representing a cell as an image also makes the model easier to interpret. Because co-expressed genes occupy nearby positions, attention maps can be mapped back to genes and read as gene-expression programs. Without any pathway supervision, scVision recovered programs that matched known biology, including proximal-tubule injury programs in kidney, an unfolded-protein-response program in ovarian pericytes, and a TREM2--DAP12 inflamed-stroma program in Crohn's disease ileum. These programs were not limited to one tissue. Across 585 matched cell-type pairs from 28 atlases, attention-derived programs from the same cell type overlapped 6.4-fold more than expected by chance. A four-gene myeloid program, IRF8, PIM1, TRAF1 and CCND2, appeared in microglia, kidney macrophages, breast macrophages and blood dendritic cells. The attention maps capture reusable cell-type programs across organs.

These findings address an important issue in single-cell foundation models. Previous evaluations have found that token-based models can struggle to outperform regularized logistic regression for cell-type annotation without fine-tuning\cite{boiarsky2023,kedzierska2023,liu2023sceval}. Our results suggest that the limitation may not be only model scale or training-set size. Instead, the way the transcriptome is represented may be critical. By giving genes biologically meaningful positions, scVision improves transfer under a stricter zero-shot evaluation. This image-based formulation also enables operations that are difficult in token models. For example, one can mask a spatial neighbourhood of genes, perturbing a co-regulated module at once, and then measure how the cell embedding changes. This provides a natural way to perform in silico perturbation at the level of gene programs rather than individual genes.

 We emphasize balanced accuracy and present-class macro-F1 throughout because these metrics measure the quality of the actual cell-type calls and give weight to rare but biologically important populations, rather than being dominated by abundant classes. Under this readout scVision is the most accurate annotator on every held-out atlas, though on the Crohn's-disease ileum and the multi-organ reference its margin over the strongest classical baselines is consistent but modest.

This study also has several limitations. All evaluations used frozen, zero-shot embeddings, which is the right setting for measuring representation quality but does not test the best possible performance after fine-tuning. The training data are human, so performance in other species remains to be tested. The current scImage uses one pan-tissue gene layout shared across all cells, which may not capture every tissue-specific relationship. Finally, attention-derived pathway programs provide useful biological clues, but they should not be treated as calibrated regulatory inference.

By turning each cell into an image, scVision brings computer-vision methods into single-cell biology. This opens several future directions. Fine-tuning the frozen backbone may improve performance when labeled data are available. Masking spatial gene neighbourhoods could support systematic in silico perturbation studies. The image formulation may also extend naturally to spatial transcriptomics and paired modalities such as surface protein or chromatin accessibility. Overall, these results suggest that progress in single-cell foundation models may come not only from larger datasets or larger models, but also from representations that better encode the relational structure of the transcriptome. Giving genes biologically meaningful positions is one such representation, and it makes the cell accessible to the tools of vision.

\section*{Methods}

\begin{table*}[t]
\centering
\caption{\textbf{List of notation.} Symbols used throughout the Methods, grouped by the scImage construction (left) and the vision transformer and its zero-shot evaluation (right).}
\label{tab:notation}
\begin{tabular}{@{}ll@{\hspace{2.5em}}ll@{}}
\toprule
\textbf{Symbol} & \textbf{Description} & \textbf{Symbol} & \textbf{Description}\\
\midrule
$G$ & Number of selected genes & $P$ & Patch side length\\
$M$ & Number of lattice pixels & $N$ & Number of patch tokens\\
$\mathbf{X}$ & scImage of a cell & $d$ & Token embedding dimension\\
$\mathbf{e},\tilde{\mathbf{e}}$ & Raw / $z$-scored expression & $H$ & Number of attention heads\\
$\mathbf{D}^{g}$ & Gene--gene distance matrix & $d_h$ & Per-head width ($d/H$)\\
$\mathbf{D}^{s}$ & Pixel--pixel distance matrix & $\mathbf{E}$ & Patch-embedding projection\\
$r_{ik}$ & Pearson correlation of genes $i,k$ & $\mathbf{p}_j$ & Positional embedding\\
$\mathbf{c}_j$ & Coordinate of pixel $j$ & $\mathbf{U}$ & Token sequence\\
$\boldsymbol{\mu},\boldsymbol{\nu}$ & Gene / pixel marginals & $\mathbf{Q},\mathbf{K},\mathbf{V}$ & Query, key, value\\
$\Pi$ & Coupling polytope & $f_{\theta},g_{\phi}$ & Encoder / decoder\\
$\mathbf{T}^{\star}$ & Optimal coupling (gene layout) & $\mathcal{M}$ & Masked-patch index set\\
$L(a,b)$ & Kullback--Leibler loss & $\mathcal{L}_{\mathrm{MAE}}$ & Reconstruction loss\\
$\epsilon$ & Entropic-regularization weight & $\mathbf{h}$ & Cell embedding\\
$\mathbf{P}$ & Soft projection matrix & $\alpha^{(h)}_j$ & Class-token$\,\to\,$patch attention\\
$i,k\,/\,j,l$ & Gene / pixel indices & $a^{(c)}_j$ & Cell-type attention profile\\
\bottomrule
\end{tabular}
\end{table*}

\subsection*{Training data and quality control}

We assembled the pretraining dataset from the CZ CELLxGENE Census\cite{czi2023},
drawing primary human cells through the Census interface and keeping only
records marked as primary data, so that a cell deposited in several studies is
counted once. These primary cells span the major published human atlases aggregated by the Census, including pan-tissue references\cite{tabulasapiens2022,eraslan2022crosstissue} and organ- and disease-focused atlases of lung\cite{sikkema2023hlca}, heart\cite{litvinukova2020heart}, breast\cite{kumar2023hbca,reed2024breast}, hypothalamus\cite{tadross2025hypomap}, gut\cite{elmentaite2021gut}, blood\cite{yazar2022onek1k,stephenson2021covid,combat2022covid}, kidney\cite{krishna2021ccrcc}, cerebellum\cite{aldinger2021cerebellum} and fallopian tube\cite{ulrich2022fallopian}. Cells were filtered by transcriptome-wide quality
control (at least 200 detected genes, at least 500 total counts and no more
than 20\% mitochondrial content) and targeted-panel assays (BD Rhapsody
Targeted, 10x Targeted Gene Expression) were excluded. The resulting dataset
comprises approximately 94 million cells. Data were handled throughout with the scanpy and anndata single-cell ecosystem\cite{wolf2018scanpy,virshup2024anndata}.

To measure generalization to new studies rather than to new cells of seen
studies, we used a study-level holdout, the cross-study holdout denoted
paradigm~A in the figures: roughly 10\% of cells, drawn from approximately 25
entirely held-out studies, form the external test pool, while the remaining
studies are partitioned into training and validation by donor, so that no donor
crosses the split. All splits use a fixed seed (42). The training split
comprises 72 million cells.

\subsection*{scImage construction by optimal transport}

Table~\ref{tab:notation} collects the notation used below. Each cell is rendered
as a single-channel image
$\mathbf{X}\in\mathbb{R}^{1\times104\times104}$, one gene per pixel (Fig.~\ref{fig:overview}), so that the
lattice carries $M=104\times104=10{,}816$ pixels and we select exactly $G=M$ genes
to fill it. The genes are the $G$ most highly variable genes under a Seurat-style
criterion, ranking each gene by its variance after a loess fit of variance against
mean expression, computed in a single chunked pass over the dataset.

We place genes on the lattice by Gromov--Wasserstein optimal
transport\cite{memoli2011gromov,peyre2019computational}, which
aligns the relational geometry of genes to that of pixels without requiring the
two spaces to be compared point by point. Co-expression defines a gene--gene
distance matrix $\mathbf{D}^{g}\in\mathbb{R}^{G\times G}$ with entries the
correlation distance $D^{g}_{ik}=1-r_{ik}$, where $r_{ik}$ is the Pearson
correlation of genes $i$ and $k$ across cells; the lattice defines a
pixel--pixel distance matrix $\mathbf{D}^{s}\in\mathbb{R}^{M\times M}$ with
$D^{s}_{jl}=\lVert\mathbf{c}_j-\mathbf{c}_l\rVert_2$ the Euclidean distance between
the two-dimensional coordinates $\mathbf{c}_j,\mathbf{c}_l$ of pixels $j$ and $l$.
With uniform marginals $\boldsymbol{\mu}=\mathbf{1}_G/G$ over genes and
$\boldsymbol{\nu}=\mathbf{1}_M/M$ over pixels, and the coupling polytope
\begin{equation}
\Pi(\boldsymbol{\mu},\boldsymbol{\nu})\overset{\mathrm{def}}{=}\bigl\{\mathbf{T}\in\mathbb{R}^{G\times M}_{\ge 0}:\ \mathbf{T}\mathbf{1}_M=\boldsymbol{\mu},\ \mathbf{T}^{\top}\mathbf{1}_G=\boldsymbol{\nu}\bigr\},
\end{equation}
the layout is the coupling that best matches the two geometries,
\begin{equation}
\mathbf{T}^{\star}=\operatorname*{arg\,min}_{\mathbf{T}\in\Pi(\boldsymbol{\mu},\boldsymbol{\nu})}\ \sum_{i,j,k,l} L\!\bigl(D^{g}_{ik},D^{s}_{jl}\bigr)\,T_{ij}\,T_{kl},
\end{equation}
penalizing pairs of genes whose co-expression distance is poorly preserved by the
spatial distance of the pixels they are sent to. The discrepancy is the Kullback--Leibler loss
$L(a,b)=\mathrm{KL}(a\,\|\,b)\overset{\mathrm{def}}{=}a\log(a/b)-a+b$, solved with no entropic
regularization ($\epsilon=0$)\cite{cuturi2013sinkhorn} for 200 iterations on a
sample of 100{,}000 cells stratified by tissue (seed 42). Optimal transport has
become a workhorse for single-cell analysis, from trajectory
inference\cite{schiebinger2019waddingtonot} to multi-omic
alignment\cite{klein2025moscot} and image-based embeddings of single-cell data\cite{islam2023genomap}; here it instead fixes a single shared layout,
computed once and applied pan-tissue to every cell.

The coupling is retained as a dense, soft projection matrix
$\mathbf{P}=M\,\mathbf{T}^{\star}\in\mathbb{R}^{G\times M}$, the plan scaled by
the number of pixels, with no Hungarian assignment or hard argmax, so that
each gene spreads a smoothly distributed mass over the lattice rather than
collapsing onto a single pixel. A cell with raw expression
$\mathbf{e}\in\mathbb{R}^{G}$ is first $z$-scored gene-wise with the layout's
stored mean $\mu^{g}_i$ and standard deviation $\sigma^{g}_i$,
$\tilde{e}_i=(e_i-\mu^{g}_i)/\sigma^{g}_i$, then projected and reshaped,
\begin{equation}
\mathbf{y}=\mathbf{P}^{\top}\tilde{\mathbf{e}}\in\mathbb{R}^{M},\qquad
\mathbf{X}=\operatorname{reshape}_{104\times104}(\mathbf{y}),
\end{equation}
with the reshape taken in column-major (Fortran) order and the image stored at
half precision as a $(1,104,104)$ tensor. Because co-expressed genes occupy
neighbouring pixels, co-regulation appears as local texture and cellular identity
as global pattern. The scImages so produced are reproducible within a cell type
and distinct across types (Supplementary Fig.~\ref{fig:sgenomaps}).

\subsection*{scVision architecture and masked-image pretraining}

scVision is a vision transformer trained as a masked
autoencoder\cite{he2022mae,dosovitskiy2021vit}, one of several self-supervised
vision paradigms that also include contrastive and self-distillation
objectives\cite{chen2020simclr,caron2021dino,oquab2024dinov2} and
alternative hierarchical backbones\cite{liu2021swin}. The scImage
$\mathbf{X}\in\mathbb{R}^{1\times104\times104}$ is partitioned into non-overlapping
$P\times P=8\times8$ patches, which tile it as a $13\times13$ grid of $N=169$
tokens, each patch flattening $P^{2}=64$ pixels, equivalently 64 genes. Writing
$\mathbf{x}^{(j)}_{p}\in\mathbb{R}^{P^{2}}$ for the flattened pixels of patch
$j$, each patch is linearly embedded and endowed with a position,
\begin{equation}
\mathbf{u}_j=\mathbf{E}\,\mathbf{x}^{(j)}_{p}+\mathbf{p}_j\in\mathbb{R}^{d},\qquad j=1,\dots,N,
\end{equation}
with a shared projection $\mathbf{E}\in\mathbb{R}^{d\times P^{2}}$ to dimension
$d=768$ and a learnable positional embedding $\mathbf{p}_j$ encoding the patch's
location on the $13\times13$ grid; prepending a learnable class token
$\mathbf{u}_0$ forms the input sequence
$\mathbf{U}^{(0)}=[\mathbf{u}_0,\mathbf{u}_1,\dots,\mathbf{u}_N]^{\top}\in\mathbb{R}^{(N+1)\times d}$.
The encoder $f_{\theta}$ is a ViT-base network ($d=768$, with $L=12$
transformer blocks, $H=12$ attention heads and an MLP ratio of 4) for
approximately 86 million parameters\cite{vaswani2017attention}. Each block
updates the sequence by pre-normalized multi-head self-attention and a
token-wise MLP. For head $h$, with per-head width $d_h=d/H=64$, the queries,
keys and values are the linear projections
$\mathbf{Q}_h=\mathbf{U}\mathbf{W}^{Q}_h$, $\mathbf{K}_h=\mathbf{U}\mathbf{W}^{K}_h$
and $\mathbf{V}_h=\mathbf{U}\mathbf{W}^{V}_h$, and tokens are mixed by scaled
dot-product attention,
\begin{equation}
\operatorname{Attn}_h(\mathbf{U})=\operatorname{softmax}\!\left(\frac{\mathbf{Q}_h\mathbf{K}_h^{\top}}{\sqrt{d_h}}\right)\mathbf{V}_h.
\end{equation}
The $H$ heads are concatenated and projected,
$\operatorname{MHSA}(\mathbf{U})=[\operatorname{Attn}_1(\mathbf{U}),\dots,\operatorname{Attn}_H(\mathbf{U})]\,\mathbf{W}^{O}$,
and, with layer normalization $\operatorname{LN}$ and residual connections, a
block maps
\begin{equation}
\begin{aligned}
\mathbf{U}'&=\mathbf{U}+\operatorname{MHSA}(\operatorname{LN}(\mathbf{U})),\\
\mathbf{U}''&=\mathbf{U}'+\operatorname{MLP}(\operatorname{LN}(\mathbf{U}')),
\end{aligned}
\end{equation}
so that the encoder $f_{\theta}$ is the composition of $L=12$ such blocks.

At each step a random subset $\mathcal{M}\subset\{1,\dots,N\}$ of patches, of size
$\lfloor 0.75\,N\rfloor$, is masked; only the visible patches are encoded, and a
lightweight decoder $g_{\phi}$ (512-dimensional, with 8 blocks and 16 heads,
approximately 14 million parameters) reconstructs the masked patches from the
encoded visible tokens together with learned mask tokens. Writing $\mathbf{z}_j$
for the per-patch-normalized pixel target of patch $j$ and $\hat{\mathbf{z}}_j$ for
its reconstruction, the objective is the mean squared error over masked patches
only,
\begin{equation}
\mathcal{L}_{\mathrm{MAE}}(\theta,\phi)=\frac{1}{|\mathcal{M}|}\sum_{j\in\mathcal{M}}\bigl\lVert\hat{\mathbf{z}}_j-\mathbf{z}_j\bigr\rVert_2^{2}.
\end{equation}
The model was optimized with AdamW ($\beta_1=0.9$, $\beta_2=0.95$), a cosine
learning-rate schedule with linear warmup, gradient clipping at a norm of 5 and
half-precision mixed-precision training, on the 72-million-cell training split.
The released backbone is the ViT-base, patch-8 model used for all evaluations.

\subsection*{Zero-shot evaluation and baselines}

All evaluations use the frozen pretrained encoder $f_{\theta^{\star}}$ with no
fine-tuning. Writing $\mathbf{t}_1,\dots,\mathbf{t}_N$ for the encoded patch tokens
of a cell, its embedding is their mean, L2-normalized,
\begin{equation}
\mathbf{h}=\frac{\bar{\mathbf{t}}}{\lVert\bar{\mathbf{t}}\rVert_2},\qquad
\bar{\mathbf{t}}=\frac{1}{N}\sum_{j=1}^{N}\mathbf{t}_j\in\mathbb{R}^{768}.
\end{equation}
Cell-type annotation is a 20-nearest-neighbour probe in cosine space against a
labelled reference bank, following the reference-based annotation paradigm of
established label-transfer
tools\cite{aran2019singler,dominguezconde2022celltypist,kiselev2018scmap}: with
unit-norm embeddings the cosine similarity is the inner product
$s(\mathbf{h},\mathbf{h}')=\mathbf{h}^{\top}\mathbf{h}'$, and the predicted label
aggregates similarity-weighted votes over the set $\mathcal{N}_{20}(\mathbf{h})$ of
20 nearest reference cells,
\begin{equation}
\hat{y}=\operatorname*{arg\,max}_{c}\sum_{r\in\mathcal{N}_{20}(\mathbf{h})} s(\mathbf{h},\mathbf{h}_r)\,\mathbb{I}[\,y_r=c\,].
\end{equation}
We report balanced accuracy, the mean per-class recall
$\mathrm{BA}=\lvert\mathcal{C}\rvert^{-1}\sum_{c\in\mathcal{C}}\mathrm{TP}_c/(\mathrm{TP}_c+\mathrm{FN}_c)$,
and present-class macro-F1, the macro average of the $F_1$ score over only the
cell types $\mathcal{C}_{\mathrm{p}}$ actually present in the held-out atlas,
\begin{equation}
\mathrm{F1}_{\mathrm{p}}=\frac{1}{\lvert\mathcal{C}_{\mathrm{p}}\rvert}\sum_{c\in\mathcal{C}_{\mathrm{p}}}\frac{2\,\mathrm{Prec}_c\,\mathrm{Rec}_c}{\mathrm{Prec}_c+\mathrm{Rec}_c},
\end{equation}
and, where stated, threshold-free macro-average precision and area under the ROC
curve.

Label efficiency was measured by $k$-shot evaluation at $k\in\{1,5,10,50\}$
labelled cells per type under five-fold cross-validation; the foundation models
share the identical nearest-neighbour probe, so that differences isolate
representation quality, while each classical baseline is additionally allowed to
fit its own native classifier on the same support. Multi-study integration was
assessed on eight held-out atlases: for each we embedded a balanced
$\sim$20{,}000-cell pool spanning multiple studies and scored batch correction and
biological conservation with the scIB v2 panel\cite{luecken2022scib}, the standard framework for ranking the many available single-cell integration methods\cite{stuart2019seurat,butler2018seurat,haghverdi2018mnn,hie2019scanorama,polanski2020bbknn,xu2021scanvi,lotfollahi2022scarches,tran2020benchmark}, comparing the
frozen scVision embedding against the token foundation models scGPT and Geneformer, the
classical baselines Harmony\cite{korsunsky2019harmony} and principal-component analysis,
and the deep-learning integration model scVI\cite{lopez2018scvi,gayoso2022scvitools}, all read from the
matched input. Unlike the foundation-model embeddings, which are read frozen, scVI was
trained separately on each atlas's raw counts in the same 10{,}816-gene space with the
dataset identity supplied as its only batch covariate (a negative-binomial variational
autoencoder, $30$ latent dimensions, two hidden layers, up to $400$ epochs with early
stopping); its latent posterior mean served as the integrated embedding. Bio-conservation is the mean of NMI, ARI,
cell-type and isolated-label silhouette and cLISI; batch correction the mean of batch
silhouette, iLISI, graph connectivity and PCR; the total weights them $0.6/0.4$.
Across-tissue confidence intervals are from $10{,}000$ bootstrap resamples and paired
method comparisons use the Wilcoxon signed-rank test over the eight tissues; a
supplementary post-hoc kBET acceptance test (pure-Python $k$-nearest-neighbour
$\chi^2$ batch test, $k_0=25$) reports per-method batch mixing.

We compared against three single-cell foundation models (Geneformer\cite{theodoris2023geneformer},
scGPT\cite{cui2024scgpt} and scFoundation\cite{hao2024scfoundation}), each
encoded under the matched protocol, and against classical and tabular learners on
the raw highly-variable-gene space: regularized logistic regression, a
multilayer perceptron, $k$-nearest neighbours, a correlation classifier, a random
forest\cite{breiman2001random}, gradient-boosted trees\cite{chen2016xgboost} and
TabNet\cite{arik2021tabnet}, implemented with
scikit-learn\cite{pedregosa2011scikit}.

\subsection*{Gene-program inference and cross-tissue recurrence}

To read gene-expression programs from the model, we used the attention that the
class token pays to each patch in the final transformer block. Writing
$\mathbf{q}_0$ for the class-token query and $\mathbf{k}_j$ for the key of patch
$j$ in head $h$ of that block, this attention is the softmax
\begin{equation}
\alpha^{(h)}_j=\frac{\exp(\mathbf{q}_0^{\top}\mathbf{k}_j/\sqrt{d_h})}{\sum_{j'=1}^{N}\exp(\mathbf{q}_0^{\top}\mathbf{k}_{j'}/\sqrt{d_h})},
\end{equation}
and the patch profile $a^{(c)}_j$ of cell type $c$ averages $\alpha^{(h)}_j$
over the $H$ heads and over the cells of type $c$. Programs were made
cell-type-specific by subtracting the per-patch mean across cell types,
\begin{equation}
\tilde{a}^{(c)}_j=a^{(c)}_j-\frac{1}{|\mathcal{C}|}\sum_{c'\in\mathcal{C}}a^{(c')}_j,
\end{equation}
which removes the shared attention-sink patches. The eight patches of largest
$\tilde{a}^{(c)}_j$ (each an $8\times8$, 64-gene spatial neighbourhood of the
scImage) define the attended program, from which the top 100 genes, in
descending attention order, were mapped back through the layout and tested for
pathway enrichment by the hypergeometric test against MSigDB
Hallmark\cite{liberzon2015hallmark,subramanian2005gsea} and, per atlas, KEGG,
Reactome and GO biological process, with Benjamini--Hochberg control of the
false-discovery rate. To confirm these programs were not artefacts of the
per-patch mean-subtraction, we recomputed them at four cell-resampling seeds
under two alternative specificity definitions, the uncentred raw profile
$a^{(c)}_j$ and a z-score dividing $\tilde{a}^{(c)}_j$ by its per-patch standard
deviation across cell types. Cross-cell-type program overlap fell monotonically
with centring stringency (mean off-diagonal Jaccard $0.27$, $0.047$ and $0.003$
for the raw, mean-subtracted and z-scored variants), and the
microglia--endothelium program identity (Jaccard $1.0$) and its p53 enrichment
(KEGG $q=0.020$, identical leading genes) were recovered under both the raw and
mean-subtracted variants but removed by the z-score, which by construction
discards patches shared across cell types, indicating that this shared
vascular--immune module reflects genuine co-attention rather than the chosen
specificity metric. These programs were likewise stable to the number of cells
sampled per type: halving or doubling the published budget of 200 cells per type
(to 100 or 400) left the mean off-diagonal Jaccard essentially unchanged
($0.046$, $0.044$ and $0.054$), and both the endothelial p53 anchor
(KEGG $q=0.020$, the same five leading genes) and oligodendrocyte-precursor RET
signalling (Reactome $q\approx0.002$, the same five genes) were recovered at every
budget; only the microglial arm of the vascular--immune module weakened at the
smallest budget, its overlap with endothelium falling to Jaccard $0.47$ and its
p53 enrichment to non-significance at 100 cells, independently mirroring the
cell-resampling hierarchy above, in which the endothelial p53 program is the most
reproducible and microglia the most sampling-sensitive member of the module. The
published top-100 program was itself unchanged for attended-patch budgets of 6, 8
and 12 patches, as expected since the 100 highest-attention genes are drawn from
the two most-attended patches, which are selected independently of the total
budget.

To test whether the same programs recur across tissues, we matched cell types
appearing in more than one of 28 atlases (585 cross-tissue pairs over 77 cell
types) and measured the Jaccard overlap
$J(\mathcal{G}_1,\mathcal{G}_2)=|\mathcal{G}_1\cap\mathcal{G}_2|/|\mathcal{G}_1\cup\mathcal{G}_2|$
of their top-100 gene programs $\mathcal{G}_1,\mathcal{G}_2$ against a null in which
cell-type labels are randomly deranged (1000 permutations, seed 20240601).

\subsection*{In silico perturbation and the disease axis}

The spatial representation supports an in silico perturbation native to its
geometry, complementing supervised perturbation-response models that learn
cellular responses from labelled perturbation
data\cite{lotfollahi2019scgen,roohani2024gears,lotfollahi2023cpa}. For a target
gene we masked the single $8\times8$ patch (64 genes) containing its co-expression
neighbourhood in the scImages of control cells, re-encoded them and measured the
mean shift in embedding,
$\Delta_{\mathrm{pred}}=\bar{\mathbf{h}}_{\mathrm{mask}}-\bar{\mathbf{h}}_{\mathrm{ctrl}}$.
We compared its cosine similarity $\cos(\Delta_{\mathrm{pred}},\Delta_{\mathrm{obs}})$
to the measured perturbation direction $\Delta_{\mathrm{obs}}$ from large-scale
CRISPR perturbation
screens\cite{norman2019perturb,adamson2016perturb,replogle2022perturbseq} against a
null in which a random patch is masked instead, by a paired Wilcoxon test.

To probe disease, we fit a logistic-regression direction
$\mathbf{w}\in\mathbb{R}^{768}$ in the frozen embedding separating disease from
control cells, scoring a cell by $\mathbf{w}^{\top}\mathbf{h}$, and evaluated
whether this axis transfers across assay and disease subtype, from dilated to
hypertrophic
cardiomyopathy\cite{chaffin2022cardiomyopathy,reichart2022cardiomyopathies,koenig2022heartfailure}, scoring
the transferred axis by area under the ROC curve.

\subsection*{Statistical analysis}

Group comparisons used the tests named with each analysis: a paired Wilcoxon test
for the perturbation null, hypergeometric tests with Benjamini--Hochberg
correction for pathway enrichment, and permutation tests for cross-tissue
recurrence. Cross-study annotation was evaluated over five folds, and every
quantitative value reported in the Results derives from a stored evaluation
record.

\section*{Data availability}

All datasets analysed in this study are publicly available. The pretraining
dataset and every held-out evaluation atlas were obtained from the CZ CELLxGENE
Census of human single-cell transcriptomes\cite{czi2023}, itself assembled from
public tissue atlases\cite{tabulasapiens2022,eraslan2022crosstissue,sikkema2023hlca};
the held-out atlases, their tissues and their cell-type resolutions are listed in
Supplementary Table~\ref{tab:balacc}. The large-scale CRISPR perturbation screens
used for the in silico perturbation analysis are the public Perturb-seq
datasets\cite{norman2019perturb,adamson2016perturb,replogle2022perturbseq} cited in
the Methods. All processed expression matrices, the shared scImage layout and the
intermediate evaluation records required to reproduce the analyses will be released with the code (see Code
availability); detailed preprocessing and rendering procedures are described in the Methods. No new human
or animal data were generated for this study.

\section*{Code availability}

The code for scVision, including data preprocessing, scImage rendering, masked-image
pretraining and the zero-shot evaluation pipelines, is reserved for now and will be
released publicly, with a persistent identifier, upon publication. Further information
and updates are available at the project page, \url{https://islamlab.org/scvision}.

\bibliography{main}

@article{boiarsky2023,
  author  = {Boiarsky, Rebecca and Singh, Nalini and Buendia, Alejandro and Getz, Gad and Sontag, David},
  title   = {A deep dive into single-cell {RNA} sequencing foundation models},
  journal = {bioRxiv},
  year    = {2023},
  doi     = {10.1101/2023.10.19.563100},
  note    = {Preprint}
}

@article{kedzierska2023,
  author  = {Kedzierska, Kasia Z. and Crawford, Lorin and Amini, Ava P. and Lu, Alex X.},
  title   = {Zero-shot evaluation reveals limitations of single-cell foundation models},
  journal = {Genome Biology},
  volume  = {26},
  number  = {1},
  pages   = {101},
  year    = {2025},
  doi     = {10.1186/s13059-025-03574-x}
}

@article{liu2023sceval,
  author  = {Liu, Tianyu and Li, Kexing and Wang, Yuge and Li, Hongyu and Zhao, Hongyu},
  title   = {Evaluating the Utilities of Foundation Models in Single-Cell Data Analysis},
  journal = {Advanced Science},
  volume  = {13},
  number  = {27},
  pages   = {e14490},
  year    = {2026},
  doi     = {10.1002/advs.202514490}
}

@article{czi2023,
  author  = {Abdulla, Shibla and Aevermann, Brian and Assis, Pedro and others},
  title   = {{CZ CELLxGENE Discover}: a single-cell data platform for scalable exploration, analysis and modeling of aggregated data},
  journal = {Nucleic Acids Research},
  volume  = {53},
  number  = {D1},
  pages   = {D886--D900},
  year    = {2025},
  doi     = {10.1093/nar/gkae1142}
}

@article{yang2022scbert,
  author  = {Yang, Fan and Wang, Wenchuan and Wang, Fang and others},
  title   = {{scBERT} as a large-scale pretrained deep language model for cell type annotation of single-cell {RNA-seq} data},
  journal = {Nature Machine Intelligence},
  volume  = {4},
  pages   = {852--866},
  year    = {2022},
  doi     = {10.1038/s42256-022-00534-z}
}

@article{theodoris2023geneformer,
  author  = {Theodoris, Christina V. and Xiao, Ling and Chopra, Anant and others},
  title   = {Transfer learning enables predictions in network biology},
  journal = {Nature},
  volume  = {618},
  pages   = {616--624},
  year    = {2023},
  doi     = {10.1038/s41586-023-06139-9}
}

@article{cui2024scgpt,
  author  = {Cui, Haotian and Wang, Chloe and Maan, Hassaan and others},
  title   = {{scGPT}: toward building a foundation model for single-cell multi-omics using generative {AI}},
  journal = {Nature Methods},
  volume  = {21},
  pages   = {1470--1480},
  year    = {2024},
  doi     = {10.1038/s41592-024-02201-0}
}

@article{hao2024scfoundation,
  author  = {Hao, Minsheng and Gong, Jing and Zeng, Xin and others},
  title   = {Large-scale foundation model on single-cell transcriptomics},
  journal = {Nature Methods},
  volume  = {21},
  pages   = {1481--1491},
  year    = {2024},
  doi     = {10.1038/s41592-024-02305-7}
}

@article{rosen2023uce,
  author  = {Rosen, Yanay and Roohani, Yusuf and Agrawal, Ayush and Samotorcan, Leon and {Tabula Sapiens Consortium} and Quake, Stephen R. and Leskovec, Jure},
  title   = {Universal cell embedding provides a foundation model for cell biology},
  journal = {Nature},
  year    = {2026},
  doi     = {10.1038/s41586-026-10689-z}
}

@article{islam2023genomap,
  author  = {Islam, Md Tauhidul and Xing, Lei},
  title   = {Cartography of Genomic Interactions Enables Deep Analysis of Single-Cell Expression Data},
  journal = {Nature Communications},
  volume  = {14},
  pages   = {679},
  year    = {2023},
  doi     = {10.1038/s41467-023-36383-6}
}

@inproceedings{vaswani2017attention,
  author    = {Vaswani, Ashish and Shazeer, Noam and Parmar, Niki and Uszkoreit, Jakob and Jones, Llion and Gomez, Aidan N. and Kaiser, {\L}ukasz and Polosukhin, Illia},
  title     = {Attention is All You Need},
  booktitle = {Advances in Neural Information Processing Systems (NeurIPS)},
  year      = {2017}
}

@inproceedings{dosovitskiy2021vit,
  author    = {Dosovitskiy, Alexey and Beyer, Lucas and Kolesnikov, Alexander and Weissenborn, Dirk and Zhai, Xiaohua and Unterthiner, Thomas and Dehghani, Mostafa and Minderer, Matthias and Heigold, Georg and Gelly, Sylvain and Uszkoreit, Jakob and Houlsby, Neil},
  title     = {An Image is Worth 16x16 Words: Transformers for Image Recognition at Scale},
  booktitle = {International Conference on Learning Representations (ICLR)},
  year      = {2021}
}

@inproceedings{he2022mae,
  author    = {He, Kaiming and Chen, Xinlei and Xie, Saining and Li, Yanghao and Doll\'{a}r, Piotr and Girshick, Ross},
  title     = {Masked Autoencoders Are Scalable Vision Learners},
  booktitle = {Proceedings of the IEEE/CVF Conference on Computer Vision and Pattern Recognition (CVPR)},
  pages     = {16000--16009},
  year      = {2022}
}

@article{memoli2011gromov,
  author  = {M\'{e}moli, Facundo},
  title   = {Gromov-Wasserstein Distances and the Metric Approach to Object Matching},
  journal = {Foundations of Computational Mathematics},
  volume  = {11},
  number  = {4},
  pages   = {417--487},
  year    = {2011},
  doi     = {10.1007/s10208-011-9093-5}
}

@inproceedings{cuturi2013sinkhorn,
  author    = {Cuturi, Marco},
  title     = {Sinkhorn Distances: Lightspeed Computation of Optimal Transport},
  booktitle = {Advances in Neural Information Processing Systems (NeurIPS)},
  year      = {2013}
}

@inproceedings{chen2016xgboost,
  author    = {Chen, Tianqi and Guestrin, Carlos},
  title     = {{XGBoost}: A Scalable Tree Boosting System},
  booktitle = {Proceedings of the 22nd ACM SIGKDD International Conference on Knowledge Discovery and Data Mining (KDD)},
  pages     = {785--794},
  year      = {2016},
  doi       = {10.1145/2939672.2939785}
}

@inproceedings{arik2021tabnet,
  author    = {Arik, Sercan \"{O}. and Pfister, Tomas},
  title     = {{TabNet}: Attentive Interpretable Tabular Learning},
  booktitle = {Proceedings of the AAAI Conference on Artificial Intelligence},
  volume    = {35},
  pages     = {6679--6687},
  year      = {2021}
}

@article{breiman2001random,
  author  = {Breiman, Leo},
  title   = {Random Forests},
  journal = {Machine Learning},
  volume  = {45},
  number  = {1},
  pages   = {5--32},
  year    = {2001},
  doi     = {10.1023/A:1010933404324}
}

@article{pedregosa2011scikit,
  author  = {Pedregosa, Fabian and Varoquaux, Ga\"{e}l and Gramfort, Alexandre and others},
  title   = {Scikit-learn: Machine Learning in {Python}},
  journal = {Journal of Machine Learning Research},
  volume  = {12},
  pages   = {2825--2830},
  year    = {2011}
}

@article{liberzon2015hallmark,
  author  = {Liberzon, Arthur and Birger, Chet and Thorvaldsd\'{o}ttir, Helga and Ghandi, Mahmoud and Mesirov, Jill P. and Tamayo, Pablo},
  title   = {The Molecular Signatures Database Hallmark Gene Set Collection},
  journal = {Cell Systems},
  volume  = {1},
  number  = {6},
  pages   = {417--425},
  year    = {2015},
  doi     = {10.1016/j.cels.2015.12.004}
}

@article{subramanian2005gsea,
  author  = {Subramanian, Aravind and Tamayo, Pablo and Mootha, Vamsi K. and others},
  title   = {Gene set enrichment analysis: A knowledge-based approach for interpreting genome-wide expression profiles},
  journal = {Proceedings of the National Academy of Sciences},
  volume  = {102},
  number  = {43},
  pages   = {15545--15550},
  year    = {2005},
  doi     = {10.1073/pnas.0506580102}
}

@article{norman2019perturb,
  author  = {Norman, Thomas M. and Horlbeck, Max A. and Replogle, Joseph M. and others},
  title   = {Exploring genetic interaction manifolds constructed from rich single-cell phenotypes},
  journal = {Science},
  volume  = {365},
  number  = {6455},
  pages   = {786--793},
  year    = {2019},
  doi     = {10.1126/science.aax4438}
}

@article{adamson2016perturb,
  author  = {Adamson, Britt and Norman, Thomas M. and Jost, Marco and others},
  title   = {A Multiplexed Single-Cell {CRISPR} Screening Platform Enables Systematic Dissection of the Unfolded Protein Response},
  journal = {Cell},
  volume  = {167},
  number  = {7},
  pages   = {1867--1882},
  year    = {2016},
  doi     = {10.1016/j.cell.2016.11.048}
}

@article{luecken2022scib,
  author  = {Luecken, Malte D. and B\"{u}ttner, Maren and Chaichoompu, Kridsadakorn and others},
  title   = {Benchmarking atlas-level data integration in single-cell genomics},
  journal = {Nature Methods},
  volume  = {19},
  number  = {1},
  pages   = {41--50},
  year    = {2022},
  doi     = {10.1038/s41592-021-01336-8}
}

@article{lopez2018scvi,
  author  = {Lopez, Romain and Regier, Jeffrey and Cole, Michael B. and Jordan, Michael I. and Yosef, Nir},
  title   = {Deep generative modeling for single-cell transcriptomics},
  journal = {Nature Methods},
  volume  = {15},
  number  = {12},
  pages   = {1053--1058},
  year    = {2018},
  doi     = {10.1038/s41592-018-0229-2}
}

@article{korsunsky2019harmony,
  author  = {Korsunsky, Ilya and Millard, Nghia and Fan, Jean and others},
  title   = {Fast, sensitive and accurate integration of single-cell data with {Harmony}},
  journal = {Nature Methods},
  volume  = {16},
  number  = {12},
  pages   = {1289--1296},
  year    = {2019},
  doi     = {10.1038/s41592-019-0619-0}
}

@article{aran2019singler,
  author  = {Aran, Dvir and Looney, Agnieszka P. and Liu, Leqian and others},
  title   = {Reference-based analysis of lung single-cell sequencing reveals a transitional profibrotic macrophage},
  journal = {Nature Immunology},
  volume  = {20},
  number  = {2},
  pages   = {163--172},
  year    = {2019},
  doi     = {10.1038/s41590-018-0276-y}
}

@article{dominguezconde2022celltypist,
  author  = {Dom\'{i}nguez Conde, C. and Xu, C. and Jarvis, L. B. and others},
  title   = {Cross-tissue immune cell analysis reveals tissue-specific features in humans},
  journal = {Science},
  volume  = {376},
  number  = {6594},
  pages   = {eabl5197},
  year    = {2022},
  doi     = {10.1126/science.abl5197}
}

@article{kiselev2018scmap,
  author  = {Kiselev, Vladimir Yu. and Yiu, Andrew and Hemberg, Martin},
  title   = {{scmap}: projection of single-cell {RNA-seq} data across data sets},
  journal = {Nature Methods},
  volume  = {15},
  number  = {5},
  pages   = {359--362},
  year    = {2018},
  doi     = {10.1038/nmeth.4644}
}

@article{stuart2019seurat,
  author  = {Stuart, Tim and Butler, Andrew and Hoffman, Paul and others},
  title   = {Comprehensive Integration of Single-Cell Data},
  journal = {Cell},
  volume  = {177},
  number  = {7},
  pages   = {1888--1902},
  year    = {2019},
  doi     = {10.1016/j.cell.2019.05.031}
}

@article{butler2018seurat,
  author  = {Butler, Andrew and Hoffman, Paul and Smibert, Peter and others},
  title   = {Integrating single-cell transcriptomic data across different conditions, technologies, and species},
  journal = {Nature Biotechnology},
  volume  = {36},
  number  = {5},
  pages   = {411--420},
  year    = {2018},
  doi     = {10.1038/nbt.4096}
}

@article{xu2021scanvi,
  author  = {Xu, Chenling and Lopez, Romain and Mehlman, Edouard and others},
  title   = {Probabilistic harmonization and annotation of single-cell transcriptomics data with deep generative models},
  journal = {Molecular Systems Biology},
  volume  = {17},
  number  = {1},
  pages   = {e9620},
  year    = {2021},
  doi     = {10.15252/msb.20209620}
}

@article{lotfollahi2022scarches,
  author  = {Lotfollahi, Mohammad and Naghipourfar, Mohsen and Luecken, Malte D. and others},
  title   = {Mapping single-cell data to reference atlases by transfer learning},
  journal = {Nature Biotechnology},
  volume  = {40},
  number  = {1},
  pages   = {121--130},
  year    = {2022},
  doi     = {10.1038/s41587-021-01001-7}
}

@article{polanski2020bbknn,
  author  = {Pola\'{n}ski, Krzysztof and Young, Matthew D. and Miao, Zhichao and others},
  title   = {{BBKNN}: fast batch alignment of single cell transcriptomes},
  journal = {Bioinformatics},
  volume  = {36},
  number  = {3},
  pages   = {964--965},
  year    = {2020},
  doi     = {10.1093/bioinformatics/btz625}
}

@article{hie2019scanorama,
  author  = {Hie, Brian and Bryson, Bryan and Berger, Bonnie},
  title   = {Efficient integration of heterogeneous single-cell transcriptomes using {Scanorama}},
  journal = {Nature Biotechnology},
  volume  = {37},
  number  = {6},
  pages   = {685--691},
  year    = {2019},
  doi     = {10.1038/s41587-019-0113-3}
}

@article{haghverdi2018mnn,
  author  = {Haghverdi, Laleh and Lun, Aaron T. L. and Morgan, Michael D. and others},
  title   = {Batch effects in single-cell {RNA}-sequencing data are corrected by matching mutual nearest neighbors},
  journal = {Nature Biotechnology},
  volume  = {36},
  number  = {5},
  pages   = {421--427},
  year    = {2018},
  doi     = {10.1038/nbt.4091}
}

@article{tran2020benchmark,
  author  = {Tran, Hoa Thi Nhu and Ang, Kok Siong and Chevrier, Marion and others},
  title   = {A benchmark of batch-effect correction methods for single-cell {RNA} sequencing data},
  journal = {Genome Biology},
  volume  = {21},
  number  = {1},
  pages   = {12},
  year    = {2020},
  doi     = {10.1186/s13059-019-1850-9}
}

@article{roohani2024gears,
  author  = {Roohani, Yusuf and Huang, Kexin and Leskovec, Jure},
  title   = {Predicting transcriptional outcomes of novel multigene perturbations with {GEARS}},
  journal = {Nature Biotechnology},
  volume  = {42},
  number  = {6},
  pages   = {927--935},
  year    = {2024},
  doi     = {10.1038/s41587-023-01905-6}
}

@article{lotfollahi2019scgen,
  author  = {Lotfollahi, Mohammad and Wolf, F. Alexander and Theis, Fabian J.},
  title   = {{scGen} predicts single-cell perturbation responses},
  journal = {Nature Methods},
  volume  = {16},
  number  = {8},
  pages   = {715--721},
  year    = {2019},
  doi     = {10.1038/s41592-019-0494-8}
}

@article{lotfollahi2023cpa,
  author  = {Lotfollahi, Mohammad and Klimovskaia Susmelj, Anna and De Donno, Carlo and others},
  title   = {Predicting cellular responses to complex perturbations in high-throughput screens},
  journal = {Molecular Systems Biology},
  volume  = {19},
  number  = {6},
  pages   = {e11517},
  year    = {2023},
  doi     = {10.15252/msb.202211517}
}

@article{replogle2022perturbseq,
  author  = {Replogle, Joseph M. and Saunders, Reuben A. and Pogson, Angela N. and others},
  title   = {Mapping information-rich genotype-phenotype landscapes with genome-scale {Perturb-seq}},
  journal = {Cell},
  volume  = {185},
  number  = {14},
  pages   = {2559--2575},
  year    = {2022},
  doi     = {10.1016/j.cell.2022.05.013}
}

@article{wolf2018scanpy,
  author  = {Wolf, F. Alexander and Angerer, Philipp and Theis, Fabian J.},
  title   = {{SCANPY}: large-scale single-cell gene expression data analysis},
  journal = {Genome Biology},
  volume  = {19},
  number  = {1},
  pages   = {15},
  year    = {2018},
  doi     = {10.1186/s13059-017-1382-0}
}

@article{gayoso2022scvitools,
  author  = {Gayoso, Adam and Lopez, Romain and Xing, Galen and others},
  title   = {A {Python} library for probabilistic analysis of single-cell omics data},
  journal = {Nature Biotechnology},
  volume  = {40},
  number  = {2},
  pages   = {163--166},
  year    = {2022},
  doi     = {10.1038/s41587-021-01206-w}
}

@article{virshup2024anndata,
  author  = {Virshup, Isaac and Rybakov, Sergei and Theis, Fabian J. and others},
  title   = {{anndata}: Access and store annotated data matrices},
  journal = {Journal of Open Source Software},
  volume  = {9},
  number  = {101},
  pages   = {4371},
  year    = {2024},
  doi     = {10.21105/joss.04371}
}

@article{tabulasapiens2022,
  author  = {{The Tabula Sapiens Consortium}},
  title   = {The {Tabula Sapiens}: A multiple-organ, single-cell transcriptomic atlas of humans},
  journal = {Science},
  volume  = {376},
  number  = {6594},
  pages   = {eabl4896},
  year    = {2022},
  doi     = {10.1126/science.abl4896}
}

@article{tabulamuris2018,
  author  = {{The Tabula Muris Consortium}},
  title   = {Single-cell transcriptomics of 20 mouse organs creates a {Tabula Muris}},
  journal = {Nature},
  volume  = {562},
  number  = {7727},
  pages   = {367--372},
  year    = {2018},
  doi     = {10.1038/s41586-018-0590-4}
}

@article{sikkema2023hlca,
  author  = {Sikkema, Lisa and Ram\'{i}rez-Su\'{a}stegui, Ciro and Strobl, Daniel C. and others},
  title   = {An integrated cell atlas of the lung in health and disease},
  journal = {Nature Medicine},
  volume  = {29},
  number  = {6},
  pages   = {1563--1577},
  year    = {2023},
  doi     = {10.1038/s41591-023-02327-2}
}

@article{litvinukova2020heart,
  author  = {Litvi\v{n}ukov\'{a}, Monika and Talavera-L\'{o}pez, Carlos and Maatz, Henrike and others},
  title   = {Cells of the adult human heart},
  journal = {Nature},
  volume  = {588},
  number  = {7838},
  pages   = {466--472},
  year    = {2020},
  doi     = {10.1038/s41586-020-2797-4}
}

@article{kumar2023hbca,
  author  = {Kumar, Tapsi and Nee, Kevin and Wei, Runmin and others},
  title   = {A spatially resolved single-cell genomic atlas of the adult human breast},
  journal = {Nature},
  volume  = {620},
  number  = {7972},
  pages   = {181--191},
  year    = {2023},
  doi     = {10.1038/s41586-023-06252-9}
}

@article{reed2024breast,
  author  = {Reed, Austin D. and Pensa, Sara and Steif, Adi and others},
  title   = {A single-cell atlas enables mapping of homeostatic cellular shifts in the adult human breast},
  journal = {Nature Genetics},
  volume  = {56},
  number  = {4},
  pages   = {652--662},
  year    = {2024},
  doi     = {10.1038/s41588-024-01688-9}
}

@article{tadross2025hypomap,
  author  = {Tadross, John A. and Steuernagel, Lukas and Dowsett, Georgina K. C. and others},
  title   = {A comprehensive spatio-cellular map of the human hypothalamus},
  journal = {Nature},
  volume  = {639},
  number  = {8055},
  pages   = {708--716},
  year    = {2025},
  doi     = {10.1038/s41586-024-08504-8}
}

@article{elmentaite2021gut,
  author  = {Elmentaite, Rasa and Kumasaka, Natsuhiko and Roberts, Kenny and others},
  title   = {Cells of the human intestinal tract mapped across space and time},
  journal = {Nature},
  volume  = {597},
  number  = {7875},
  pages   = {250--255},
  year    = {2021},
  doi     = {10.1038/s41586-021-03852-1}
}

@article{yazar2022onek1k,
  author  = {Yazar, Seyhan and Alquicira-Hernandez, Jose and Wing, Kristof and others},
  title   = {Single-cell {eQTL} mapping identifies cell type-specific genetic control of autoimmune disease},
  journal = {Science},
  volume  = {376},
  number  = {6589},
  pages   = {eabf3041},
  year    = {2022},
  doi     = {10.1126/science.abf3041}
}

@article{stephenson2021covid,
  author  = {Stephenson, Emily and Reynolds, Gary and Botting, Rachel A. and others},
  title   = {Single-cell multi-omics analysis of the immune response in {COVID-19}},
  journal = {Nature Medicine},
  volume  = {27},
  number  = {5},
  pages   = {904--916},
  year    = {2021},
  doi     = {10.1038/s41591-021-01329-2}
}

@article{combat2022covid,
  author  = {{COvid-19 Multi-omics Blood ATlas (COMBAT) Consortium}},
  title   = {A blood atlas of {COVID-19} defines hallmarks of disease severity and specificity},
  journal = {Cell},
  volume  = {185},
  number  = {5},
  pages   = {916--938},
  year    = {2022},
  doi     = {10.1016/j.cell.2022.01.012}
}

@article{krishna2021ccrcc,
  author  = {Krishna, Chirag and DiNatale, Renzo G. and Kuo, Fengshen and others},
  title   = {Single-cell sequencing links multiregional immune landscapes and tissue-resident {T} cells in {ccRCC} to tumor topology and therapy efficacy},
  journal = {Cancer Cell},
  volume  = {39},
  number  = {5},
  pages   = {662--677},
  year    = {2021},
  doi     = {10.1016/j.ccell.2021.03.007}
}

@article{eraslan2022crosstissue,
  author  = {Eraslan, G\"{o}kcen and Drokhlyansky, Eugene and Anand, Shankara and others},
  title   = {Single-nucleus cross-tissue molecular reference maps toward understanding disease gene function},
  journal = {Science},
  volume  = {376},
  number  = {6594},
  pages   = {eabl4290},
  year    = {2022},
  doi     = {10.1126/science.abl4290}
}

@article{aldinger2021cerebellum,
  author  = {Aldinger, Kimberly A. and Thomson, Zachary and Phelps, Ian G. and others},
  title   = {Spatial and cell type transcriptional landscape of human cerebellar development},
  journal = {Nature Neuroscience},
  volume  = {24},
  number  = {8},
  pages   = {1163--1175},
  year    = {2021},
  doi     = {10.1038/s41593-021-00872-y}
}

@article{ulrich2022fallopian,
  author  = {Ulrich, Nicole D. and Shen, Yu-chi and Ma, Qianyi and others},
  title   = {Cellular heterogeneity of human fallopian tubes in normal and hydrosalpinx disease states identified using {scRNA-seq}},
  journal = {Developmental Cell},
  volume  = {57},
  number  = {7},
  pages   = {914--929},
  year    = {2022},
  doi     = {10.1016/j.devcel.2022.02.017}
}

@inproceedings{caron2021dino,
  author    = {Caron, Mathilde and Touvron, Hugo and Misra, Ishan and others},
  title     = {Emerging Properties in Self-Supervised Vision Transformers},
  booktitle = {Proceedings of the IEEE/CVF International Conference on Computer Vision (ICCV)},
  pages     = {9650--9660},
  year      = {2021}
}

@article{oquab2024dinov2,
  author  = {Oquab, Maxime and Darcet, Timoth\'{e}e and Moutakanni, Th\'{e}o and others},
  title   = {{DINOv2}: Learning Robust Visual Features without Supervision},
  journal = {Transactions on Machine Learning Research},
  year    = {2024}
}

@inproceedings{chen2020simclr,
  author    = {Chen, Ting and Kornblith, Simon and Norouzi, Mohammad and Hinton, Geoffrey},
  title     = {A Simple Framework for Contrastive Learning of Visual Representations},
  booktitle = {Proceedings of the 37th International Conference on Machine Learning (ICML)},
  year      = {2020}
}

@inproceedings{liu2021swin,
  author    = {Liu, Ze and Lin, Yutong and Cao, Yue and others},
  title     = {{Swin Transformer}: Hierarchical Vision Transformer using Shifted Windows},
  booktitle = {Proceedings of the IEEE/CVF International Conference on Computer Vision (ICCV)},
  year      = {2021}
}

@article{schiebinger2019waddingtonot,
  author  = {Schiebinger, Geoffrey and Shu, Jian and Tabaka, Marcin and others},
  title   = {Optimal-Transport Analysis of Single-Cell Gene Expression Identifies Developmental Trajectories in Reprogramming},
  journal = {Cell},
  volume  = {176},
  number  = {4},
  pages   = {928--943},
  year    = {2019},
  doi     = {10.1016/j.cell.2019.01.006}
}

@article{peyre2019computational,
  author  = {Peyr\'{e}, Gabriel and Cuturi, Marco},
  title   = {Computational Optimal Transport: With Applications to Data Science},
  journal = {Foundations and Trends in Machine Learning},
  volume  = {11},
  number  = {5--6},
  pages   = {355--607},
  year    = {2019},
  doi     = {10.1561/2200000073}
}

@article{klein2025moscot,
  author  = {Klein, Dominik and Palla, Giovanni and Lange, Marius and others},
  title   = {Mapping cells through time and space with {moscot}},
  journal = {Nature},
  volume  = {638},
  pages   = {1065--1075},
  year    = {2025},
  doi     = {10.1038/s41586-024-08453-2}
}

@article{chaffin2022cardiomyopathy,
  author  = {Chaffin, Mark and Papangeli, Irinna and Simonson, Bridget and others},
  title   = {Single-nucleus profiling of human dilated and hypertrophic cardiomyopathy},
  journal = {Nature},
  volume  = {608},
  pages   = {174--180},
  year    = {2022},
  doi     = {10.1038/s41586-022-04817-8}
}

@article{reichart2022cardiomyopathies,
  author  = {Reichart, Daniel and Lindberg, Eric L. and Maatz, Henrike and others},
  title   = {Pathogenic variants damage cell composition and single cell transcription in cardiomyopathies},
  journal = {Science},
  volume  = {377},
  number  = {6606},
  pages   = {eabo1984},
  year    = {2022},
  doi     = {10.1126/science.abo1984}
}

@article{koenig2022heartfailure,
  author  = {Koenig, Andrew L. and Shchukina, Irina and Amrute, Junedh and others},
  title   = {Single-cell transcriptomics reveals cell-type-specific diversification in human heart failure},
  journal = {Nature Cardiovascular Research},
  volume  = {1},
  pages   = {263--280},
  year    = {2022},
  doi     = {10.1038/s44161-022-00028-6}
}

@article{macosko2015dropseq,
  author  = {Macosko, Evan Z. and Basu, Anindita and Satija, Rahul and others},
  title   = {Highly parallel genome-wide expression profiling of individual cells using nanoliter droplets},
  journal = {Cell},
  volume  = {161},
  number  = {5},
  pages   = {1202--1214},
  year    = {2015},
  doi     = {10.1016/j.cell.2015.05.002}
}

@article{klein2015indrops,
  author  = {Klein, Allon M. and Mazutis, Linas and Akartuna, Ilke and others},
  title   = {Droplet barcoding for single-cell transcriptomics applied to embryonic stem cells},
  journal = {Cell},
  volume  = {161},
  number  = {5},
  pages   = {1187--1201},
  year    = {2015},
  doi     = {10.1016/j.cell.2015.04.044}
}

@article{cao2019organogenesis,
  author  = {Cao, Junyue and Spielmann, Malte and Qiu, Xiaojie and others},
  title   = {The single-cell transcriptional landscape of mammalian organogenesis},
  journal = {Nature},
  volume  = {566},
  number  = {7745},
  pages   = {496--502},
  year    = {2019},
  doi     = {10.1038/s41586-019-0969-x}
}

@article{tirosh2016melanoma,
  author  = {Tirosh, Itay and Izar, Benjamin and Prakadan, Sanjay M. and others},
  title   = {Dissecting the multicellular ecosystem of metastatic melanoma by single-cell {RNA}-seq},
  journal = {Science},
  volume  = {352},
  number  = {6282},
  pages   = {189--196},
  year    = {2016},
  doi     = {10.1126/science.aad0501}
}

@article{lecun2015deeplearning,
  author  = {LeCun, Yann and Bengio, Yoshua and Hinton, Geoffrey},
  title   = {Deep learning},
  journal = {Nature},
  volume  = {521},
  number  = {7553},
  pages   = {436--444},
  year    = {2015},
  doi     = {10.1038/nature14539}
}

@article{tanay2017scaling,
  author  = {Tanay, Amos and Regev, Aviv},
  title   = {Scaling single-cell genomics from phenomenology to mechanism},
  journal = {Nature},
  volume  = {541},
  number  = {7637},
  pages   = {331--338},
  year    = {2017},
  doi     = {10.1038/nature21350}
}

\clearpage
\makeatletter
\setlength{\@fptop}{0pt}
\setlength{\@fpsep}{14pt plus 1fil}
\makeatother
\setcounter{figure}{0}
\renewcommand{\thefigure}{S\arabic{figure}}
\setcounter{table}{0}
\renewcommand{\thetable}{S\arabic{table}}
\section*{Supplementary Information}

This supplement reports the full per-atlas benchmark panels behind the headline numbers, the significance testing of the cross-study comparison, and the integration, gene-program and confusion analyses. Throughout, scVision is evaluated zero-shot under the study-level holdout (paradigm~A); $\Delta$ denotes scVision's frozen-probe score minus the best competing single-cell, classical or tabular baseline, and we report ties and losses alongside wins. Data files label the model \texttt{TabVisionFM}.

\FloatBarrier

\subsection*{scImage construction and reproducibility}

\begin{figure}[H]\centering
\includegraphics[width=0.92\linewidth]{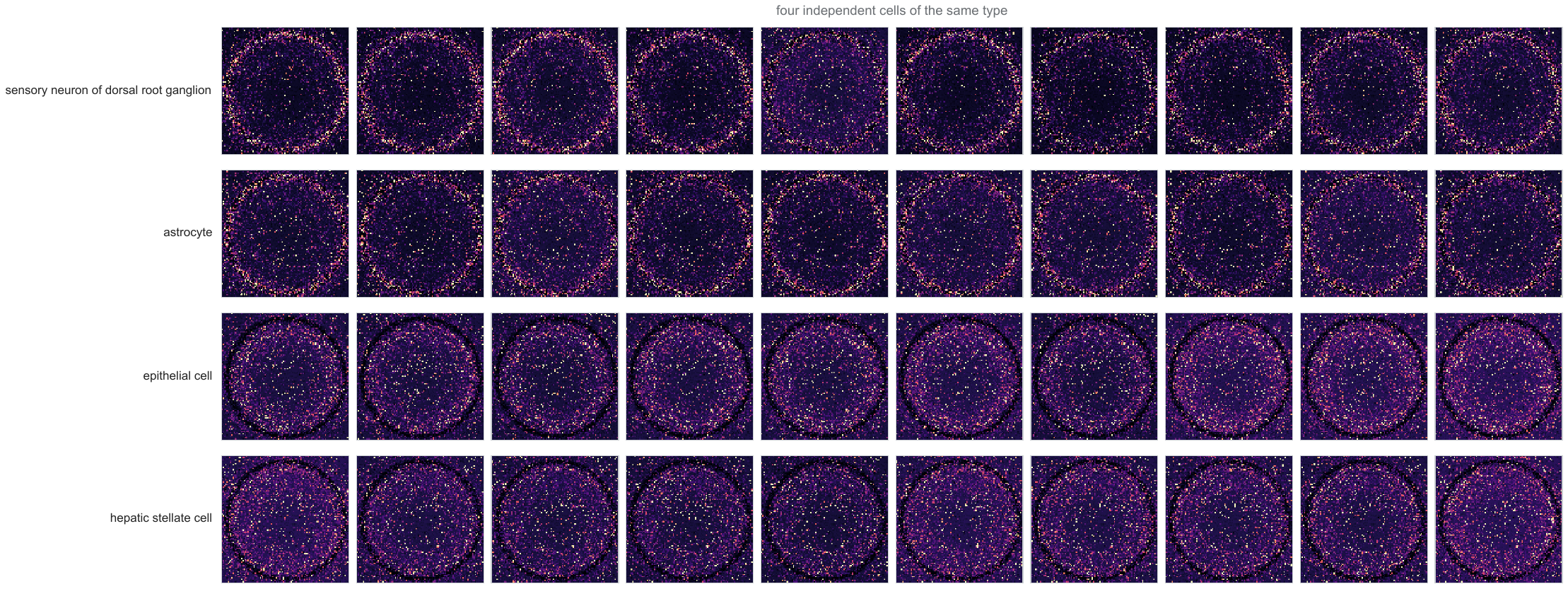}
\caption{\textbf{scImages are reproducible within a cell type and distinct across types.} Each $104\times104$ image places the 10{,}816 most informative genes at optimal-transport coordinates, so co-expressed genes become spatial neighbours; intensity is the cell's $z$-scored expression projected onto that fixed layout. Rows are cell types; columns are four independent cells of that type drawn from different tissues. The within-row similarity (same spatial texture) against the across-row contrast is what the vision encoder reads as identity. Intensity is clipped to the 2nd--98th percentile per panel for display.}
\label{fig:sgenomaps}
\end{figure}

\FloatBarrier

\subsection*{Zero-shot cell-type annotation: full panels}

\begin{table*}[tbp]\centering
\caption{\textbf{Balanced accuracy under cross-study holdout (paradigm~A), all probes and atlases.} Five-fold mean balanced accuracy for scVision's frozen 20-NN, logistic-regression and MLP probes against three single-cell foundation models and the classical/tabular baselines. Bold marks the row-best method; $\Delta$ is scVision minus the best of scGPT, scFoundation, logistic regression, MLP, XGBoost and random forest (it excludes TabNet, HVG-kNN and Pearson). scVision leads on most atlases, but the panel is honest about exceptions\,---\,scVision trails on the obstructive-nephropathy 20-NN block.}
\label{tab:balacc}
\scriptsize
\setlength{\tabcolsep}{3.0pt}
\resizebox{\textwidth}{!}{%
\begin{tabular}{lrrrrrrrrrrrr}
\toprule
Atlas & scVision & scGPT & scFound. & Genef. & LogReg & MLP & XGB & RF & TabNet & HVG-kNN & Pears. & $\Delta$ \\
\midrule
\multicolumn{13}{l}{\textit{20-NN probe}} \\
Retina & \textbf{0.649} & 0.510 & 0.598 & 0.468 & 0.610 & 0.624 & 0.593 & 0.509 & 0.529 & 0.283 & 0.297 & +0.025 \\
Kidney cortex & \textbf{0.468} & 0.377 & 0.399 & 0.153 & 0.397 & 0.331 & 0.302 & 0.300 & 0.096 & 0.157 & 0.027 & +0.069 \\
Ovary & \textbf{0.830} & 0.778 & 0.793 & 0.589 & 0.701 & 0.668 & 0.668 & 0.542 & 0.516 & 0.097 & 0.305 & +0.036 \\
Obstructive neph. & 0.482 & 0.334 & 0.358 & 0.143 & \textbf{0.503} & 0.475 & 0.440 & 0.455 & 0.120 & 0.221 & 0.087 & -0.021 \\
Focal cort. dys. & \textbf{0.648} & 0.637 & 0.462 & 0.566 & 0.493 & 0.489 & 0.505 & 0.390 & 0.277 & 0.385 & 0.406 & +0.011 \\
\multicolumn{13}{l}{\footnotesize scVision best ($\Delta>0$): 4/5 ; mean $\Delta=+0.024$} \\
\midrule
\multicolumn{13}{l}{\textit{Logistic-regression probe}} \\
Retina & \textbf{0.680} & 0.538 & 0.664 & 0.680 & 0.610 & 0.624 & 0.593 & 0.509 & 0.529 & 0.283 & 0.297 & +0.016 \\
Kidney cortex & 0.444 & 0.368 & 0.384 & \textbf{0.448} & 0.397 & 0.331 & 0.302 & 0.300 & 0.096 & 0.157 & 0.027 & +0.047 \\
Ovary & 0.718 & \textbf{0.852} & 0.760 & 0.623 & 0.701 & 0.668 & 0.668 & 0.542 & 0.516 & 0.097 & 0.305 & -0.134 \\
Obstructive neph. & \textbf{0.510} & 0.362 & 0.457 & 0.460 & 0.503 & 0.475 & 0.440 & 0.455 & 0.120 & 0.221 & 0.087 & +0.007 \\
Focal cort. dys. & \textbf{0.584} & 0.502 & 0.525 & 0.528 & 0.493 & 0.489 & 0.505 & 0.390 & 0.277 & 0.385 & 0.406 & +0.059 \\
\multicolumn{13}{l}{\footnotesize scVision best ($\Delta>0$): 4/5 ; mean $\Delta=-0.001$} \\
\midrule
\multicolumn{13}{l}{\textit{MLP probe}} \\
Retina & \textbf{0.695} & 0.598 & 0.645 & 0.654 & 0.610 & 0.624 & 0.593 & 0.509 & 0.529 & 0.283 & 0.297 & +0.050 \\
Kidney cortex & \textbf{0.470} & 0.416 & 0.368 & 0.447 & 0.397 & 0.331 & 0.302 & 0.300 & 0.096 & 0.157 & 0.027 & +0.055 \\
Ovary & \textbf{0.822} & 0.761 & 0.725 & 0.643 & 0.701 & 0.668 & 0.668 & 0.542 & 0.516 & 0.097 & 0.305 & +0.061 \\
Obstructive neph. & \textbf{0.542} & 0.453 & 0.476 & 0.493 & 0.503 & 0.475 & 0.440 & 0.455 & 0.120 & 0.221 & 0.087 & +0.039 \\
Focal cort. dys. & \textbf{0.597} & 0.579 & 0.484 & 0.597 & 0.493 & 0.489 & 0.505 & 0.390 & 0.277 & 0.385 & 0.406 & +0.018 \\
\multicolumn{13}{l}{\footnotesize scVision best ($\Delta>0$): 5/5 ; mean $\Delta=+0.045$} \\
\bottomrule
\end{tabular}}
\end{table*}

\begin{table*}[tbp]\centering
\caption{\textbf{Present-class macro-F1 under cross-study holdout (paradigm~A).} As Table~\ref{tab:balacc} but for the present-class macro-F1 (averaged only over classes present in a fold, the harder and more honest score). Dashes mark baseline/atlas combinations that were not run at full supervision. Macro-specificity (one-vs-rest) was near ceiling for every method and atlas (range 0.978--1.000); it is omitted as a separate table.}
\label{tab:f1}
\scriptsize
\setlength{\tabcolsep}{3.0pt}
\resizebox{\textwidth}{!}{%
\begin{tabular}{lrrrrrrrrrrrr}
\toprule
Atlas & scVision & scGPT & scFound. & Genef. & LogReg & MLP & XGB & RF & TabNet & HVG-kNN & Pears. & $\Delta$ \\
\midrule
\multicolumn{13}{l}{\textit{20-NN probe}} \\
Retina & \textbf{0.712} & 0.565 & 0.636 & 0.525 & 0.688 & 0.683 & 0.641 & 0.549 & 0.579 & 0.353 & 0.331 & +0.024 \\
Kidney cortex & \textbf{0.560} & 0.468 & 0.481 & 0.226 & -- & -- & -- & -- & 0.132 & -- & -- & +0.079 \\
Ovary & \textbf{0.890} & 0.859 & 0.861 & 0.700 & -- & -- & -- & -- & 0.661 & -- & -- & +0.029 \\
Obstructive neph. & 0.538 & 0.399 & 0.432 & 0.209 & \textbf{0.575} & 0.542 & 0.526 & 0.488 & 0.146 & 0.295 & 0.106 & -0.037 \\
Focal cort. dys. & \textbf{0.714} & 0.706 & 0.532 & 0.643 & -- & -- & -- & -- & 0.328 & -- & -- & +0.007 \\
\multicolumn{13}{l}{\footnotesize scVision best ($\Delta>0$): 4/5 ; mean $\Delta=+0.020$} \\
\midrule
\multicolumn{13}{l}{\textit{MLP probe}} \\
Retina & \textbf{0.750} & 0.675 & 0.710 & 0.703 & 0.688 & 0.683 & 0.641 & 0.549 & 0.579 & 0.353 & 0.331 & +0.040 \\
Kidney cortex & \textbf{0.553} & 0.510 & 0.447 & 0.509 & -- & -- & -- & -- & 0.132 & -- & -- & +0.043 \\
Ovary & \textbf{0.893} & 0.846 & 0.818 & 0.773 & -- & -- & -- & -- & 0.661 & -- & -- & +0.047 \\
Obstructive neph. & \textbf{0.601} & 0.534 & 0.539 & 0.551 & 0.575 & 0.542 & 0.526 & 0.488 & 0.146 & 0.295 & 0.106 & +0.026 \\
Focal cort. dys. & \textbf{0.665} & 0.652 & 0.544 & 0.658 & -- & -- & -- & -- & 0.328 & -- & -- & +0.013 \\
\multicolumn{13}{l}{\footnotesize scVision best ($\Delta>0$): 5/5 ; mean $\Delta=+0.034$} \\
\bottomrule
\end{tabular}}
\end{table*}

\begin{table*}[tbp]\centering
\caption{\textbf{Macro-recall (sensitivity) under cross-study holdout (paradigm~A).} Per-class recall averaged over present classes. Recall is the metric most sensitive to rare cell types. Bold marks the row-best; $\Delta$ as in Table~\ref{tab:balacc}.}
\label{tab:recall}
\scriptsize
\setlength{\tabcolsep}{3.0pt}
\resizebox{\textwidth}{!}{%
\begin{tabular}{lrrrrrrrrrrrr}
\toprule
Atlas & scVision & scGPT & scFound. & Genef. & LogReg & MLP & XGB & RF & TabNet & HVG-kNN & Pears. & $\Delta$ \\
\midrule
\multicolumn{13}{l}{\textit{20-NN probe}} \\
Retina & 0.238 & 0.147 & \textbf{0.246} & 0.138 & 0.212 & 0.205 & 0.194 & 0.182 & 0.169 & 0.060 & 0.077 & -0.008 \\
Kidney cortex & \textbf{0.091} & 0.069 & 0.071 & 0.030 & 0.060 & 0.061 & 0.051 & 0.078 & 0.018 & 0.024 & 0.005 & +0.012 \\
Ovary & \textbf{0.338} & 0.237 & 0.317 & 0.155 & 0.151 & 0.157 & 0.240 & 0.254 & 0.115 & 0.008 & 0.036 & +0.021 \\
Obstructive neph. & 0.068 & 0.044 & 0.050 & 0.015 & 0.047 & 0.048 & 0.064 & \textbf{0.076} & 0.021 & 0.025 & 0.009 & -0.008 \\
Focal cort. dys. & \textbf{0.177} & 0.172 & 0.135 & 0.163 & 0.126 & 0.130 & 0.130 & 0.148 & 0.101 & 0.083 & 0.097 & +0.005 \\
\multicolumn{13}{l}{\footnotesize scVision best ($\Delta>0$): 3/5 ; mean $\Delta=+0.004$} \\
\midrule
\multicolumn{13}{l}{\textit{MLP probe}} \\
Retina & \textbf{0.293} & 0.218 & 0.225 & 0.263 & 0.212 & 0.205 & 0.194 & 0.182 & 0.169 & 0.060 & 0.077 & +0.068 \\
Kidney cortex & 0.092 & 0.075 & 0.056 & \textbf{0.093} & 0.060 & 0.061 & 0.051 & 0.078 & 0.018 & 0.024 & 0.005 & +0.014 \\
Ovary & \textbf{0.286} & 0.262 & 0.249 & 0.155 & 0.151 & 0.157 & 0.240 & 0.254 & 0.115 & 0.008 & 0.036 & +0.024 \\
Obstructive neph. & 0.074 & 0.053 & 0.046 & 0.046 & 0.047 & 0.048 & 0.064 & \textbf{0.076} & 0.021 & 0.025 & 0.009 & -0.002 \\
Focal cort. dys. & \textbf{0.181} & 0.177 & 0.125 & 0.168 & 0.126 & 0.130 & 0.130 & 0.148 & 0.101 & 0.083 & 0.097 & +0.004 \\
\multicolumn{13}{l}{\footnotesize scVision best ($\Delta>0$): 4/5 ; mean $\Delta=+0.022$} \\
\bottomrule
\end{tabular}}
\end{table*}

\begin{table}[H]\centering
\caption{\textbf{Head-to-head significance on cell-type identity (paradigm~A, balanced accuracy).} For each baseline, the per-fold balanced-accuracy difference (scVision's best probe minus the baseline) was pooled across the four deeply-evaluated atlases (ovary, kidney cortex, focal cortical dysplasia, cystic fibrosis) and tested with a paired Wilcoxon signed-rank test; $p$ values are Holm-adjusted across the eleven baselines. Win/tie/loss counts are per atlas. scVision wins every comparison in aggregate, but the counts are honest about the one loss\,---\,scFoundation takes cystic fibrosis. Per-atlas overall winners: Ovary: scVision (20-NN), Kidney cortex: scVision, Focal cort. dysplasia: scVision (20-NN), Cystic fibrosis: scFoundation.}
\label{tab:signif}
\footnotesize
\setlength{\tabcolsep}{5pt}
\begin{tabular}{lrccc c}
\toprule
Baseline & mean $\Delta$ & W/T/L & Wilcoxon $p$ & Holm $p$ & sig. \\
\midrule
scGPT & +0.084 & 3/1/0 & $3.3\times10^{-4}$ & $6.6\times10^{-4}$ & yes \\
scFoundation & +0.045 & 3/0/1 & 0.0013 & 0.0013 & yes \\
Geneformer & +0.089 & 4/0/0 & $5.3\times10^{-5}$ & $1.6\times10^{-4}$ & yes \\
MLP & +0.128 & 4/0/0 & $7.6\times10^{-6}$ & $8.4\times10^{-5}$ & yes \\
LogReg & +0.114 & 4/0/0 & $7.6\times10^{-6}$ & $8.4\times10^{-5}$ & yes \\
XGBoost & +0.137 & 4/0/0 & $1.5\times10^{-5}$ & $1.4\times10^{-4}$ & yes \\
Random Forest & +0.191 & 4/0/0 & $1.5\times10^{-5}$ & $1.4\times10^{-4}$ & yes \\
TabNet & +0.317 & 4/0/0 & $1.5\times10^{-5}$ & $1.4\times10^{-4}$ & yes \\
HVG-kNN & +0.384 & 4/0/0 & $1.5\times10^{-5}$ & $1.4\times10^{-4}$ & yes \\
Pearson & +0.382 & 4/0/0 & $1.5\times10^{-5}$ & $1.4\times10^{-4}$ & yes \\
\bottomrule
\end{tabular}
\end{table}

\begin{figure*}[tbp]\centering
\includegraphics[width=\textwidth]{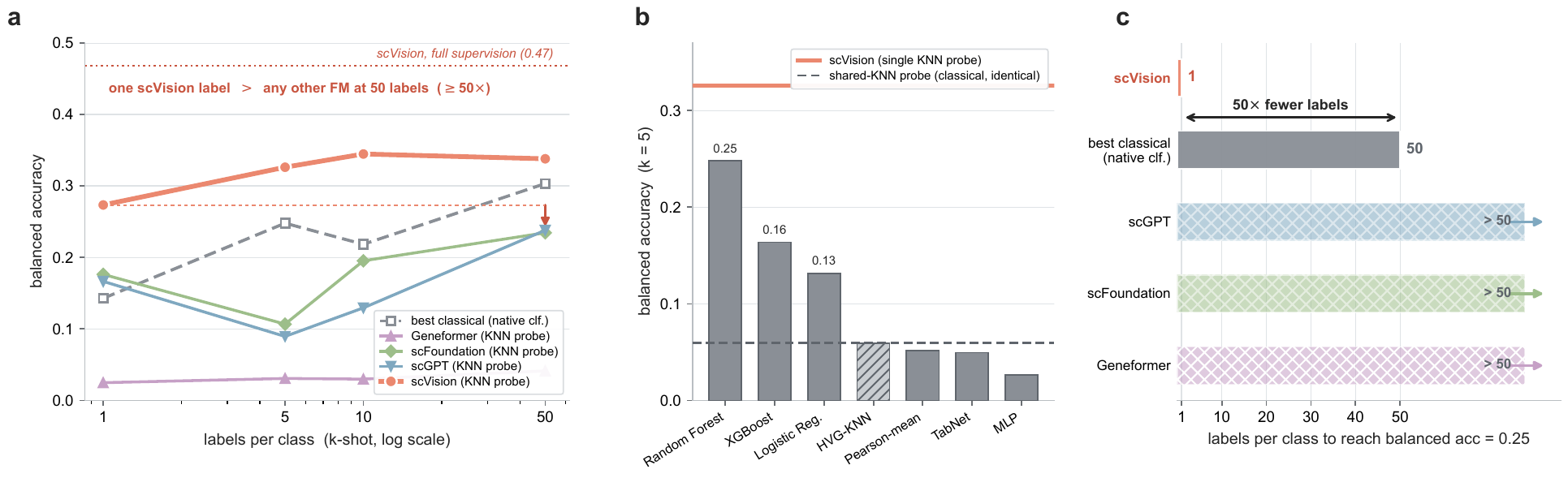}
\caption{\textbf{Label efficiency on kidney cortex (paradigm~A).} Balanced accuracy versus labelled cells per class ($k\in\{1,5,10,50\}$). All foundation models share one 20-NN probe, isolating feature-space quality; classical baselines use their own native classifier (the steelman envelope). (a)~scVision leads at every budget, and a single scVision label exceeds every other foundation model given fifty. (b)~Even when each classical baseline fits its own best classifier on a 5-shot support, all remain below scVision's single probe. (c)~Annotation budget to reach balanced accuracy $0.25$.}
\label{fig:sfewshot}
\end{figure*}

\FloatBarrier

\subsection*{Multi-study integration}

\begin{figure*}[tbp]\centering
\includegraphics[width=\textwidth]{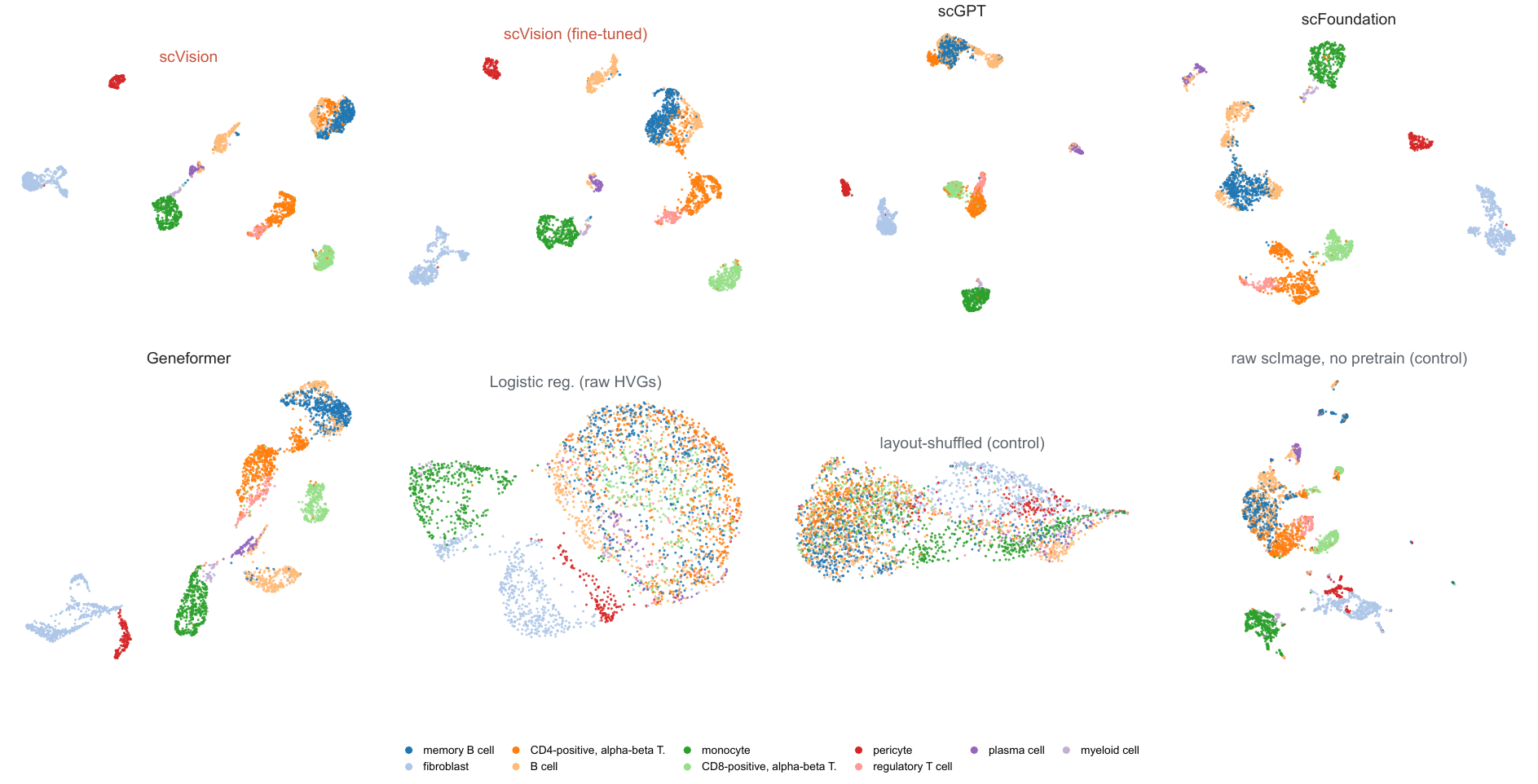}
\caption{\textbf{Frozen-embedding integration of the Crohn ileum atlas, by method.} UMAP of each method's representation of the same 3{,}121 cells, coloured by cell type (top-18 by frequency; rarer types grey). scVision and the token foundation models resolve the major lineages; the two controls collapse that structure\,---\,shuffling the scImage layout before encoding (\emph{layout-shuffled}) and probing the un-pretrained raw scImage (\emph{raw scImage}) both lose the cell-type geometry, showing the separation comes from the learned spatial arrangement, not the gene set alone. Coordinates from \texttt{crohn\_umap\_per\_method.npz}.}
\label{fig:scrohn}
\end{figure*}

\begin{figure}[tbp]\centering
\includegraphics[width=\linewidth]{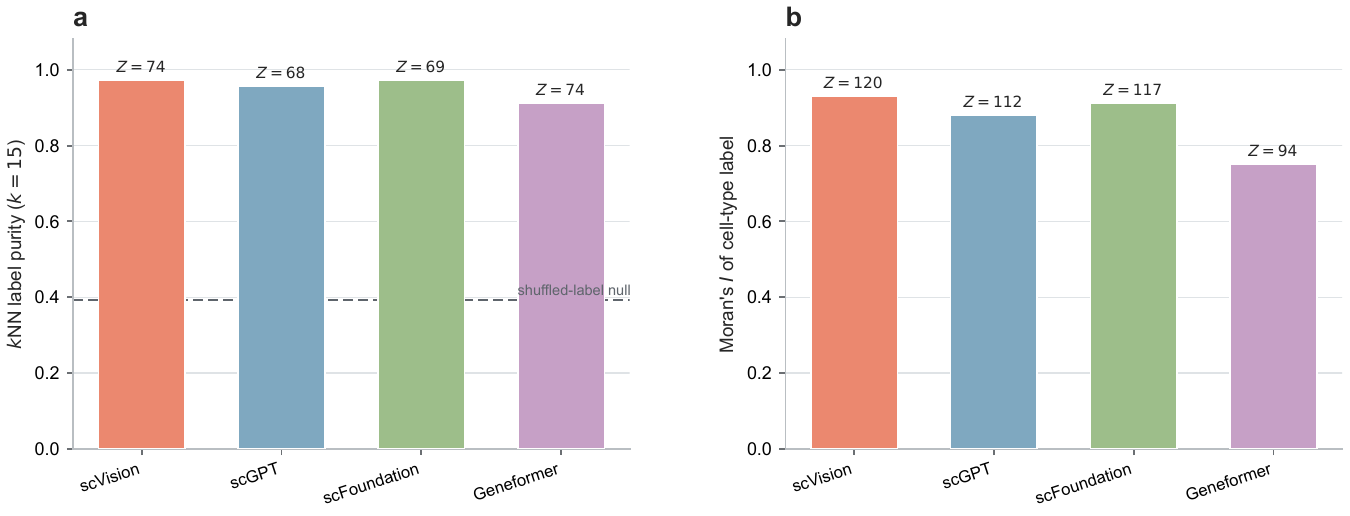}
\caption{\textbf{Quantitative integration quality on the ovary cross-donor pool.} (a)~$k$NN label purity ($k=15$) and (b)~Moran's $I$ of the cell-type label on the embedding graph, for four foundation models, each annotated with its $Z$ score against a shuffled-label null. scVision attains the highest purity and label autocorrelation, with scFoundation close behind. Values from \texttt{ovary\_umap4\_quality.json}.}
\label{fig:sovaryq}
\end{figure}

\begin{table}[H]\centering
\caption{\textbf{Embedding-space integration quality (ovary cross-donor pool).} From a frozen embedding of 536 ovary cells spanning four coarse identities, $k$NN label purity ($k=15$) and Moran's $I$ spatial autocorrelation of the cell-type label on the embedding graph, each with a shuffled-label null reported as a $Z$ score. scVision attains the highest purity and the strongest label autocorrelation, with scFoundation a close second. Classes: endothelial cell ($n{=}57$), leukocyte ($n{=}133$), pericyte ($n{=}46$), stromal cell of ovary ($n{=}7052$).}
\label{tab:integ}
\footnotesize
\setlength{\tabcolsep}{8pt}
\begin{tabular}{lrrrr}
\toprule
Method & $k$NN purity & purity $Z$ & Moran's $I$ & Moran $Z$ \\
\midrule
scVision & \textbf{0.973} & 73.603 & \textbf{0.930} & \textbf{119.8} \\
scGPT & 0.957 & 68.324 & 0.880 & 111.9 \\
scFoundation & 0.972 & 68.509 & 0.913 & 116.6 \\
Geneformer & 0.911 & \textbf{73.625} & 0.752 & 93.8 \\
\bottomrule
\end{tabular}
\end{table}

\begin{figure}[tbp]\centering
\includegraphics[width=\linewidth]{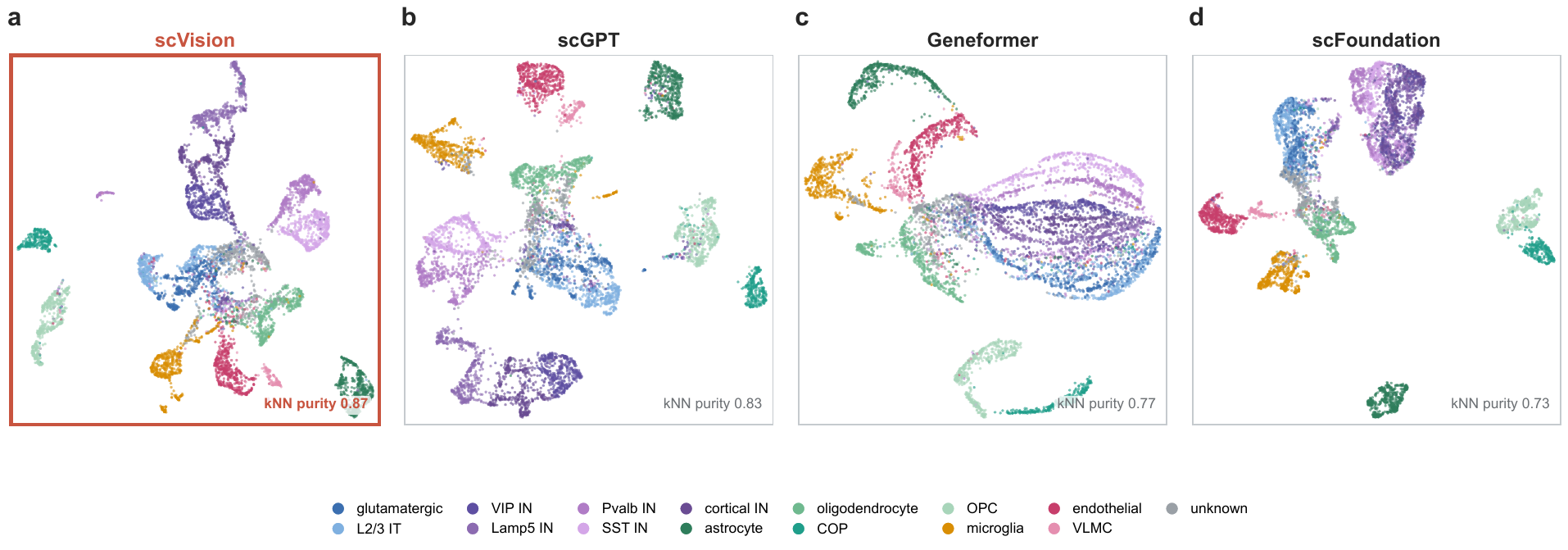}
\caption{\textbf{Frozen-embedding UMAP of the focal cortical dysplasia atlas.} scVision's representation of held-out epilepsy-surgery tissue, coloured by annotated cell type, recovers the neuronal, glial and vascular compartments without any task-specific training.}
\label{fig:sfcd}
\end{figure}

\clearpage

\FloatBarrier

\subsection*{Gene programs and cross-tissue recurrence}

\begin{figure*}[tp]\centering
\includegraphics[width=\textwidth]{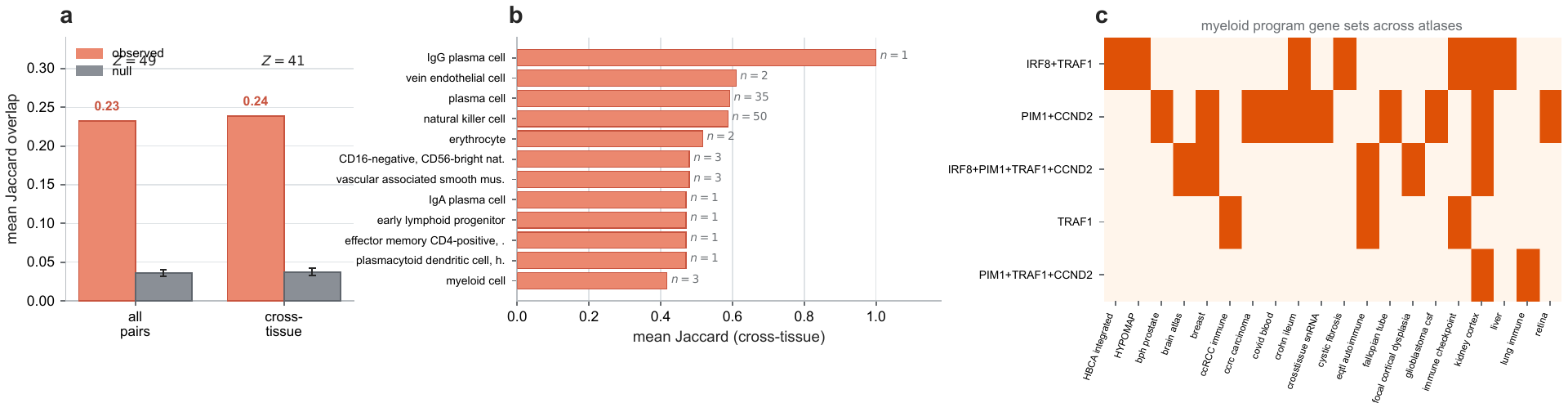}
\caption{\textbf{Attention-derived gene programs recur across tissues.} (a)~Observed mean Jaccard overlap of the top-100 attention genes between atlas pairs sharing a cell type, against a label-derangement null (error bar, null s.d.; $Z$ annotated). (b)~The most reproducible cross-tissue programs are immune and stromal identities ($n$, contributing atlas pairs). (c)~A myeloid program (IRF8/TRAF1/PIM1/CCND2) recurs across many atlases and organs (filled cell, the gene set appears in that atlas).}
\label{fig:srecur}
\end{figure*}

\begin{table}[H]\centering
\caption{\textbf{Recurrence of attention-derived gene programs across tissues.} The top-100 genes of each cell type's last-block attention map were compared between every pair of atlases sharing that cell type. Observed mean Jaccard overlap is contrasted with a per-pair label-derangement null over 1000 permutations (seed 20240601). Cross-tissue pairs (different organs) overlap far more than chance ($Z=41.3$). The most reproducible programs are immune and stromal identities shared organ-wide; single-pair entries ($n_{\mathrm{pairs}}=1$) reflect one atlas pair and are listed for completeness only.}
\label{tab:recur}
\footnotesize
\setlength{\tabcolsep}{4pt}
\begin{tabular}{lrrrrrrr}
\toprule
Set & $n$ pairs & cell types & obs.\ $J$ & null $J$ & null sd & $Z$ & $p$ \\
\midrule
\multicolumn{8}{l}{\textit{Recurrence statistics (top-100 attention genes; 1000 permutations)}} \\
All cell-type pairs & 731 & 106 & 0.232 & 0.036 & 0.004 & 48.7 & $1.0\times10^{-3}$ \\
Cross-tissue pairs & 585 & 77 & 0.239 & 0.037 & 0.005 & 41.3 & $1.0\times10^{-3}$ \\
\midrule
\multicolumn{8}{l}{\textit{Most reproducible cross-tissue cell types (mean pairwise Jaccard of the top-100 gene set)}} \\
\multicolumn{6}{l}{IgG plasma cell} & 1.000 & $n{=}1$ \\
\multicolumn{6}{l}{vein endothelial cell} & 0.610 & $n{=}2$ \\
\multicolumn{6}{l}{plasma cell} & 0.592 & $n{=}35$ \\
\multicolumn{6}{l}{natural killer cell} & 0.587 & $n{=}50$ \\
\multicolumn{6}{l}{erythrocyte} & 0.517 & $n{=}2$ \\
\multicolumn{6}{l}{CD16-negative, CD56-bright natural killer cell, human} & 0.480 & $n{=}3$ \\
\multicolumn{6}{l}{vascular associated smooth muscle cell} & 0.480 & $n{=}3$ \\
\multicolumn{6}{l}{IgA plasma cell} & 0.471 & $n{=}1$ \\
\multicolumn{6}{l}{early lymphoid progenitor} & 0.471 & $n{=}1$ \\
\multicolumn{6}{l}{effector memory CD4-positive, alpha-beta T cell} & 0.471 & $n{=}1$ \\
\multicolumn{6}{l}{plasmacytoid dendritic cell, human} & 0.471 & $n{=}1$ \\
\multicolumn{6}{l}{myeloid cell} & 0.418 & $n{=}3$ \\
\bottomrule
\end{tabular}
\end{table}

\begin{table}[H]\centering
\caption{\textbf{Gene programs are enriched across independent pathway databases.} Top: hypergeometric enrichment (Benjamini--Hochberg $q<0.05$) of the specificity-subtracted cross-tissue programs, evaluated against four independent databases; columns give the number of cell-type tests, significant hits, and distinct cell types with at least one hit. Bottom: the same test applied per win-atlas. Enrichment reproduces beyond the Hallmark collection used in the main text, indicating the attention programs are not an artifact of a single annotation source.}
\label{tab:enrich}
\footnotesize
\setlength{\tabcolsep}{6pt}
\begin{tabular}{lrrr}
\toprule
Database & tests & sig.\ hits & cell types \\
\midrule
\multicolumn{4}{l}{\textit{Cross-tissue recurrent programs (subtract specificity), $q<0.05$}} \\
MSigDB Hallmark & 300 & 13 & 7 \\
Reactome & 300 & 10 & 5 \\
KEGG & 300 & 15 & 6 \\
GO biological process & 300 & 3 & 2 \\
\midrule
\multicolumn{4}{l}{\textit{Per-win-atlas extension: significant hits ($q<0.05$) by database}} \\
Atlas & Reactome & KEGG & GO-BP \\
Ovary & 0 & 0 & 1 \\
Kidney cortex & 4 & 4 & 9 \\
Cystic fibrosis & 6 & 2 & 7 \\
Focal cort. dys. & 3 & 3 & 0 \\
\bottomrule
\end{tabular}
\end{table}

\begin{figure}[tbp]\centering
\includegraphics[width=\linewidth]{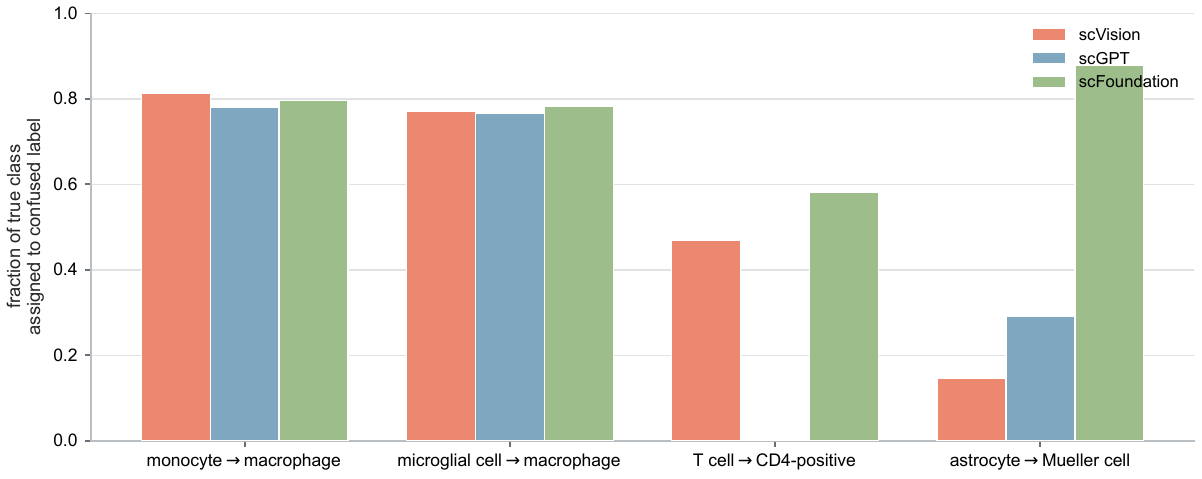}
\caption{\textbf{Residual errors are ontology collapses shared by all methods.} On retina, the dominant confusions are biology-driven lineage merges\,---\, monocyte and microglia called as macrophage, the broad \emph{T cell} label split into CD4/CD8 subsets, astrocyte versus Mueller glia. scVision shares these with the token models rather than making idiosyncratic mistakes; the bars give the fraction of each true class assigned to the confused label (from \texttt{retina\_confusion\_top\_pairs.json}).}
\label{fig:sconf}
\end{figure}

\FloatBarrier

\subsection*{Ablation: pretraining objective}

\begin{figure}[H]\centering
\includegraphics[width=0.82\linewidth]{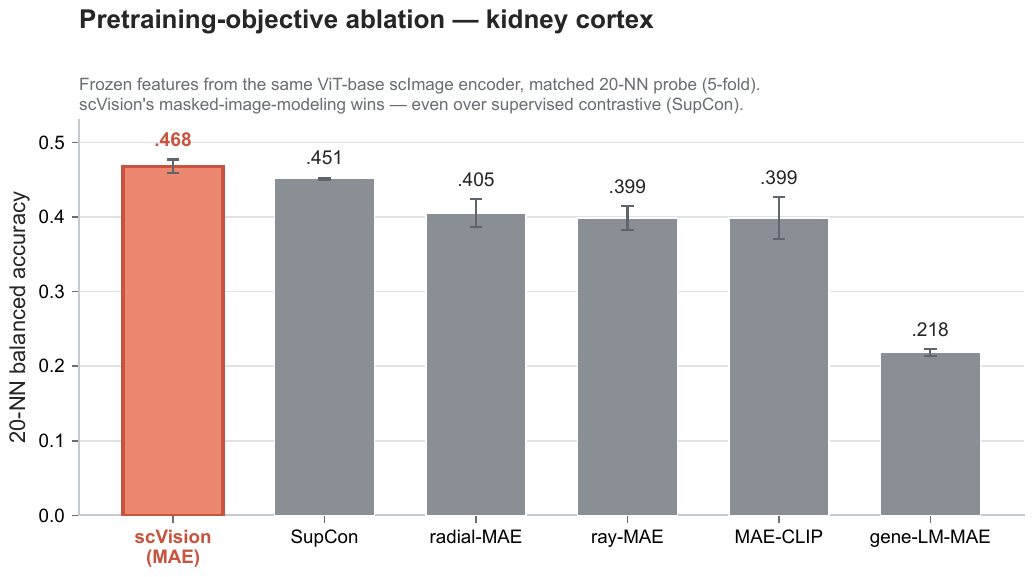}
\caption{\textbf{Pretraining-objective ablation: masked-image-modeling gives the best frozen scImage features.} The frozen embedding of the \emph{same} ViT-base scImage encoder, pretrained under six self-supervised / pretraining objectives, is probed on the kidney-cortex atlas with the identical matched 20-NN protocol (five-fold, paradigm~A; every encoder loaded with zero missing keys). scVision's masked-image-modeling attains the highest balanced accuracy (0.468), ahead of supervised contrastive (SupCon, 0.451, which additionally uses cell-type labels), the MAE-masking and MAE+CLIP variants (0.399--0.405), and MAE with a gene-language model (0.218, where the gene-text conditioning is actively harmful as a frozen encoder). Bars are five-fold means $\pm$ s.d.; values from \texttt{results/ablations/ssl\_kidney}.}
\label{fig:s_ssl_ablation}
\end{figure}

\FloatBarrier

\subsection*{Per-dataset cross-model benchmarks across diverse held-out tissues}

Beyond the main-text atlases, we benchmark scVision against the three single-cell foundation models (scGPT, scFoundation and Geneformer) on 15 additional held-out cell atlases spanning diverse organs\,---\,brain, retina, breast, kidney, liver, prostate, lymph node, adipose, heart, eye, uterus and blood. Each atlas is a study-level holdout that scVision never saw during pre-training; to broaden tissue coverage beyond the curated collections we draw several directly from otherwise-unused CELLxGENE Census test studies. Every model is evaluated apples-to-apples on the same seeded held-out cells (identical cells, cell-type labels and class vocabulary across models), with a single shared frozen-embedding 20-nearest-neighbour probe (the decisive metric of the main text; paradigm~A, five-fold cross-validation), a shared $k$-shot protocol, and identical $k$NN-purity and Moran's~$I$ geometry metrics. Within each figure the only variable is the learned representation.

On the decisive matched 20-NN balanced accuracy, \textbf{scVision's frozen embedding ranks first on all 15 atlases}, with scFoundation the most frequent runner-up; each per-atlas margin over the best competing model is reported verbatim from the on-disk five-fold metrics in the captions below. For every atlas we show (a)~the frozen-embedding UMAP of each model, coloured by native cell type (coral frame marks scVision), with the matched balanced accuracy and 2-D $k$NN purity inset; (b)~balanced accuracy and present-class macro-F1; (c)~label efficiency at $k\in\{1,5,10,50\}$ labels per class; and (d)~2-D $k$NN purity and Moran's~$I$.

\begin{table}[H]\centering\small
\caption{\textbf{Per-dataset matched 20-NN benchmark summary.} Held-out cell atlases (study-level holdout, paradigm~A), ordered by scVision's margin over the best competing single-cell foundation model. Columns: held-out cells scored, native cell types, scVision's frozen-embedding 20-NN balanced accuracy (five-fold mean on identical matched cells), and $\Delta$, scVision minus the best of scGPT, scFoundation and Geneformer. scVision is first on every atlas.}\label{tab:sperdataset}
\begin{tabular}{lrrcc}
\toprule
Atlas & Cells & Cell types & scVision (20-NN) & $\Delta$ vs best FM \\
\midrule
Adipose tissue & 2{,}573 & 7 & 0.512 & $+$0.173 \\
Eye & 1{,}266 & 5 & 0.563 & $+$0.145 \\
Obstructive nephropathy & 2{,}000 & 4 & 0.482 & $+$0.124 \\
Uterus & 1{,}197 & 4 & 0.337 & $+$0.088 \\
Cross-tissue (snRNA) & 3{,}881 & 11 & 0.522 & $+$0.088 \\
Lymph node & 1{,}720 & 5 & 0.713 & $+$0.083 \\
Kidney cortex & 1{,}916 & 7 & 0.468 & $+$0.069 \\
Retina & 2{,}775 & 11 & 0.649 & $+$0.050 \\
Brain atlas & 6{,}500 & 13 & 0.596 & $+$0.027 \\
Liver & 1{,}698 & 8 & 0.451 & $+$0.021 \\
Clear-cell RCC, immune & 3{,}500 & 7 & 0.324 & $+$0.014 \\
Heart & 2{,}029 & 5 & 0.490 & $+$0.013 \\
Human breast cell atlas & 2{,}167 & 10 & 0.243 & $+$0.012 \\
Benign prostatic hyperplasia & 2{,}337 & 7 & 0.539 & $+$0.011 \\
Blood & 5{,}476 & 12 & 0.560 & $+$0.003 \\
\bottomrule
\end{tabular}
\end{table}

\begin{figure*}[tbp]\centering
\includegraphics[width=\textwidth]{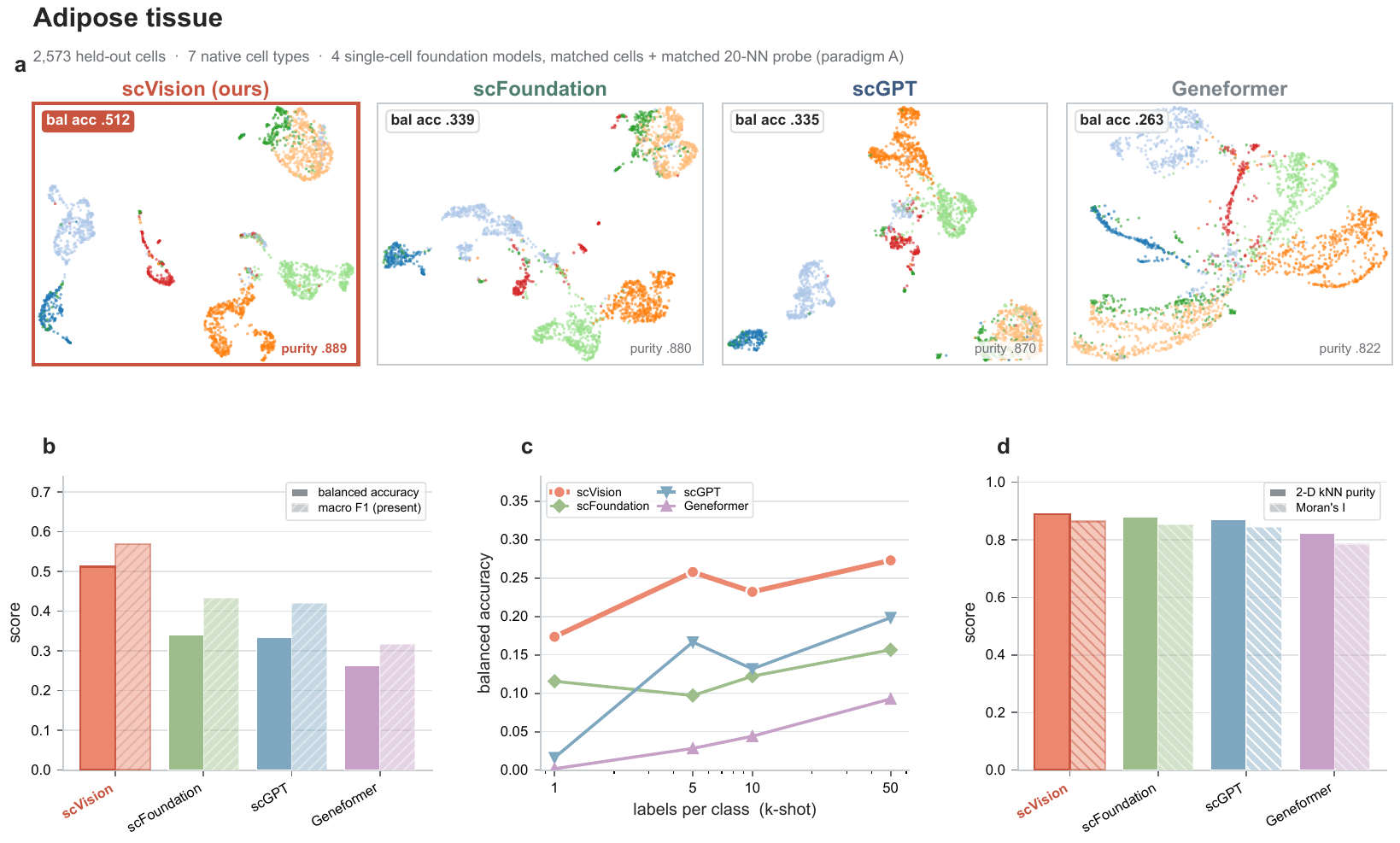}
\caption{\textbf{Adipose tissue: matched four-model benchmark.} 2{,}573 held-out cells across 7 native cell types, evaluated apples-to-apples across scVision, scFoundation, scGPT and Geneformer (identical cells and labels, one shared 20-NN probe, paradigm~A). (a)~Frozen-embedding UMAP per model, cells coloured by native cell type; badges give the matched balanced accuracy (top-left) and 2-D $k$NN purity (bottom-right), the coral frame marking scVision. (b)~Balanced accuracy and present-class macro-F1. (c)~Label efficiency ($k\in\{1,5,10,50\}$ labels per class). (d)~2-D $k$NN purity and Moran's $I$. scVision attains the highest matched balanced accuracy (0.512 vs.\ 0.339 for scFoundation). scVision also leads 2-D $k$NN purity (0.889).}
\label{fig:s_cs_adiposeti_nl_10}
\end{figure*}

\begin{figure*}[tbp]\centering
\includegraphics[width=\textwidth]{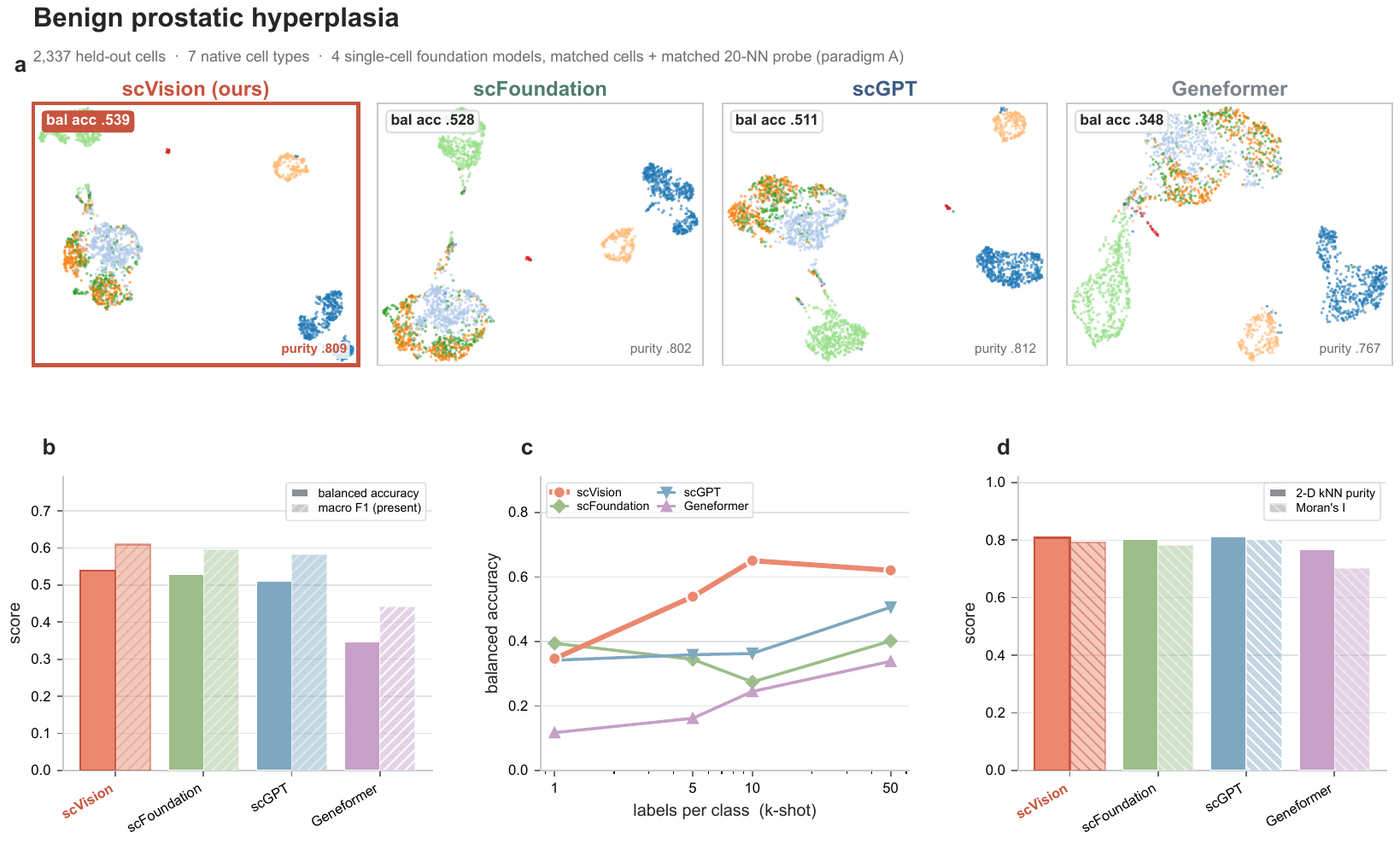}
\caption{\textbf{Benign prostatic hyperplasia: matched four-model benchmark.} 2{,}337 held-out cells across 7 native cell types, evaluated apples-to-apples across scVision, scFoundation, scGPT and Geneformer (identical cells and labels, one shared 20-NN probe, paradigm~A). (a)~Frozen-embedding UMAP per model, cells coloured by native cell type; badges give the matched balanced accuracy (top-left) and 2-D $k$NN purity (bottom-right), the coral frame marking scVision. (b)~Balanced accuracy and present-class macro-F1. (c)~Label efficiency ($k\in\{1,5,10,50\}$ labels per class). (d)~2-D $k$NN purity and Moran's $I$. scVision attains the highest matched balanced accuracy (0.539 vs.\ 0.528 for scFoundation). On 2-D geometry scGPT edges scVision on $k$NN purity (0.812 vs.\ 0.809).}
\label{fig:s_bph_prostate}
\end{figure*}

\begin{figure*}[tbp]\centering
\includegraphics[width=\textwidth]{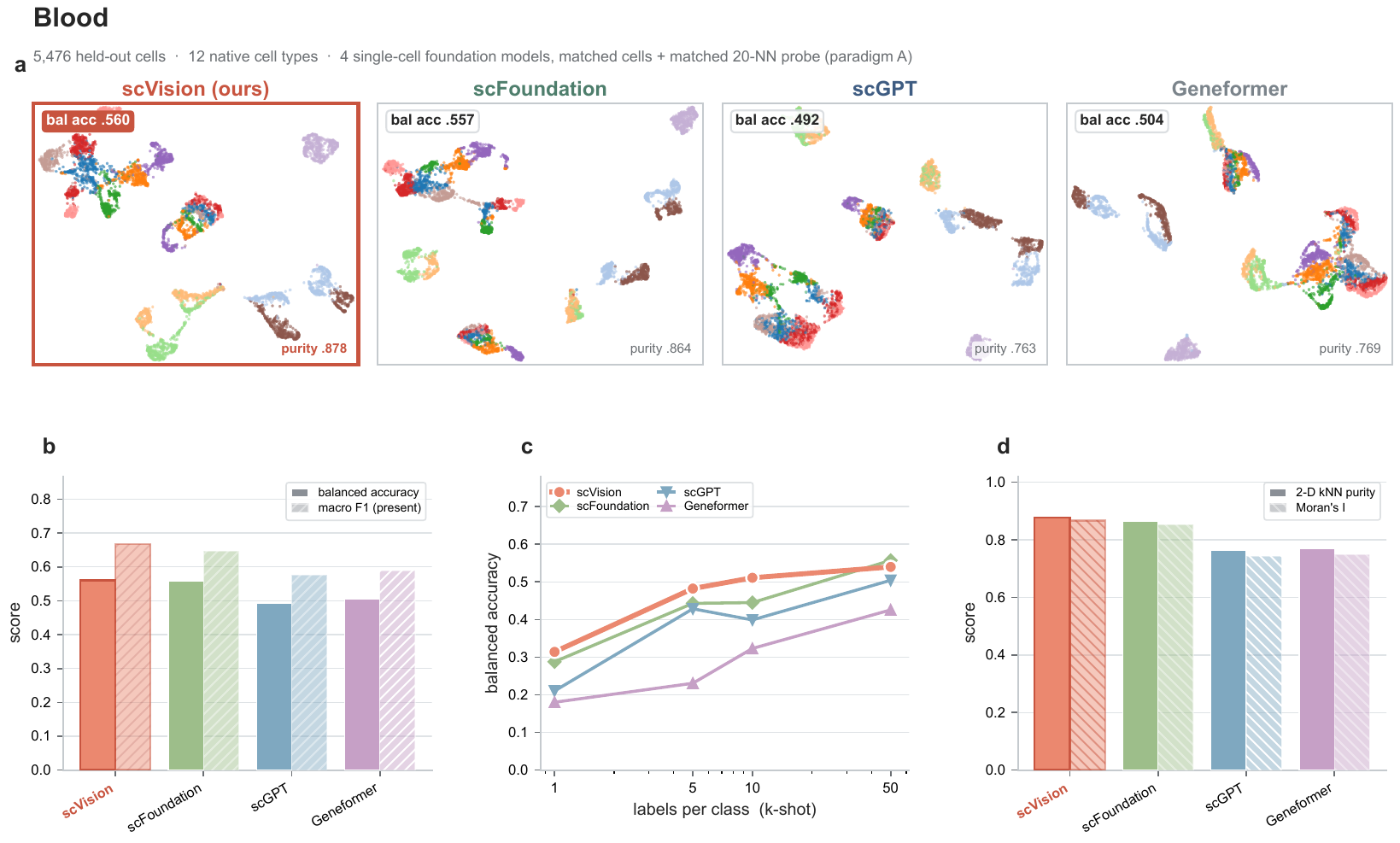}
\caption{\textbf{Blood: matched four-model benchmark.} 5{,}476 held-out cells across 12 native cell types, evaluated apples-to-apples across scVision, scFoundation, scGPT and Geneformer (identical cells and labels, one shared 20-NN probe, paradigm~A). (a)~Frozen-embedding UMAP per model, cells coloured by native cell type; badges give the matched balanced accuracy (top-left) and 2-D $k$NN purity (bottom-right), the coral frame marking scVision. (b)~Balanced accuracy and present-class macro-F1. (c)~Label efficiency ($k\in\{1,5,10,50\}$ labels per class). (d)~2-D $k$NN purity and Moran's $I$. scVision attains the highest matched balanced accuracy (0.560 vs.\ 0.557 for scFoundation). scVision also leads 2-D $k$NN purity (0.878).}
\label{fig:s_cs_blood_nl_3}
\end{figure*}

\begin{figure*}[tbp]\centering
\includegraphics[width=\textwidth]{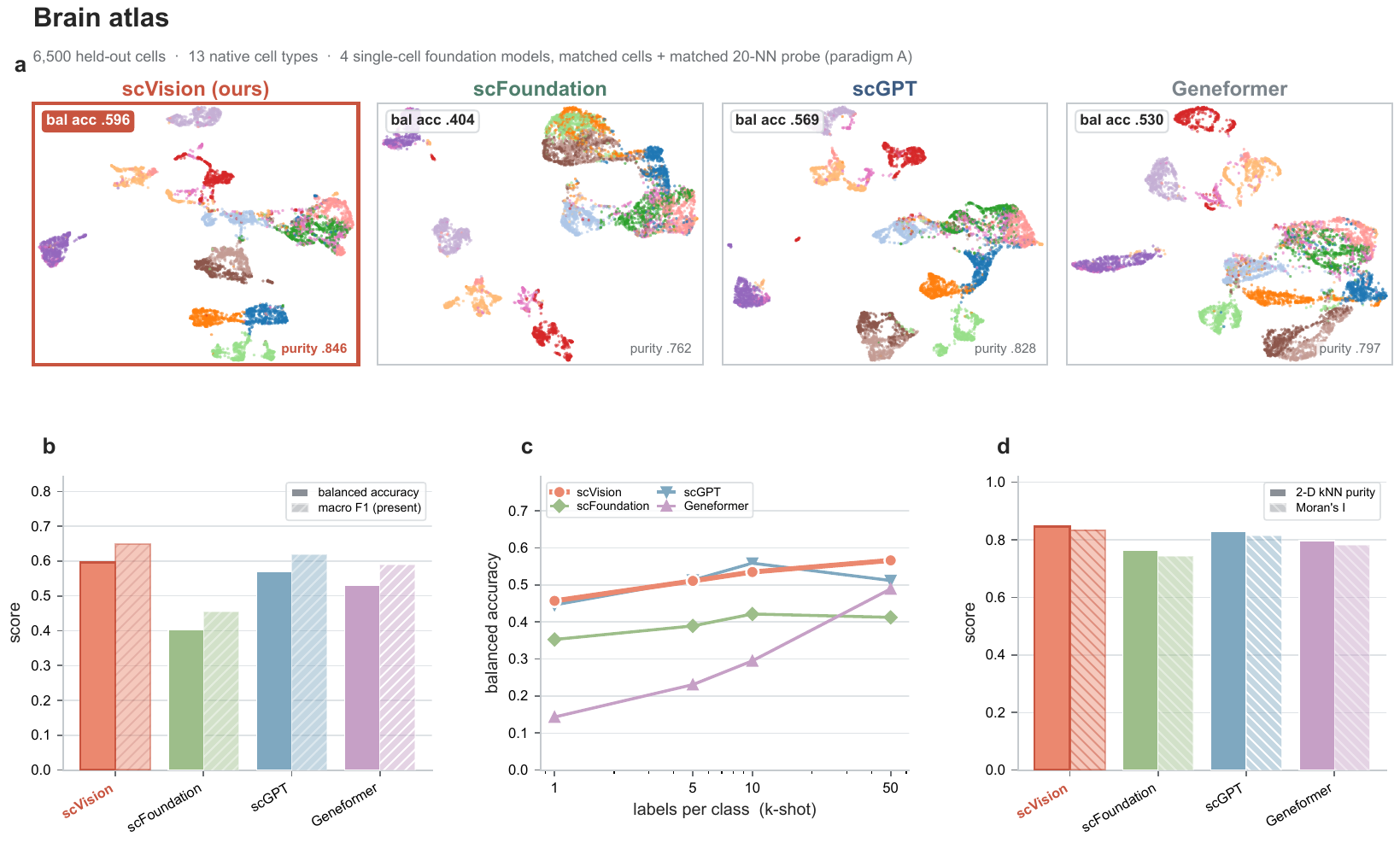}
\caption{\textbf{Brain atlas: matched four-model benchmark.} 6{,}500 held-out cells across 13 native cell types, evaluated apples-to-apples across scVision, scFoundation, scGPT and Geneformer (identical cells and labels, one shared 20-NN probe, paradigm~A). (a)~Frozen-embedding UMAP per model, cells coloured by native cell type; badges give the matched balanced accuracy (top-left) and 2-D $k$NN purity (bottom-right), the coral frame marking scVision. (b)~Balanced accuracy and present-class macro-F1. (c)~Label efficiency ($k\in\{1,5,10,50\}$ labels per class). (d)~2-D $k$NN purity and Moran's $I$. scVision attains the highest matched balanced accuracy (0.596 vs.\ 0.569 for scGPT). scVision also leads 2-D $k$NN purity (0.846).}
\label{fig:s_brain_atlas}
\end{figure*}

\begin{figure*}[tbp]\centering
\includegraphics[width=\textwidth]{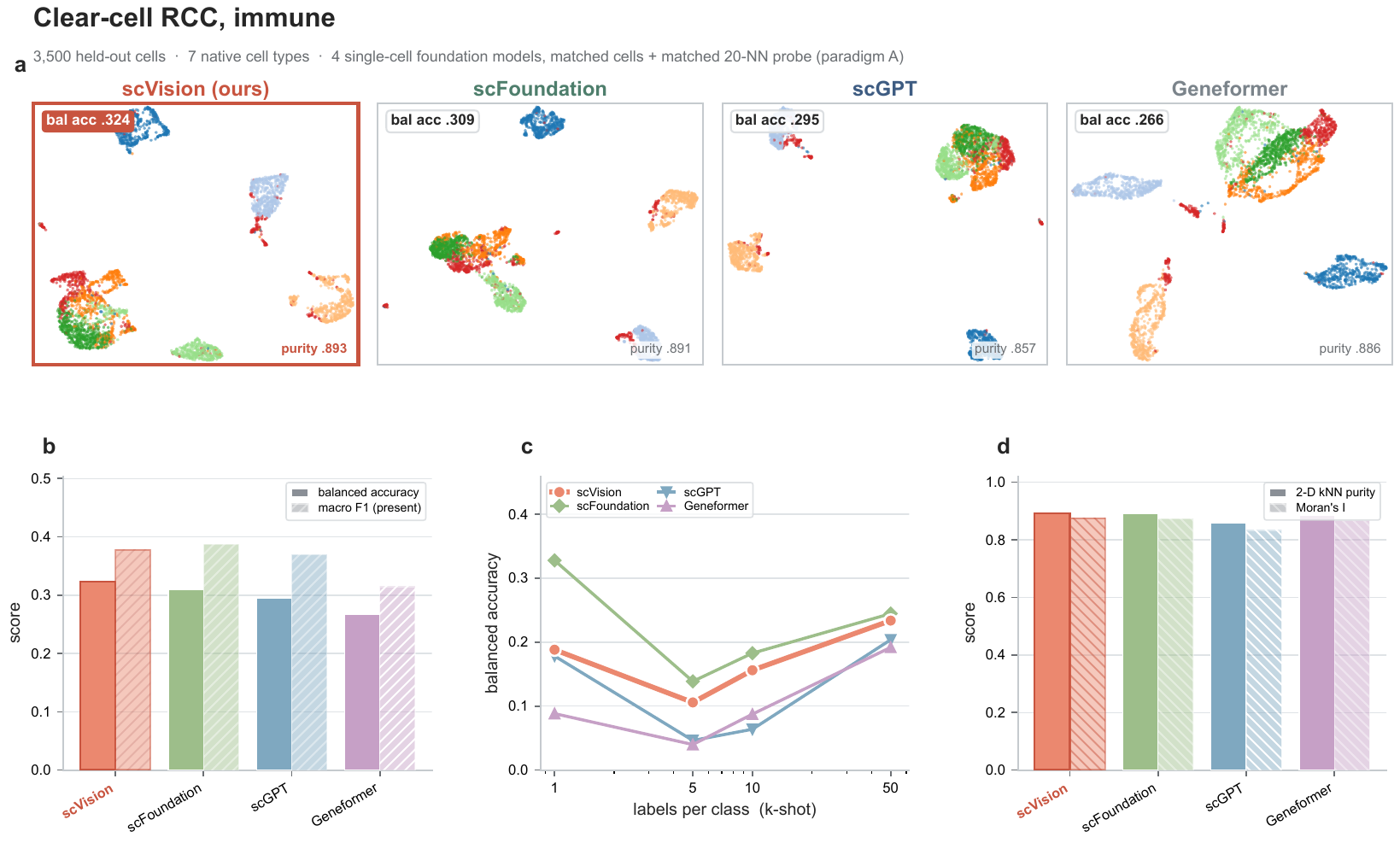}
\caption{\textbf{Clear-cell RCC, immune: matched four-model benchmark.} 3{,}500 held-out cells across 7 native cell types, evaluated apples-to-apples across scVision, scFoundation, scGPT and Geneformer (identical cells and labels, one shared 20-NN probe, paradigm~A). (a)~Frozen-embedding UMAP per model, cells coloured by native cell type; badges give the matched balanced accuracy (top-left) and 2-D $k$NN purity (bottom-right), the coral frame marking scVision. (b)~Balanced accuracy and present-class macro-F1. (c)~Label efficiency ($k\in\{1,5,10,50\}$ labels per class). (d)~2-D $k$NN purity and Moran's $I$. scVision attains the highest matched balanced accuracy (0.324 vs.\ 0.309 for scFoundation). scVision also leads 2-D $k$NN purity (0.893).}
\label{fig:s_ccRCC_immune}
\end{figure*}

\begin{figure*}[tbp]\centering
\includegraphics[width=\textwidth]{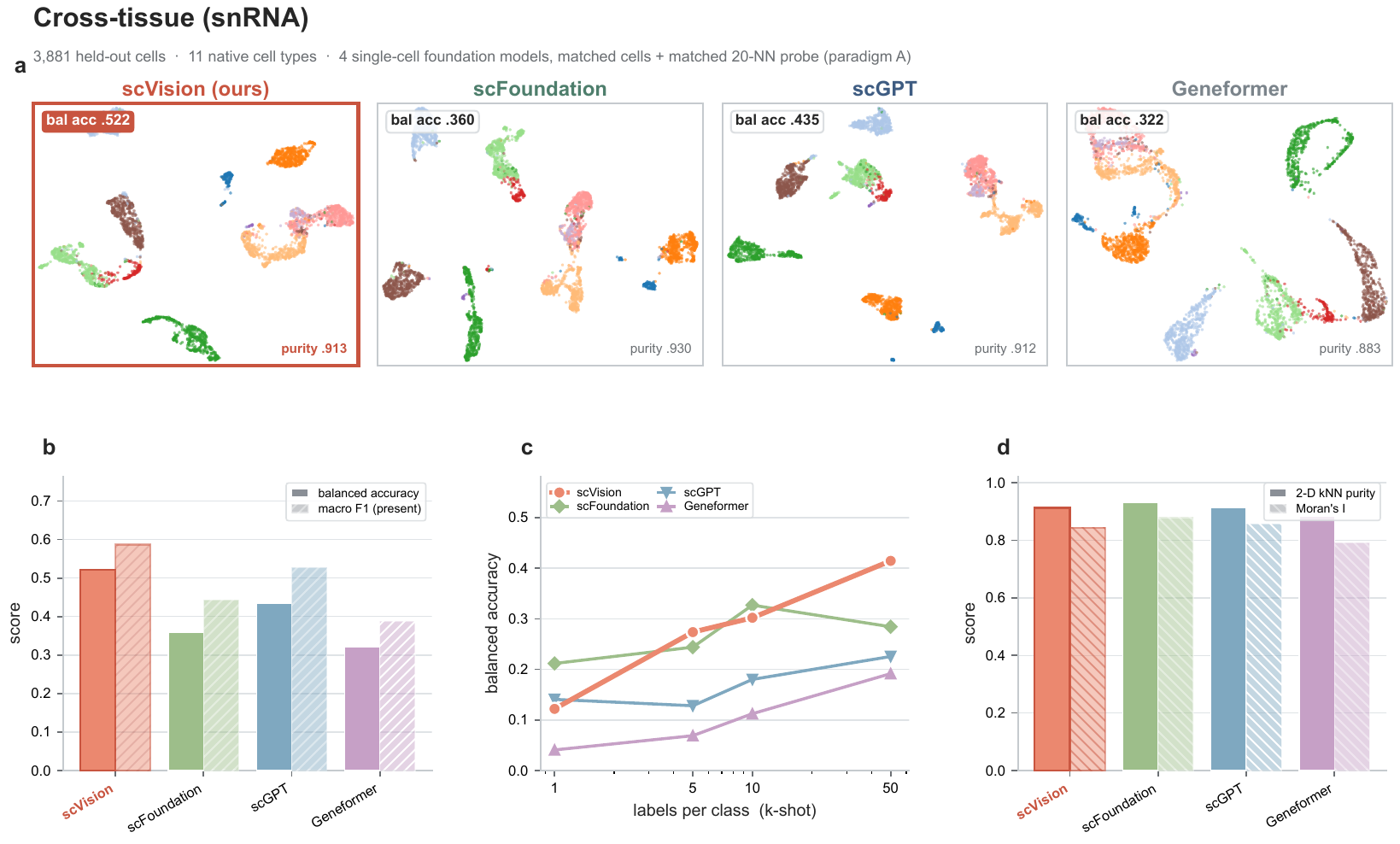}
\caption{\textbf{Cross-tissue (snRNA): matched four-model benchmark.} 3{,}881 held-out cells across 11 native cell types, evaluated apples-to-apples across scVision, scFoundation, scGPT and Geneformer (identical cells and labels, one shared 20-NN probe, paradigm~A). (a)~Frozen-embedding UMAP per model, cells coloured by native cell type; badges give the matched balanced accuracy (top-left) and 2-D $k$NN purity (bottom-right), the coral frame marking scVision. (b)~Balanced accuracy and present-class macro-F1. (c)~Label efficiency ($k\in\{1,5,10,50\}$ labels per class). (d)~2-D $k$NN purity and Moran's $I$. scVision attains the highest matched balanced accuracy (0.522 vs.\ 0.435 for scGPT). On 2-D geometry scFoundation edges scVision on $k$NN purity (0.930 vs.\ 0.913).}
\label{fig:s_crosstissue_snRNA}
\end{figure*}

\begin{figure*}[tbp]\centering
\includegraphics[width=\textwidth]{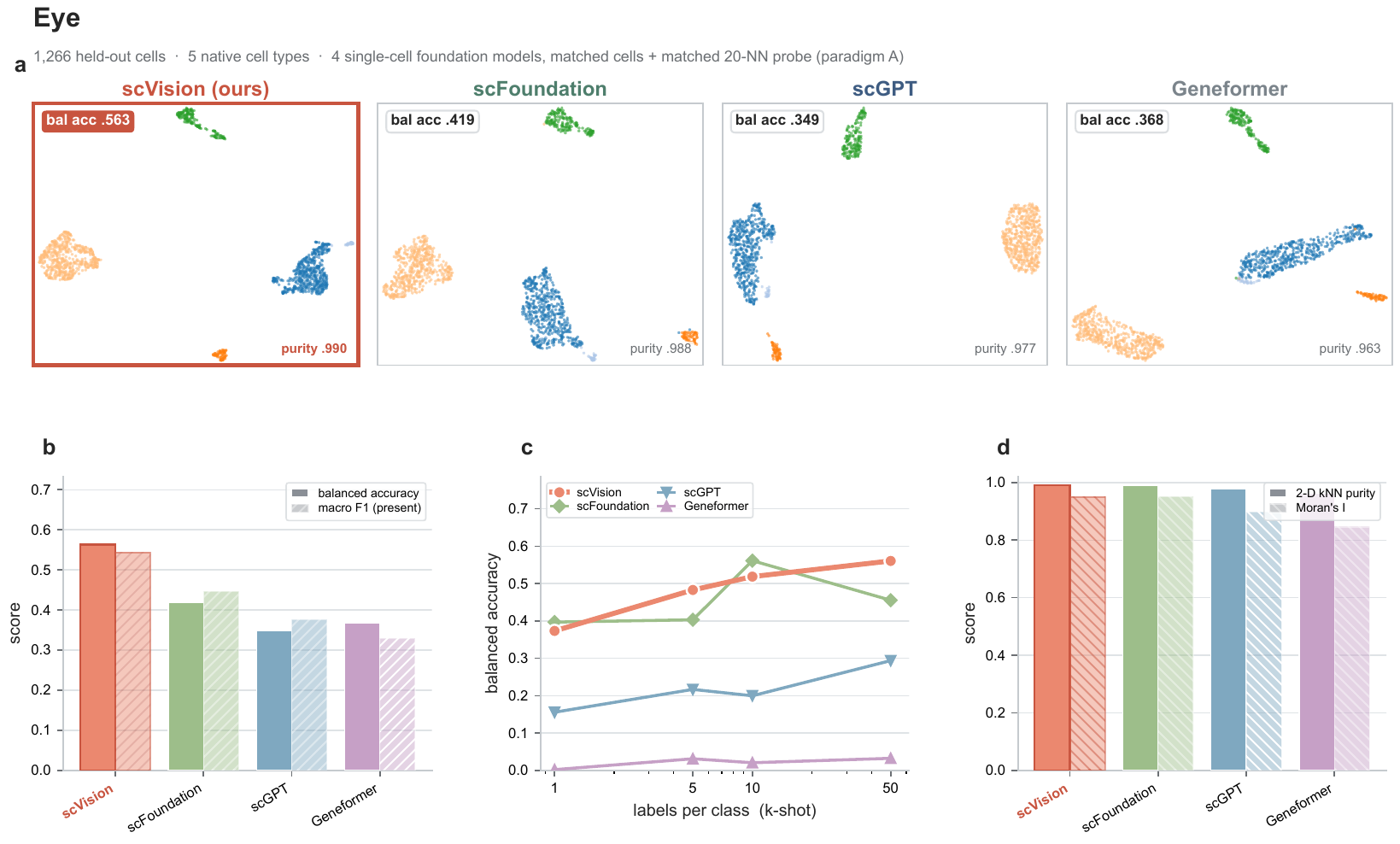}
\caption{\textbf{Eye: matched four-model benchmark.} 1{,}266 held-out cells across 5 native cell types, evaluated apples-to-apples across scVision, scFoundation, scGPT and Geneformer (identical cells and labels, one shared 20-NN probe, paradigm~A). (a)~Frozen-embedding UMAP per model, cells coloured by native cell type; badges give the matched balanced accuracy (top-left) and 2-D $k$NN purity (bottom-right), the coral frame marking scVision. (b)~Balanced accuracy and present-class macro-F1. (c)~Label efficiency ($k\in\{1,5,10,50\}$ labels per class). (d)~2-D $k$NN purity and Moran's $I$. scVision attains the highest matched balanced accuracy (0.563 vs.\ 0.419 for scFoundation). scVision also leads 2-D $k$NN purity (0.990).}
\label{fig:s_cs_eye_nl_9}
\end{figure*}

\begin{figure*}[tbp]\centering
\includegraphics[width=\textwidth]{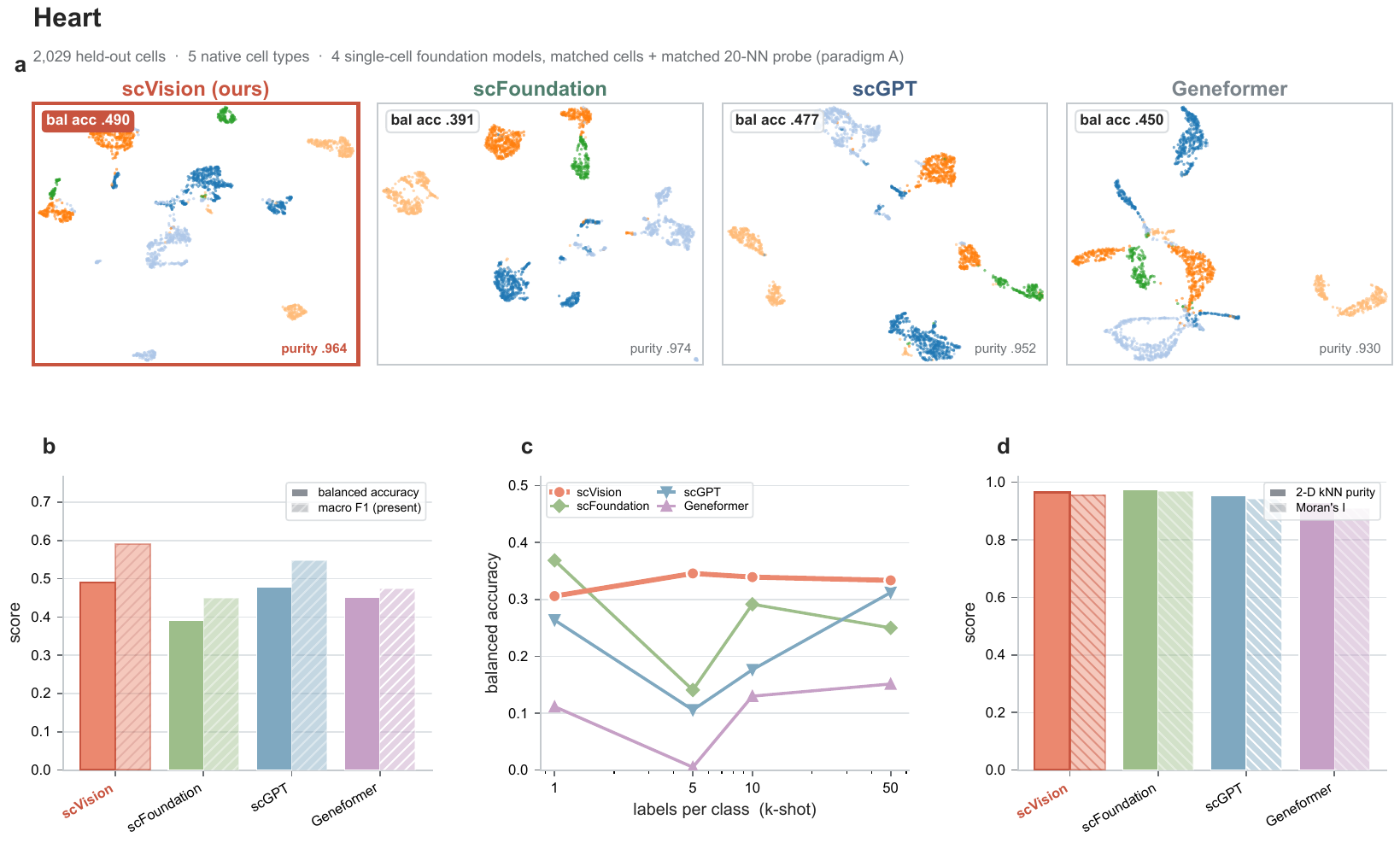}
\caption{\textbf{Heart: matched four-model benchmark.} 2{,}029 held-out cells across 5 native cell types, evaluated apples-to-apples across scVision, scFoundation, scGPT and Geneformer (identical cells and labels, one shared 20-NN probe, paradigm~A). (a)~Frozen-embedding UMAP per model, cells coloured by native cell type; badges give the matched balanced accuracy (top-left) and 2-D $k$NN purity (bottom-right), the coral frame marking scVision. (b)~Balanced accuracy and present-class macro-F1. (c)~Label efficiency ($k\in\{1,5,10,50\}$ labels per class). (d)~2-D $k$NN purity and Moran's $I$. scVision attains the highest matched balanced accuracy (0.490 vs.\ 0.477 for scGPT). On 2-D geometry scFoundation edges scVision on $k$NN purity (0.974 vs.\ 0.964).}
\label{fig:s_cs_heart_nl_6}
\end{figure*}

\begin{figure*}[tbp]\centering
\includegraphics[width=\textwidth]{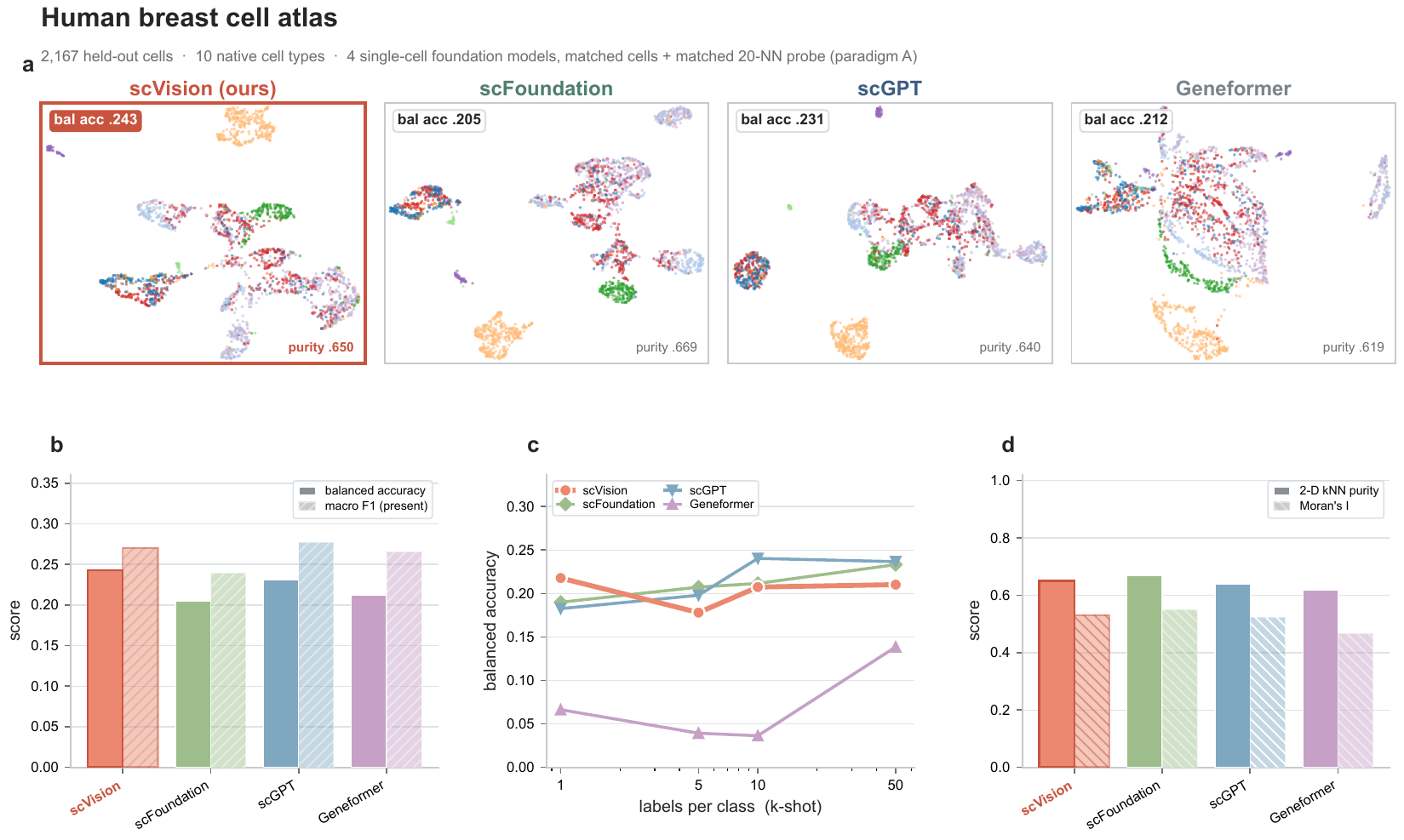}
\caption{\textbf{Human breast cell atlas: matched four-model benchmark.} 2{,}167 held-out cells across 10 native cell types, evaluated apples-to-apples across scVision, scFoundation, scGPT and Geneformer (identical cells and labels, one shared 20-NN probe, paradigm~A). (a)~Frozen-embedding UMAP per model, cells coloured by native cell type; badges give the matched balanced accuracy (top-left) and 2-D $k$NN purity (bottom-right), the coral frame marking scVision. (b)~Balanced accuracy and present-class macro-F1. (c)~Label efficiency ($k\in\{1,5,10,50\}$ labels per class). (d)~2-D $k$NN purity and Moran's $I$. scVision attains the highest matched balanced accuracy (0.243 vs.\ 0.231 for scGPT). On 2-D geometry scFoundation edges scVision on $k$NN purity (0.669 vs.\ 0.650).}
\label{fig:s_HBCA_integrated}
\end{figure*}

\begin{figure*}[tbp]\centering
\includegraphics[width=\textwidth]{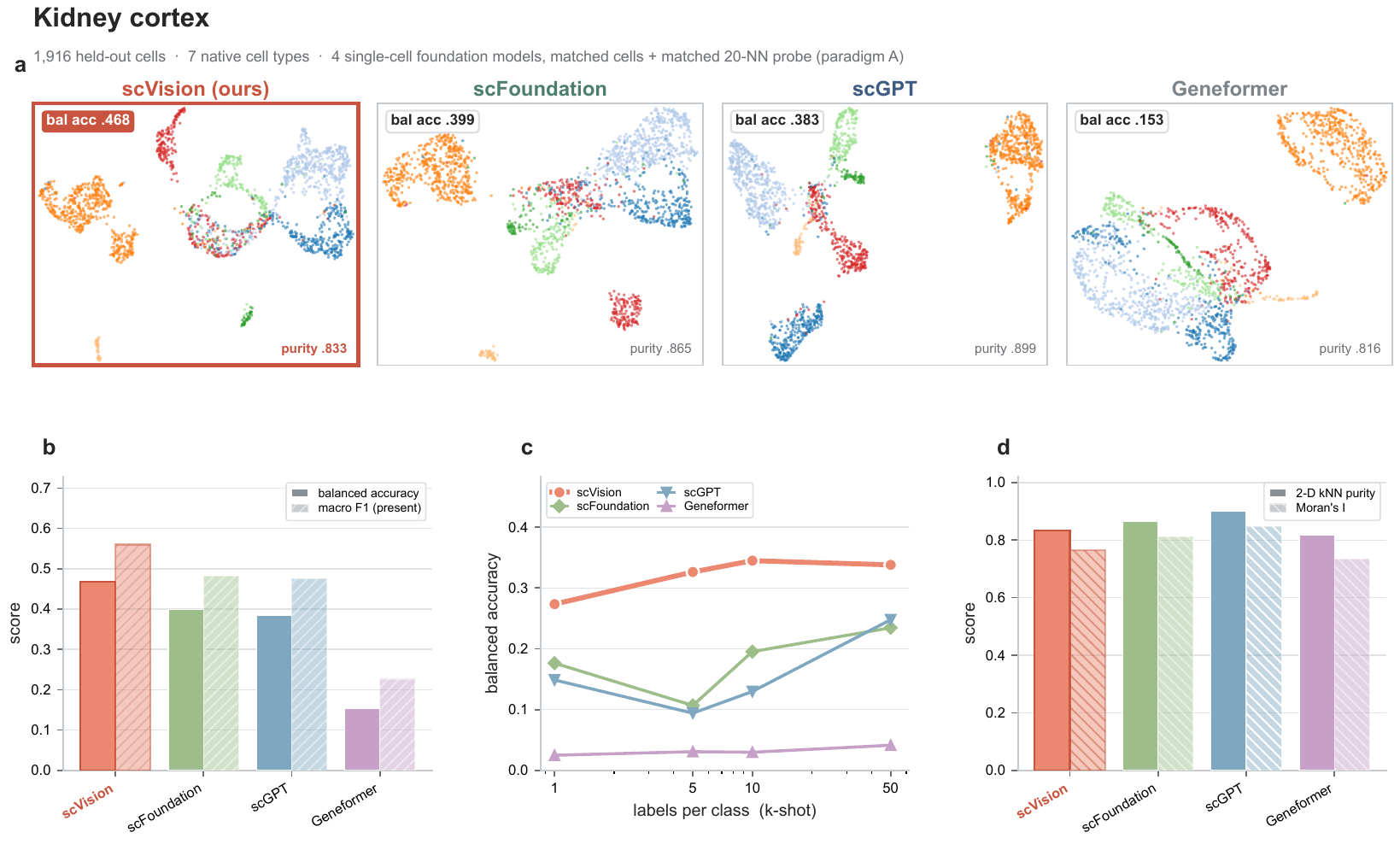}
\caption{\textbf{Kidney cortex: matched four-model benchmark.} 1{,}916 held-out cells across 7 native cell types, evaluated apples-to-apples across scVision, scFoundation, scGPT and Geneformer (identical cells and labels, one shared 20-NN probe, paradigm~A). (a)~Frozen-embedding UMAP per model, cells coloured by native cell type; badges give the matched balanced accuracy (top-left) and 2-D $k$NN purity (bottom-right), the coral frame marking scVision. (b)~Balanced accuracy and present-class macro-F1. (c)~Label efficiency ($k\in\{1,5,10,50\}$ labels per class). (d)~2-D $k$NN purity and Moran's $I$. scVision attains the highest matched balanced accuracy (0.468 vs.\ 0.399 for scFoundation). On 2-D geometry scGPT edges scVision on $k$NN purity (0.899 vs.\ 0.833).}
\label{fig:s_kidney_cortex}
\end{figure*}

\begin{figure*}[tbp]\centering
\includegraphics[width=\textwidth]{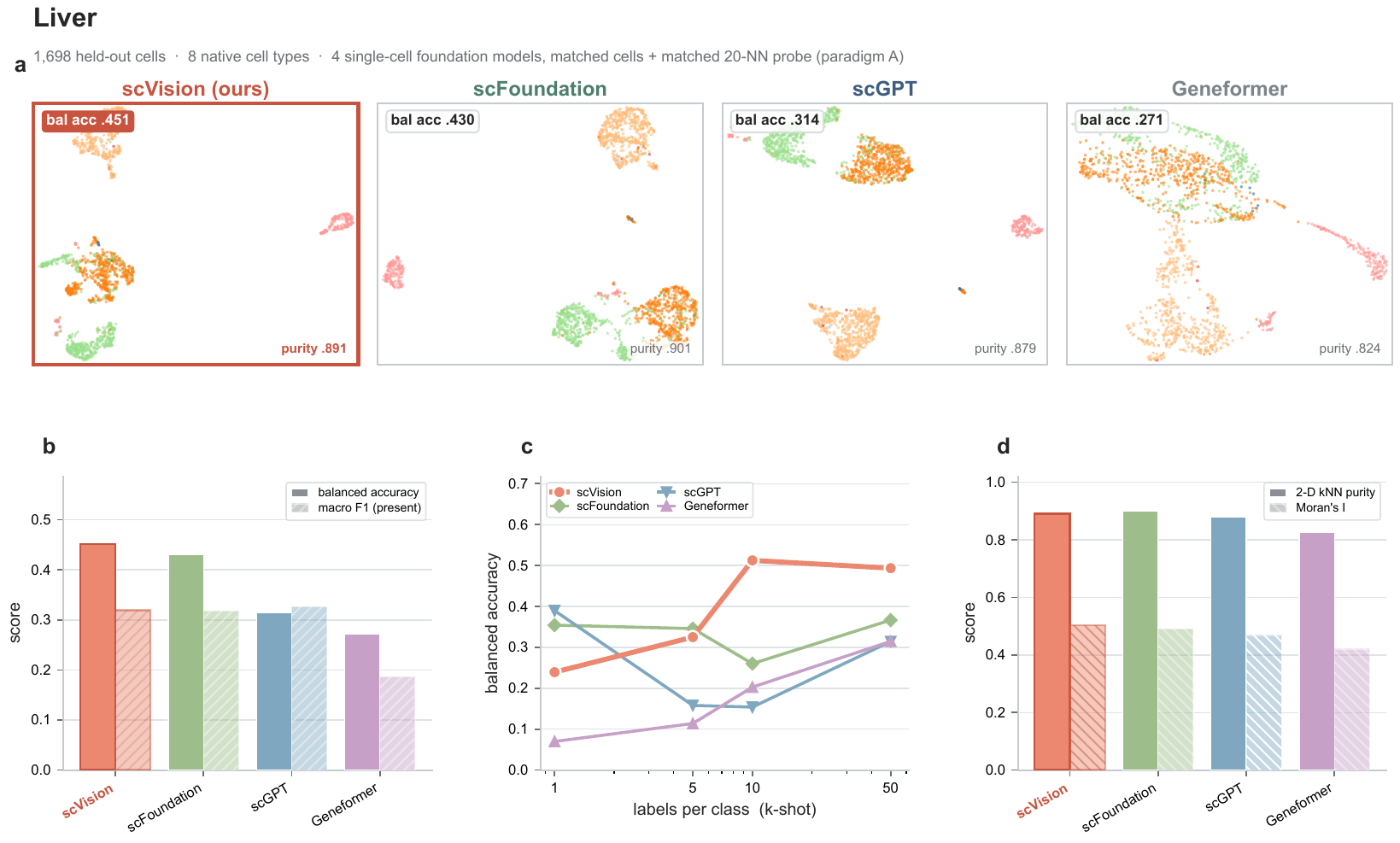}
\caption{\textbf{Liver: matched four-model benchmark.} 1{,}698 held-out cells across 8 native cell types, evaluated apples-to-apples across scVision, scFoundation, scGPT and Geneformer (identical cells and labels, one shared 20-NN probe, paradigm~A). (a)~Frozen-embedding UMAP per model, cells coloured by native cell type; badges give the matched balanced accuracy (top-left) and 2-D $k$NN purity (bottom-right), the coral frame marking scVision. (b)~Balanced accuracy and present-class macro-F1. (c)~Label efficiency ($k\in\{1,5,10,50\}$ labels per class). (d)~2-D $k$NN purity and Moran's $I$. scVision attains the highest matched balanced accuracy (0.451 vs.\ 0.430 for scFoundation). On 2-D geometry scFoundation edges scVision on $k$NN purity (0.901 vs.\ 0.891).}
\label{fig:s_liver}
\end{figure*}

\begin{figure*}[tbp]\centering
\includegraphics[width=\textwidth]{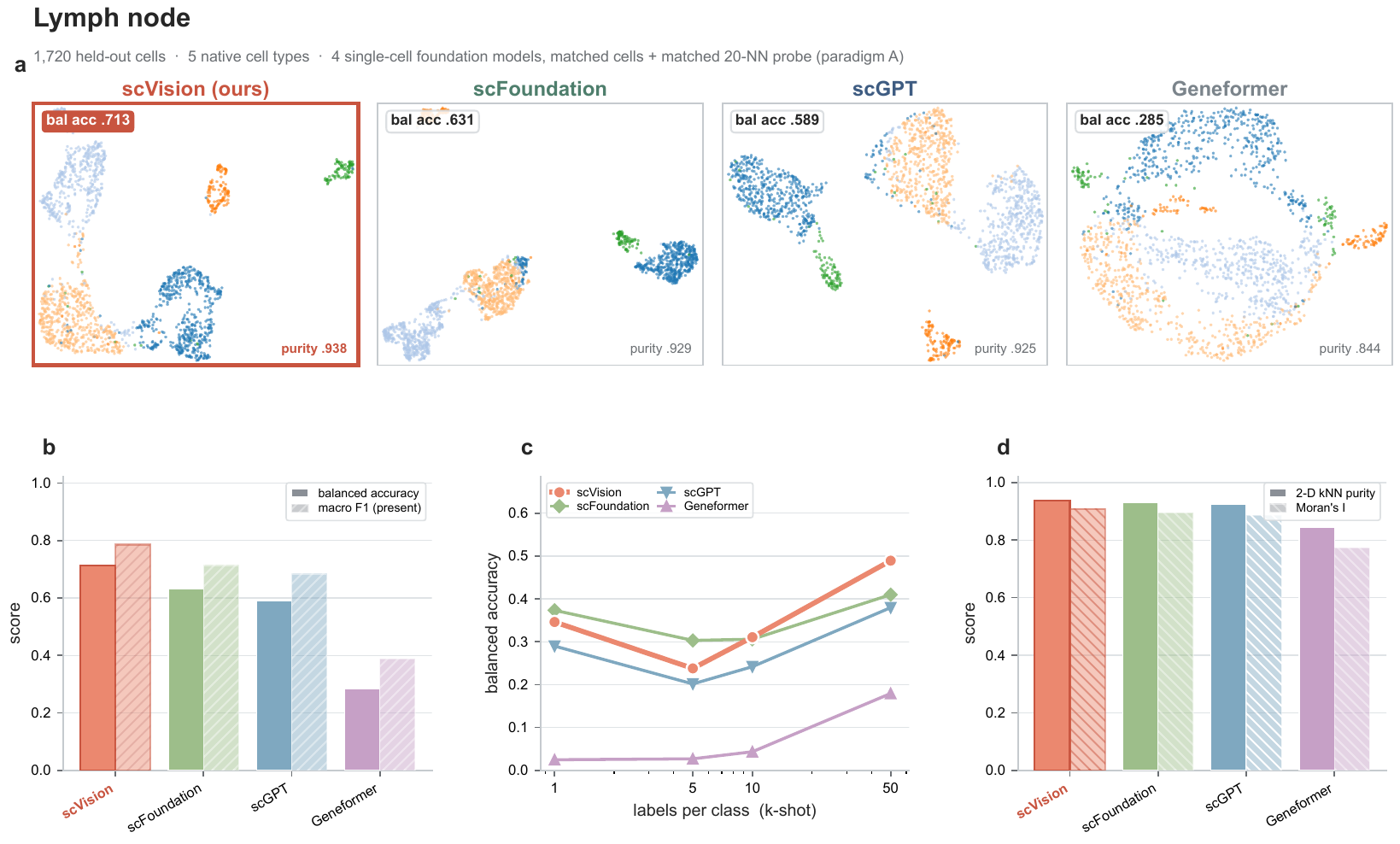}
\caption{\textbf{Lymph node: matched four-model benchmark.} 1{,}720 held-out cells across 5 native cell types, evaluated apples-to-apples across scVision, scFoundation, scGPT and Geneformer (identical cells and labels, one shared 20-NN probe, paradigm~A). (a)~Frozen-embedding UMAP per model, cells coloured by native cell type; badges give the matched balanced accuracy (top-left) and 2-D $k$NN purity (bottom-right), the coral frame marking scVision. (b)~Balanced accuracy and present-class macro-F1. (c)~Label efficiency ($k\in\{1,5,10,50\}$ labels per class). (d)~2-D $k$NN purity and Moran's $I$. scVision attains the highest matched balanced accuracy (0.713 vs.\ 0.631 for scFoundation). scVision also leads 2-D $k$NN purity (0.938).}
\label{fig:s_cs_lymphnode_nl_1}
\end{figure*}

\begin{figure*}[tbp]\centering
\includegraphics[width=\textwidth]{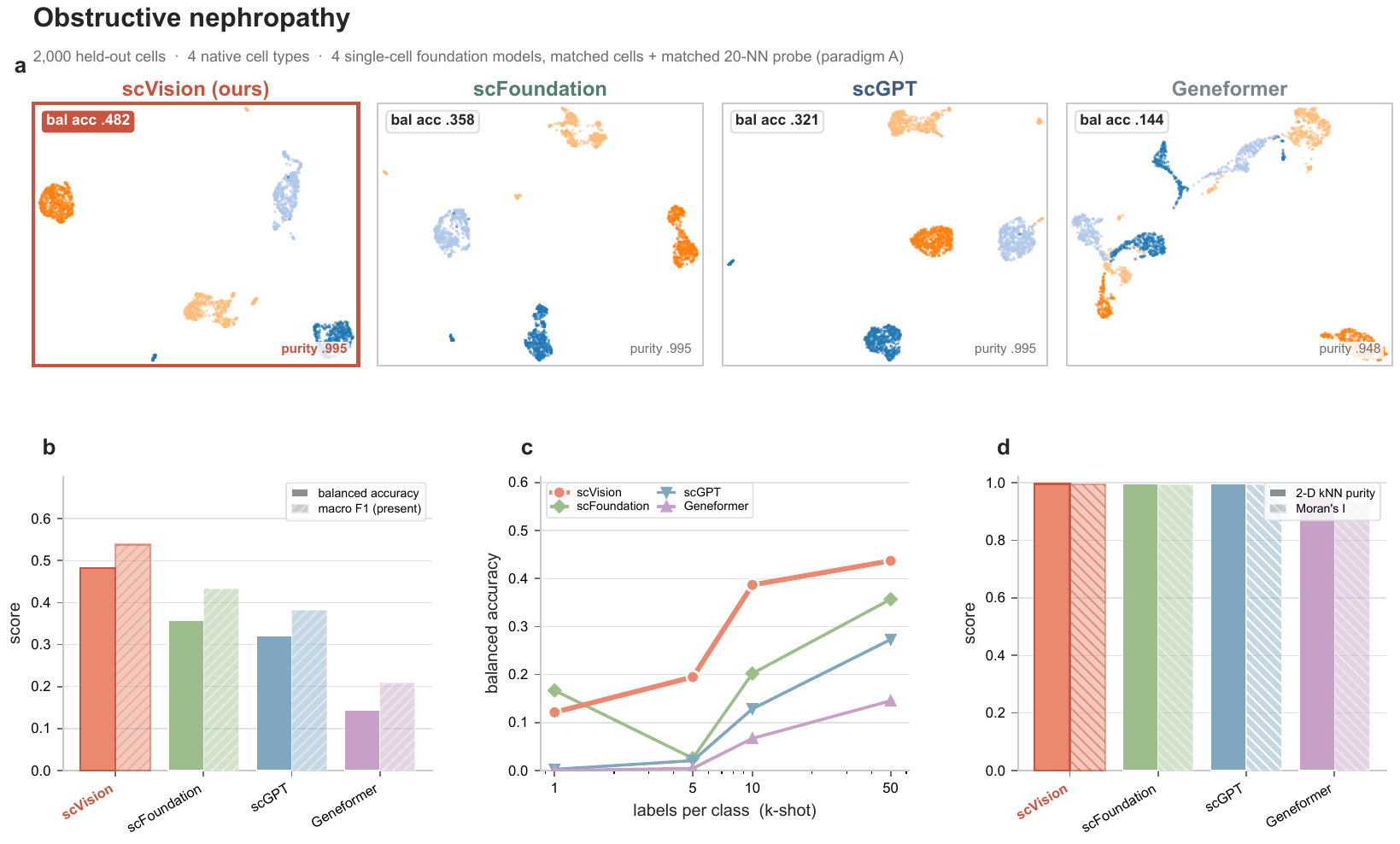}
\caption{\textbf{Obstructive nephropathy: matched four-model benchmark.} 2{,}000 held-out cells across 4 native cell types, evaluated apples-to-apples across scVision, scFoundation, scGPT and Geneformer (identical cells and labels, one shared 20-NN probe, paradigm~A). (a)~Frozen-embedding UMAP per model, cells coloured by native cell type; badges give the matched balanced accuracy (top-left) and 2-D $k$NN purity (bottom-right), the coral frame marking scVision. (b)~Balanced accuracy and present-class macro-F1. (c)~Label efficiency ($k\in\{1,5,10,50\}$ labels per class). (d)~2-D $k$NN purity and Moran's $I$. scVision attains the highest matched balanced accuracy (0.482 vs.\ 0.358 for scFoundation). scVision also leads 2-D $k$NN purity (0.995).}
\label{fig:s_obstructive_nephropathy}
\end{figure*}

\begin{figure*}[tbp]\centering
\includegraphics[width=\textwidth]{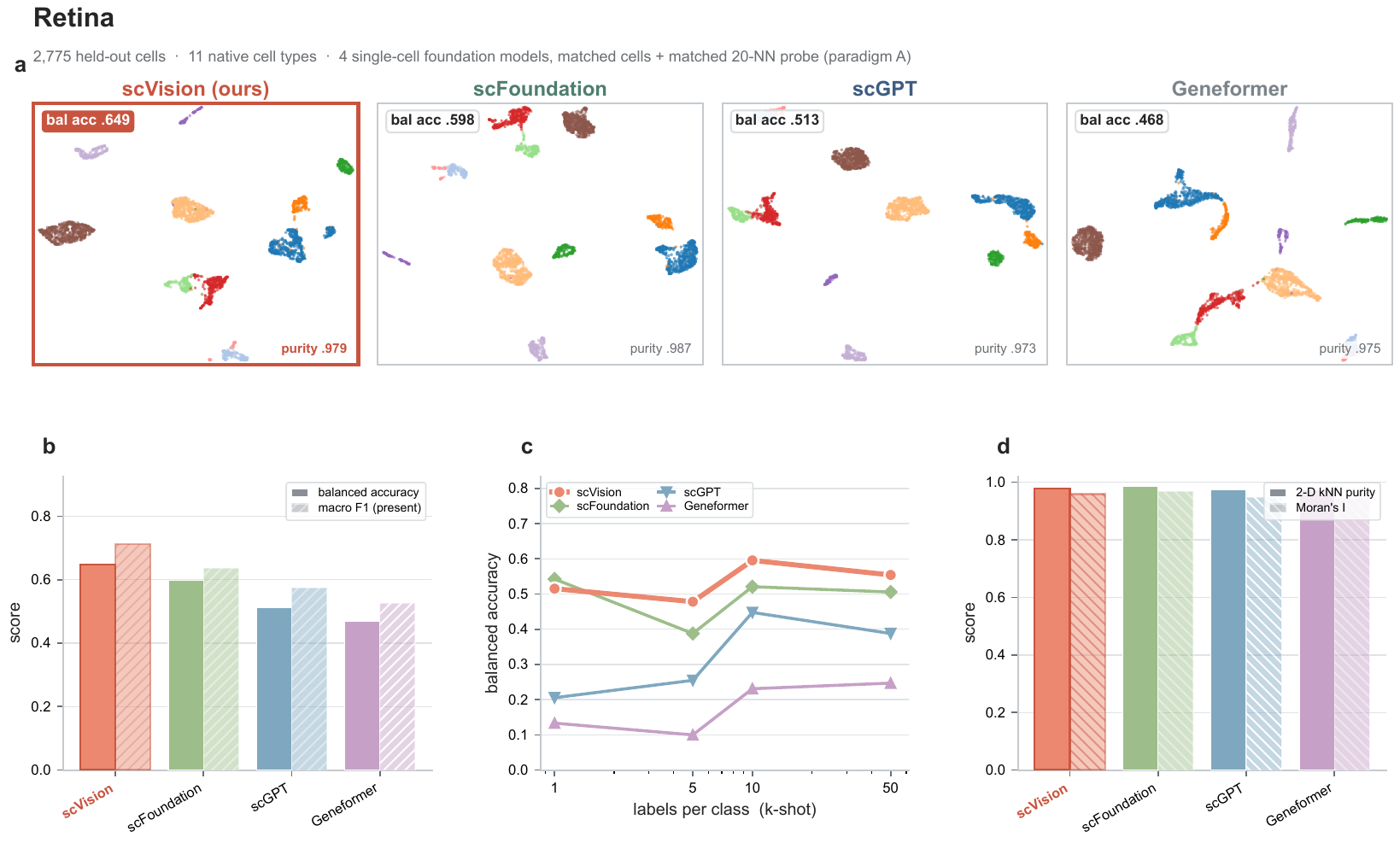}
\caption{\textbf{Retina: matched four-model benchmark.} 2{,}775 held-out cells across 11 native cell types, evaluated apples-to-apples across scVision, scFoundation, scGPT and Geneformer (identical cells and labels, one shared 20-NN probe, paradigm~A). (a)~Frozen-embedding UMAP per model, cells coloured by native cell type; badges give the matched balanced accuracy (top-left) and 2-D $k$NN purity (bottom-right), the coral frame marking scVision. (b)~Balanced accuracy and present-class macro-F1. (c)~Label efficiency ($k\in\{1,5,10,50\}$ labels per class). (d)~2-D $k$NN purity and Moran's $I$. scVision attains the highest matched balanced accuracy (0.649 vs.\ 0.598 for scFoundation). On 2-D geometry scFoundation edges scVision on $k$NN purity (0.987 vs.\ 0.979).}
\label{fig:s_retina}
\end{figure*}

\begin{figure*}[tbp]\centering
\includegraphics[width=\textwidth]{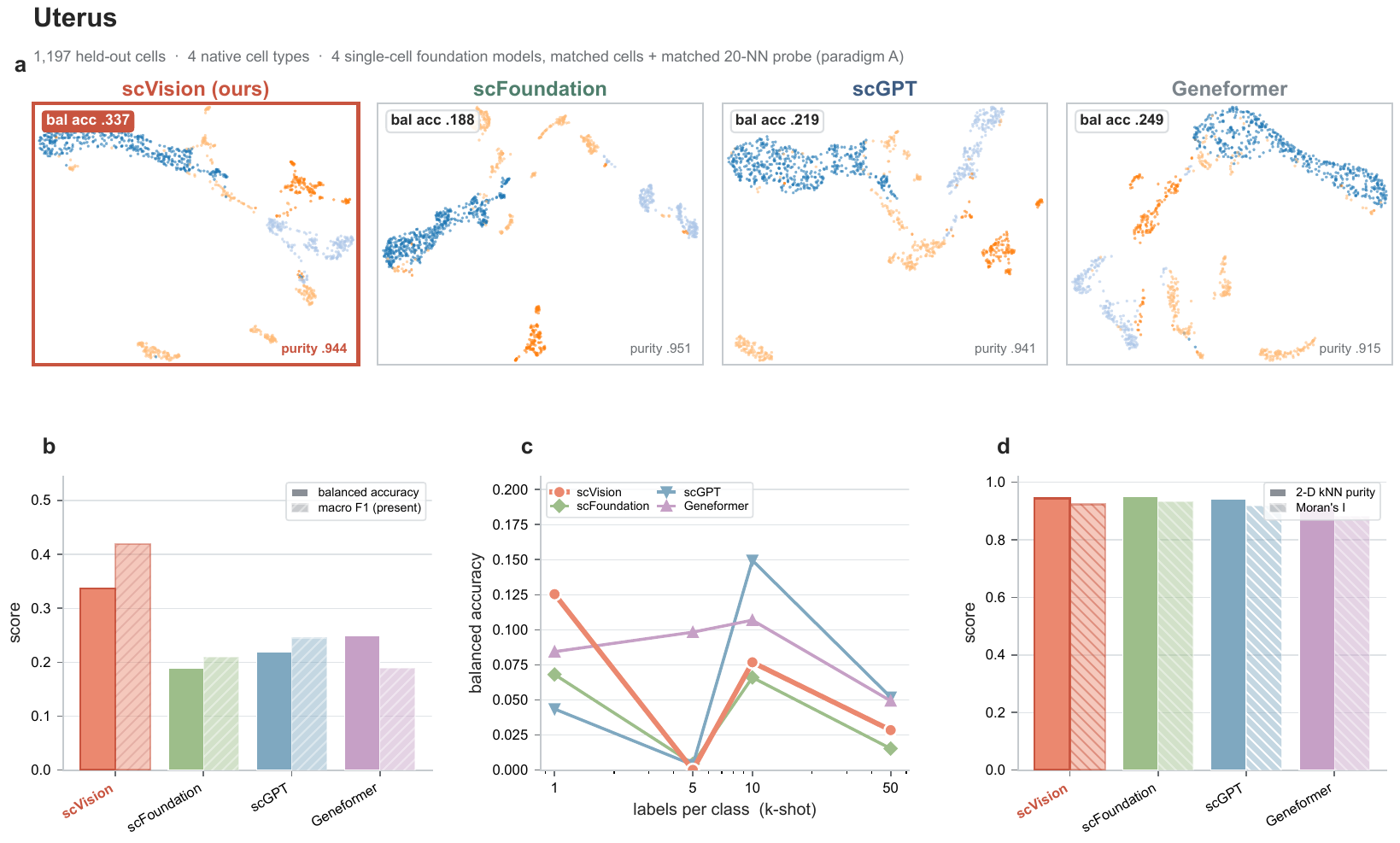}
\caption{\textbf{Uterus: matched four-model benchmark.} 1{,}197 held-out cells across 4 native cell types, evaluated apples-to-apples across scVision, scFoundation, scGPT and Geneformer (identical cells and labels, one shared 20-NN probe, paradigm~A). (a)~Frozen-embedding UMAP per model, cells coloured by native cell type; badges give the matched balanced accuracy (top-left) and 2-D $k$NN purity (bottom-right), the coral frame marking scVision. (b)~Balanced accuracy and present-class macro-F1. (c)~Label efficiency ($k\in\{1,5,10,50\}$ labels per class). (d)~2-D $k$NN purity and Moran's $I$. scVision attains the highest matched balanced accuracy (0.337 vs.\ 0.249 for Geneformer). On 2-D geometry scFoundation edges scVision on $k$NN purity (0.951 vs.\ 0.944).}
\label{fig:s_cs_uterus_nl_2}
\end{figure*}

\end{document}